\documentclass[12pt]{iopart}
\pdfoutput=1
\usepackage{color}
\usepackage{amssymb}
\usepackage[export]{adjustbox}
\usepackage{graphicx} 

\def\beqra{\begin{eqnarray}}
\def\eeqra{\end{eqnarray}}
\def\beq{\begin{equation}}
\def\eeq{\end{equation}}
\def\Hc{{\cal H}}

\def\etain{\eta_{in}}

\def\bV0{{\bf{V_0}}}
\def\re#1{(\ref{#1})}

\def\agt{~\mbox{\raisebox{-.6ex}{$\stackrel{>}{\sim}$}}~}
\def\alt{~\mbox{\raisebox{-.6ex}{$\stackrel{<}{\sim}$}}~}

\begin{document}

\title[Structure formation beyond shell-crossing]{Structure formation beyond shell-crossing: nonperturbative expansions and late-time attractors}

\author{Massimo Pietroni}
\vskip 0.3 cm
\address{
Dipartimento di Scienze Matematiche, Fisiche ed Informatiche dell'Universit\`a di Parma, Parco Area delle Scienze 7/a, 43124 Parma, Italy\\
and INFN, Sezione di Padova, via Marzolo 8, I-35131, Padova, Italy\\
}

\begin{abstract}
Structure formation in 1+1 dimensions is considered, with emphasis on the effects of shell-crossing. The breakdown of the perturbative expansion beyond shell-crossing is discussed, and it is shown, in a simple example, that the perturbative series can be extended to a {\it transseries} including nonperturbative terms. The latter converges to the exact result well beyond the range of validity of perturbation theory. The crucial role of the divergences induced by shell-crossing is discussed. They provide constraints on the structure of the transseries and act as a bridge between the perturbative and the nonperturbative sectors.
Then, we  show that the dynamics  in the deep multistreaming regime is governed by attractors.  In the case of simple initial conditions, these attractors coincide with the asymptotic configurations of the adhesion model, but in general they may differ. These results are applied to a cosmological setting, and an algorithm to build the attractor solution starting from the Zel'dovich approximation is developed. Finally, this algorithm is applied to the search of `haloes' and the results are compared with those obtained from the exact dynamical equations.
\end{abstract}

\maketitle
%\pacs{PACS: 98.80.-k, 95.35.+d }
\section{Introduction}
Cosmological structure formation is a nonlinear process. It can be described either in Eulerian space, where the  evolution equations contain terms quadratic in the deviations from a Friedmann-Robertson-Walker (FRW) Universe (the density contrast, the peculiar velocity, and all higher moments of the matter distribution functions), or in Lagrangian space, where the force is a nonlinear functional of the displacement field. These nonlinear terms are, in general, always present, even at the earliest stages of the perturbation growth. However, their effect is small at early times, and therefore it is believed that it can be treated perturbatively at sufficiently early times and large scales (for reviews on cosmological perturbation theory (PT) see \cite{PT} and \cite{Bernardeau:2013oda}). Indeed, nonlinear and nonperturbative are not synonyms, and one can expect that nonlinear quantities, such as, for instance, the variance of the density field, its power spectrum (PS), and so on, can be represented by a perturbative expansion in some parameter. 

On the other hand, it is well known that PT expansions, both Eulerian and Lagrangian, are doomed to fail at some point. In Eulerian PT, one truncates the full Boltzmann hierarchy to the continuity and the Euler equations, by setting to zero the velocity dispersion. This {\em single stream approximation} corresponds to assigning  a unique velocity value to any space-time point, and therefore breaks down after {\em shell-crossing}, when multiple streams of matter coexist in the same region of space.
In Lagrangian coordinates, the force becomes intrinsically nonlocal after shell-crossing, and therefore it cannot be represented as a local expansion in the displacement field and its derivatives.

In order to extend the PT approaches beyond shell-crossing, some information on the effect of multistreaming has to be provided. This is the approach followed in  the ``Coarse Grained PT" of  \cite{Pietroni:2011iz, Manzotti:2014loa}, recently implemented in the Time Renormalization Group (TRG) evolution equations \cite{Peloso:2016qdr, Noda:2017tfh,Nishimichi:2017gdq}. The idea there is to keep the full Boltzmann hierarchy, and to treat the higher moments as external sources to be computed from N-body simulations. These sources carry the fully nonperturbative information encoded in the small (UV) scales, including the shell-crossing effects. Earlier work along similar lines can be found in \cite{Pueblas:2008uv}.
A related approach is the Effective Field Theory of the Large Scale Structure \cite{Carrasco:2012cv}, in which the source terms are expanded in terms of the long wavelength fields via a derivative expansion, whose coefficients are then fit by comparing with simulations.

Besides these ``effective approaches'', the impact of shell-crossing, also in relation to the range of validity  of PT approaches (including resummations) deserves more consideration. In \cite{Valageas:2010rx}  the effect of shell-crossing on the PS was estimated by comparing the prediction of two dynamics which were exactly coincident before shell-crossing and drastically different afterwards.  Assuming gaussian initial conditions and Zel'dovich dynamics before shell-crossing, the difference is due to field configurations populating the tails of the distribution, and is therefore nonperturbative in the PT expansion parameter, the variance of the field. Valageas showed that these nonperturbative effects are subdominant with respect to PT corrections in a sizable range of wavenumbers $k$, rapidly increasing with redshift.

More recently, cosmological PT and shell-crossing have been studied in the context of 1+1 dimensional gravity \cite{McQuinn:2015tva,Taruya:2017ohk,McDonald:2017ths,Pajer:2017ulp, Rampf:2017jan}. This setup has the advantage that the Zel'dovich dynamics is exact before shell-crossing and therefore one can hope to single out and model the effects of multistreaming more neatly. 
In particular, in \cite{McQuinn:2015tva} and \cite{Pajer:2017ulp} the behavior of the PT expansion was considered, and it was shown that it converges to the Zel'dovich approximation before shell-crossing. The importance of the nonperturbative corrections was emphasized in both papers. In \cite{Taruya:2017ohk} and \cite{McDonald:2017ths} semi-analytical approaches to go beyond shell-crossing were proposed,  showing a clear improvement with respect to PT after first shell-crossing. It would be very interesting to extend the regime of validity of these approaches beyond second shell-crossing, possibly by some sort of resummation as envisaged in \cite{Taruya:2017ohk}. Finally, in \cite{Rampf:2017jan}, the analysis was extended beyond 1+1 dimensions in a controlled perturbative expansion valid up to shell-crossing.

In this paper we further investigate structure formation in 1+1 gravity providing, we believe, some insight on two aspects. First, we study the impact of shell-crossing on the Eulerian PT expansion. We regulate the divergence of the density contrast induced by the singularity of the Lagrangian to Eulerian mapping after shell-crossing and study the analyticity properties of the density field with respect to the regulator. Then, by taking cosmological averages of the density field, we show the crucial role of the divergent behavior in linking the PT expansion with the nonperturbative contributions. The two sectors are then, maybe surprisingly, intimately related, at least in our toy example. 

Then we take into account the real dynamics, and consider the regime of deep multistreaming, with the aim of finding a ``resummation'' of the shell-crossing effects. We show explicitly that the evolution equations have an attractor in that regime. For simple initial conditions it coincides with the prescription of the ``adhesion model" for structure formation \cite{Gurbatov:1989az,Dubrulle:1994psg,Bernardeau:2009ab,Valageas:2010uh}, while for cosmological initial conditions this identification is not so immediate. We show how to compute this attractor from the Zel'dovich approximation, and discuss how to implement this procedure in a cosmological context, for instance, to predict halo positions and sizes.

The paper is organised as follows. In section \ref{npterms} we classify the divergences induced by shell-crossing on the Eulerian density contrast, and then focus on the statistical distribution of maxima of this field. We derive an expansion for this quantity, which can be written as the sum of perturbative and nonperturbative terms, including a divergent one, and show that it converges to the exact result well beyond the  PT range. In section \ref{exdyn} we introduce the exact equations of motion for the displacement field, and show the non-local nature of the force term after shell-crossing. Then, in section \ref{numsol} we solve the equation numerically for a simple set of initial condition and in section \ref{pattr} we show analytically the existence of an attractor in the multistreaming regime, and discuss its properties. In section \ref{cosmosym} we present a cosmological simulation (in 1+1 dimensions!) and implement an algorithm to simulate the attractor solution starting from the Zel'dovich approximation. We show how this algorithm can be used to predict the location and sizes of ``haloes" in the real simulation. Finally, in section \ref{out} we summarize our findings and discuss some possible future developments.

\section{Shell-crossing and non-perturbative terms}
\label{npterms}
\subsection{Density contrast after shell-crossing}
The mapping between the initial (Lagrangian) position of a given particle, $q$, and its later (Eulerian) position  $x$ at a conformal time $\tau$ is given by
\beq
x(q,\tau)=q+\Psi(q,\tau)\,,
\eeq
$\Psi(q,\tau)$ is the displacement field, $\Psi(q,\tau=0)=0$.
Assuming a uniform density at $\tau=0$, the density contrast in Eulerian space, $\delta(x,\tau)=-1+\rho(x,\tau)/\bar\rho$, is given by
\beq
\delta(x,\tau)=\int dq\;\delta_D\left(q+\Psi(q,\tau) - x\right)-1\,.
\label{0map}
\eeq
Shell-crossing induces divergences in Eulerian space. These are, however, unphysical, as they are  regulated in any physical realization, either by  pressure, if dark matter is not perfectly cold,  or by the finite space resolution. Therefore, in order to discuss the impact of shell-crossing on the Eulerian quantities, and on the PT expansion, it is more physical to consider a smoothened version of \re{0map},
\beq
\bar\delta(\bar x,\tau;\sigma)\equiv \int\frac{dx}{\sqrt{2\pi\sigma^2}}\; e^{-\frac{(x-\bar x)^2}{2 \sigma^2}} \,\delta(x,\tau)= \int\frac{dq}{\sqrt{2\pi\sigma^2}}\,e^{-\frac{( q+\Psi(q,\tau)-\bar x)^2}{2 \sigma^2}} -1\,,
\label{Gmap}
\eeq
and to study the analyticity properties in the $\sigma^2\to 0$ limit. The integral in \re{Gmap} is dominated by regions of $q$ around the values $q_i(\bar x)$, such that 
\beq
 x(q,\tau)-\bar x=  q+\Psi(q,\tau)-\bar x=0\,.
\label{S}
\eeq  
We can now classify the different types of small $\sigma^2$ behavior by considering the derivatives of of $\Psi(q,\tau)$ with respect to $q$ at the roots of eq.~\re{S}, that we will indicate with $q_i(\bar x)$, and will number in increasing values by the natural index $i$, that is $q_i(\bar x)\le q_2(\bar x)\le q_3(\bar x) \cdots$.

\begin{figure}[t]
\centering 
\includegraphics[width=.65\textwidth,clip]{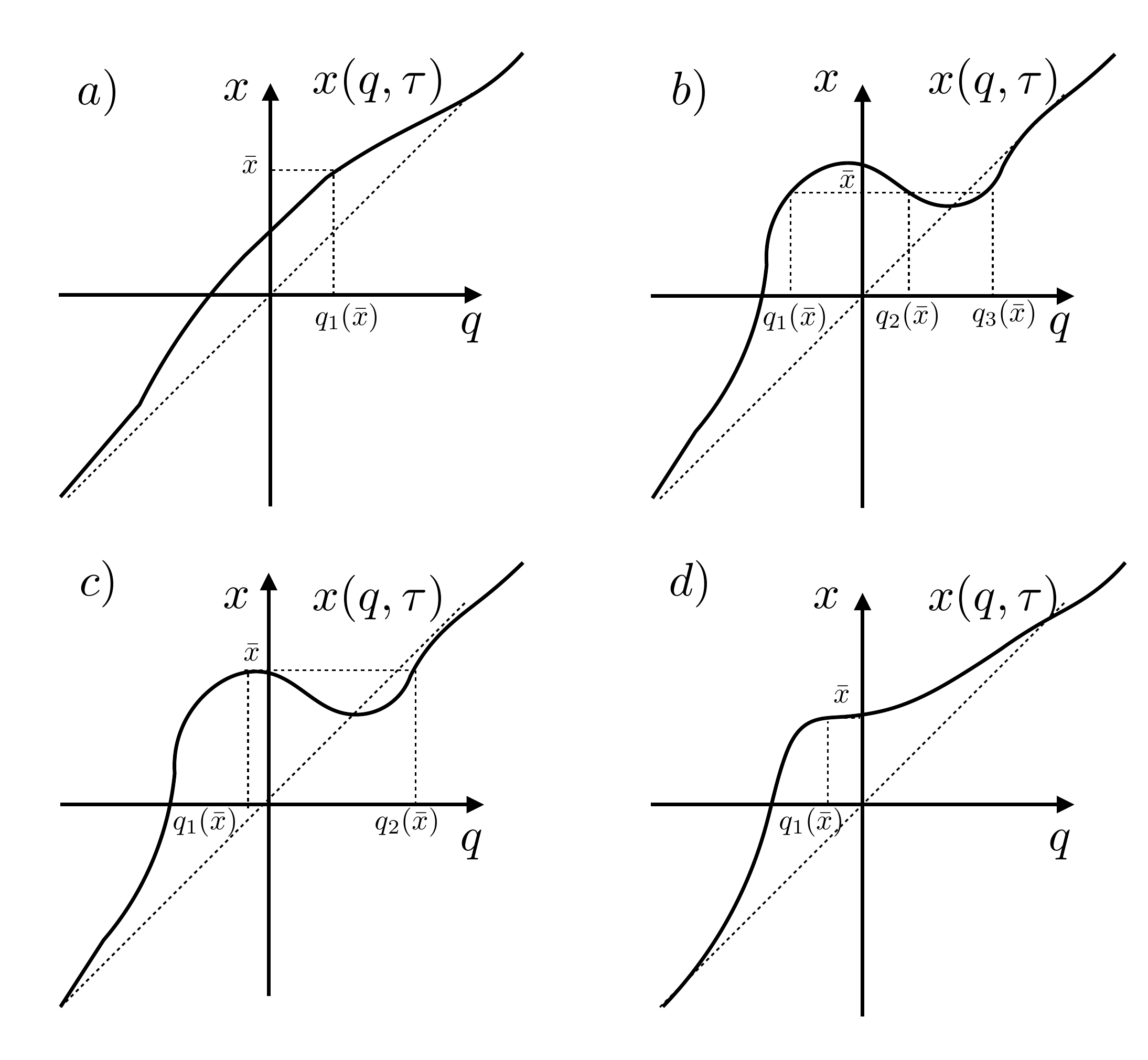}
\caption{Various examples of correspondence between Lagrangian and Eulerian space. $a)$ shows a situation in which there is a single stream everywere in Eulerian space. In $b)$ there are three streams in $\bar x$, corresponding to the three roots $q_i(\bar x)$. In $c)$, $x'(q,\tau)=0$ in $q_1$: this is the border between   case $a)$ and $b)$. Finally, in panel $d)$, point $q_1(\bar x)$ corresponds to a point of first shell crossing $x''(q_1(\bar x),\tau)=0$, $x'''(q_1(\bar x),\tau)>0$ .}
\label{configs}
\end{figure}

The first case to be considered is when (primes denote derivatives with respect to $q$),\\
{\bf 1) $ x'(q_i(\bar x),\tau)  \neq 0$,  for any $q_i(\bar x)$ (as in Fig.~\ref{configs} a) and b)):}

\noindent
Around each of the $q_i's(\bar x)$, we can compute the integral in the steepest descent approximation, 
\beqra
&&\bar\delta(\bar x,\tau;\sigma)\simeq \sum_i \int\frac{dq}{\sqrt{2\pi\sigma^2}}\; e^{-\frac{\left(1+ \Psi'(q_i(\bar x),\tau) \right) ^2}{2 \sigma^2} \left(q-q_i(\bar x)\right)^2} -1\,,\nonumber\\
&&\qquad\quad\;\;\,= \sum_i \frac{1}{\left| 1+ \Psi'(q_i(\bar x),\tau) \right|}-1\,.
\label{1d}
\eeqra
The approximation at the first line is exact in the $\sigma^2\to 0$ limit, that is, when $\sigma$ is much smaller than the distance between roots, and leads to the $\sigma^2$-independent result of the last line.

The requirement that $x(q,\tau) $ is a continuous function of $q$ taking all values from $-\infty$ to $+\infty$
guarantees that for any $\bar x$ there are an odd number of $q_i(\bar x)$ roots, corresponding to the number of streams in $\bar x$, and that, by ordering the $q_i(\bar x)$'s in increasing values, the signs of $x'(q_i(\bar x),\tau)= 1+ \Psi'(q_i(\bar x),\tau)$ are alternating, starting from positive. 

In the {\it single stream} case, Fig.~\ref{configs} a), there is only one root, $q(\bar x)$, therefore $1+ \Psi'(q(\bar x),\tau) >0$ and the density contrast is given by
\beq
\bar\delta(\bar x,\tau;\sigma) = -\frac{ \Psi'(q(\bar x),\tau)}{1+ \Psi'(q(\bar x),\tau)}\,.
\label{ssc}
\eeq
The denominator can be expanded perturbatively in $\Psi'(q(\bar x),\tau)$, and the series is guaranteed to converge since $\left| \Psi'(q(\bar x),\tau)\right| <1$. Moreover, the single stream hypothesis ensures that the mapping between Lagrangian and Eulerian space can be inverted perturbatively, giving the PT expansion for $\bar\delta(\bar x,\tau;\sigma)$ \cite{McQuinn:2015tva,Pajer:2017ulp}.

Next, we consider the case in which at one of the $q_i(\bar x)$ roots one  has  \\
{\bf 2) $ x'(q_i(\bar x),\tau)=0$, $x''(q_i(\bar x),\tau)\neq 0$ (see root $q_1$ in Fig.~\ref{configs} c)):}

\noindent
Around that point, we have
\beqra
&&\bar\delta(\bar x,\tau;\sigma)\simeq  \int\frac{dq}{\sqrt{2\pi\sigma^2}}\; e^{-\frac{\left(\Psi''(q_i(\bar x),\tau) \right) ^2}{8\, \sigma^2} \left(q-q_i(\bar x)\right)^4} +``\mathrm{other\;roots}"-1\,,\nonumber\\
&&\qquad\quad\;\;\, = \frac{\Gamma\left(\frac{1}{4}\right)}{\pi^{1/2}|\Psi''(q_i(\bar x),\tau)|^{1/2} \left(8 \,\sigma^2\right)^{1/4}}+``\mathrm{other\;roots}"-1\,,
\label{d2}
\eeqra
where $\Gamma(x)$ is Euler's gamma function, and with ``other roots" we indicate the contribution from the remaining (at least, one) roots. 

Comparing to \re{1d} we notice some very relevant differences. First of all, the divergence in the $\sigma^2 \to 0$ limit signals that the mapping is singular at $q_i(\bar x)$, whereas it is invertible on a finite interval around any of the roots in \re{1d}. Moreover, the non-analyticity in $\sigma^2$ signals the onset of non-locality in lagrangian space, as the present situation occurs when two new streams are reaching $\bar x$. Finally, eq.~\re{d2} clearly does not admit a PT expansion in powers of $\Psi$ or its derivatives. 

The third case we will discuss is: \\
{\bf 3) $x'(q_i(\bar x),\tau)=x''(q_i(\bar x),\tau)= 0$, $x^{\prime\prime\prime}(q_i(\bar x),\tau)\neq 0$  (as in Fig.~\ref{configs} d))}:

\noindent
This case (if $x^{\prime\prime\prime}(q_i(\bar x),\tau)>0$) corresponds to $q_i(\bar x)$ being a point of first shell-crossing. 
Now, 
\beqra
&&\bar\delta(\bar x,\tau;\sigma)\simeq  \int\frac{dq}{\sqrt{2\pi\sigma^2}}\; e^{-\frac{\left(\Psi'''(q_i(\bar x),\tau) \right) ^2}{72\, \sigma^2} \left(q-q_i(\bar x)\right)^6} +``\mathrm{other\;roots}"-1\,,\nonumber\\
&&\qquad\quad\;\;\, = \frac{\Gamma\left(\frac{1}{6}\right)}{\pi^{1/2}|\Psi'''(q_i(\bar x),\tau)|^{1/3} \left(9 \,\sigma^2\right)^{1/3}}+``\mathrm{other\;roots}"-1\,,
\label{d4}
\eeqra
which, as expected, is more singular than \re{d2} in the $\sigma^2\to 0$ limit. 

Increasing the degree of ``flatness'' of $x(q,\tau)$ in Lagrangian space leads to more and more singular behavior  for the density contrast in Eulerian space, which, in turn, decrees the failure of the Eulerian PT expansion.

In  order to perform a concrete computation we will consider the case in which the point $x=q=0$ is a maximum of the initial density field, 
\beq
\left.\frac{\partial \delta(x,\tau_{in})}{\partial x}\right|_{x=0} =0\,,\qquad \left.\frac{\partial^2 \delta(x,\tau_{in})}{\partial x^2}\right|_{x=0} < 0\,.
\eeq
 In this case,  the initial displacement field has vanishing second derivative in $q=0$ and can be written as (see eq.~\re{ssc}),
\beq
\Psi(q,\tau_{in}) = -qA_{in}+\frac{q^3}{\bar q^2} \,,
\eeq
where $\bar q$ is a fixed length scale and, when $A_{in} \ll 1$, we can identify,
\beqra
&& A_{in}= \delta(x=0,\tau_{in})\,,\nonumber\\
&& \frac{1}{\bar q^2}= -\frac{1}{6}\delta''(x=0,\tau_{in})\,.
\label{2dc}
\eeqra
In the Zel'dovich approximation (see next section) the displacement field evolves as 
\beq
\Psi(q,\tau)=\frac{D(\tau)}{D(\tau_{in})}\Psi(q,\tau_{in}) = - A(\tau)q+ \frac{q^3}{\bar q^2(\tau)}\,,
\label{qpsi}
\eeq
where $D(\tau)$ is the linear growth factor ($D(\tau)=a(\tau)$ in the Einstein de Sitter cosmology) and we have defined
\beqra
&&A(\tau)=\frac{D(\tau)}{D(\tau_{in})} A_{in} = \delta_L(x=0,\tau) \,,\nonumber\\
&&\frac{1}{\bar q^2(\tau)} =\frac{D(\tau)}{D(\tau_{in})}\frac{1}{\bar q^2} = -\frac{1}{6}\delta_L''(x=0,\tau)\,,
\label{At}
\eeqra
where $\delta_L(x=0,\tau) $ is the linearly evolved density field.

The root equation \re{S}
 in $x=0$ has now one real solution, $q_1(0)=0$ for $A(\tau)<1$, and three solutions,
\beq
q_{1,3}(0)= \mp\; \bar q(\tau) \,\sqrt{A(\tau)-1}\,,\qquad\qquad q_2(0)=0\,,
\eeq
for $A(\tau)>1$. For $A=1$ the three real solutions coincide.

The regularized density contrast, \re{Gmap}, is now
\beqra
&&\!\!\!\!\!\!\!\!\!\!\!\!\!\!\!\!\!\!\!\!\! \!\!\!\!\!\!\!\!\!\!\!\!\!\!\bar\delta(\bar x, \tau;\sigma)=\int\frac{dq}{\sqrt{2 \pi \sigma^2}} \;e^{-\frac{1}{2\sigma^2}\left(q(1-A) + \frac{q^3}{\bar q^2}-\bar x\right)^2}-1 = \int\frac{dy}{\sqrt{2 \pi}} \;e^{-\frac{1}{2}\left(y(1-A) +  \epsilon^2 y^3-\bar x\right)^2}-1\,,\nonumber\\
&&\!\!\!\!\!\!\!\!\!\!\!\!\!\!\,\,\equiv\Delta(\bar x, A(\tau);\epsilon(\tau))
\label{intA}
\eeqra
where we have defined the effective regularization parameter, given by the ration between the smoothing scale in Eulerian space, $\sigma^2$, and the scale setting the distance between the roots in Lagrangian space, $\bar q^2$, 
\beq
\epsilon^2(\tau)\equiv \sigma^2/\bar q^2(\tau)\,.
\eeq Notice that, if one sends the scale $\bar q$ to infinity, the density contrast in $\bar x=0$ diverges, at $A=1$, even if the smoothing parameter, $\sigma^2$, is kept fixed. 
In other words, the would-be short-distance (local) divergence in Eulerian space around $\bar x=0$ in controlled by the large distance (nonlocal) structure in Lagrangian space. 

Performing the integral in \re{intA} in the $\epsilon\to 0$ limit gives, in $\bar x=0$,
\beqra
&&\!\!\!\!\!\qquad\qquad\quad\,\frac{1}{1-A} -1 = \frac{A}{1-A}\qquad\qquad \qquad\qquad\qquad\quad\;\;\;(A<1)\nonumber\\
&&\!\!\!\!\!\!\!\!\!\!\!\!\! \!\!\!\!\!\!  \!\!\!\!\! \Delta(\bar x=0, A;\epsilon)= \,\quad\frac{\Gamma\left(\frac{1}{6}\right)}{ 2^{1/3} 3 \,\pi^{1/2}} \frac{1}{(\epsilon^2)^{1/3}}-1 \qquad\qquad\qquad\qquad\qquad\quad\;\; (A=1)\nonumber\\
&&\qquad\quad\quad\;\,\frac{1}{2(A-1)}+\frac{1}{A-1}+\frac{1}{2(A-1)}-1=\frac{3-A}{A-1}\qquad (A>1)\,,
\label{ds}
\eeqra
where, for the $A>1$ case, we have shown explicitly the contribution of each root. Before shell-crossing ($A<1$) the solution can be expanded perturbatively in the ``time" parameter $A$ around $A=0$, and the series is guaranteed to converge has it has convergence radius $|A|=1$.  For $A=1$ we recover the result \re{d4}, and 
the singular dependence $( \Psi^{\prime\prime\prime}(0,\tau_{sc})\sigma^2)^{-1/3}\sim(\epsilon^2) ^{-1/3}$.

\subsection{Non-perturbative cosmological expansion}
To see how the singular behavior at shell-crossing manifests itself in cosmological averages, and its impact on the cosmological PT expansion, we perform a Gaussian average of  \re{intA} with respect to $A$. The quantity we obtain corresponds to the average of the nonlinear field evaluated at positions corresponding to a maximum of the linear density field. Moreover, to simplify the computation, we keep the second derivative of the field fixed at the value \re{At}. We get,
\beqra
&&\langle \Delta(\bar x=0, A;\epsilon)
\rangle_{\sigma_A}=\int\frac{d A}{\sqrt{2\pi \sigma_A^2}}\;e^{-\frac{A^2}{2\sigma_A^2}} \int\frac{dy}{\sqrt{2 \pi}} \;e^{-\frac{1}{2}\left(y(1-A) +  \epsilon^2 y^3\right)^2}-1\,,\nonumber\\
&&\qquad\qquad\quad\quad\;\;\;\;\,\,\,= \int\frac{dy}{\sqrt{2 \pi}} \frac{e^{-\frac{y^2 (1+\epsilon^2 y^2)^2}{2\left(1+y^2 \sigma_A^2\right)}}}{\sqrt{1+y^2 \sigma_A^2}}-1\,,
\label{it}
\eeqra
where, from \re{At} we have
\beq
\sigma_A^2 = \langle \delta_L^2(\tau) \rangle\,,
\eeq
which will play the role of the PT expansion parameter in what follows.
Being constrained to run over maxima of the density field, the average \re{it} does not vanish. 

A power series  in $\sigma_A^2$,
\beq
\sum_{n=1}^{N_\mathrm{max}} c_{2n}(\epsilon) \sigma_A^{2n}\,,
\label{PTd}
\eeq
clearly is not enough to represent \re{it}. By Taylor  expanding in $\sigma_A^2$ around $\sigma_A^2=0$, one easily realizes that he $c_{2n}(\epsilon)$'s, are all finite for $\epsilon \to 0$, as the corresponding integrals are all cut off by the $e^{-y^2/2}$ factor. On the other hand, eq.~\re{it} diverges logarithmically for $\epsilon=0$, with a coefficient which is nonperturbative (and non analytic) in $\sigma_A^2$. This can be seen by considering the regime $\epsilon^2\ll \sigma_A^2$, and evaluating the contribution to the integral from the region $ 1/\sigma_A^2 \ll y^2 \ll 1/\epsilon^2$, in which it can be approximated as
\beq
\sim 2  e^{-\frac{1}{2 \sigma_A^2}} \int_{1/\sigma_A}^{1/\epsilon}\frac{dy}{\sqrt{2 \pi y^2 \sigma_A^2}}\sim \log(\sigma_A^2/\epsilon^2)\frac{e^{-\frac{1}{2 \sigma_A^2}} }{\sqrt{2\pi \sigma_A^2}}\,.
\label{logdiv1}
\eeq
Indeed, as shown in \ref{transerie}, the integral can be represented with a  {\it transseries} with the following structure (for an introduction to transseries and resurgence, see \cite{Aniceto:2018bis} and references therein),
\beqra
&& \!\!\!\!\!\!  \!\!\!\!\!\!  \!\!\!\!\!\! \!\!\!\!\!\!  \!\!\!\! \langle \Delta(\bar x=0, A;\epsilon)
\rangle_{\sigma_A}\simeq \sum_{n=1}^{N_{\mathrm{max}}} c_{2n}(\epsilon) \sigma_A^{2n} \nonumber\\
&&\qquad\quad +\frac{e^{-\frac{1}{2 \sigma_A^2}}}{ \sqrt{2\pi \sigma_A^2}}  \left(\log(\sigma_A^2/\epsilon^2)+ C_0(\epsilon)+\sum_{m=1}^{N_{\mathrm{max}}} d_{2m}(\epsilon)\sigma_A^{2m}\right) +\cdots\,,
\label{tss}
\eeqra
where all the $c_{2n}(\epsilon)$ and $d_{2m}(\epsilon)$ are regular in the $\epsilon \to 0$ limit.  The perturbative coefficients are, for $\epsilon=0$, 
\beq
c_{2n}(0)=(2n-1)!!\,.
\label{c2n}
\eeq
The remaining terms are all nonperturbative, as, due to the $\exp(-1/2\sigma_A^2)$ factor, they go to zero faster than any power of $\sigma_A^2$ as $\sigma_A^2\to 0$, and therefore cannot be captured at any order of the Taylor expansion.
Among the nonperturbative coefficients, $C_0(\epsilon)$ depends on the particular $\epsilon$-regularization procedure and cannot be computed in the general case (as it can be reabsorbed in the logarithmic term by a redefinition of $\epsilon$).  On the other hand, the other coefficients can be computed analytically by taking the appropriate number of derivatives with respect to $\sigma_A$, evaluated in $\sigma_A=0$, of  the expression,
\beq
\sum_{m=1}^{N_{\mathrm{max}}} d_{2m}(0) \sigma_A^{2m}=-\,e^{\frac{1}{2\sigma_A^2}}\sum_{n=0}^{N_{\mathrm{max}}} ( 2 \sigma_A^2)^{n+1/2}\;\Gamma\left(n+\frac{1}{2},\frac{1}{2\,\sigma_A^2}\right)\,,
\label{largecoeff}
\eeq
where $\Gamma(n,x)=\int_x^\infty dt \;t^{n-1}e^{-t}$ is the incomplete Gamma function. The dots in eq.~\re{tss} represent nonperturbative terms that are more suppressed than $\exp(-1/2\sigma_A^2)$, see \ref{transerie}.

 Notice that the perturbative and the nonperturbative sectors of the expansion are intimately related, as they originate from the same integral (see eq~\re{A5}). Indeed, the relation can be made more explicit by writing them collectively as
\beqra
&&\sum_{n=1}^{N_{\mathrm{max}}} \left(c_{2n}(0)  +\frac{e^{-\frac{1}{2 \sigma_A^2}}}{ \sqrt{2\pi \sigma_A^2}}   d_{2n}(0)\right)\sigma_A^{2n}  \nonumber\\
&&= \frac{1}{\sqrt{\pi}} \sum_{n=0}^{N_{\mathrm{max}}} (2\sigma_A^2)^n\left[\Gamma\left(n+\frac{1}{2}\right) -\Gamma\left(n+\frac{1}{2},\frac{1}{2\,\sigma_A^2}\right)  \right]-1\,,
\label{sumnpt}
\eeqra
with $\Gamma(n+1/2)=\sqrt{\pi}(2n-1)!!/2^n$. As $\Gamma(n,0)=\Gamma(n)$ we see that the coefficients of the nonperturbative series coincide with those of the perturbative one in the $\sigma_A\to \infty$ limit. The reason for this remarkable connection is clear: in the $\epsilon/\sigma_A \to 0$ limit the expression \re{tss} has to diverge as the full integral \re{it}, that is, as dictated by the logarithm in \re{logdiv1} or \re{tss}, and the remaining terms in the expansion should conspire as to give a finite quantity. In other terms, the divergences induced by shell-crossing provide a  bridge between the perturbative and the nonperturbative sectors. The PT expansion  misses the latter, and therefore is doomed to fail after shell-crossing.

This perturbative/nonperturbative connection is an example of {\it resurgent} behavior of a perturbative expansion, which appears almost ubiquitously in physical problems \cite{Aniceto:2018bis}. 

In order to explore this point further, we ask ourselves to what extent we can recover information on the full function, eq.~\re{it}, from the simple knowledge of its PT expansion, namely \re{PTd}, by using methods typical of the resurgence approach such as Borel summation. We anticipate that this attempt will be unsuccessful, therefore, the reader not interested in the details of the Borel procedure can safely skip this part and jump to the text after eq.~\re{CCB}.  

As a first step, we define
\beq
\tilde g(z)\equiv \sum_{n=0}^\infty c_{2n}(0)\; z^{-n-1}\,,
\eeq
where the coefficients are given in \re{c2n}.
$\tilde g(z)$  is related to the PT expansion \re{PTd} for $\epsilon=0$ by 
\beq
\sum_{n=1}^\infty c_{2n}(0) \sigma_A^{2n}= \sigma_A^{-2}\,\tilde g(\sigma_A^{-2})-1\,.
\eeq
We then define the Borel transform of $\tilde g(z)$ as
\beq
\hat  g(\xi) = \sum_{n=0}^\infty c_{2n}(0)\frac{\xi^n}{n!}\,,
\eeq
which, due to the $1/n!$ factors has a finite radius of convergence, in which it converges to 
\beq
\hat  g(\xi) = \frac{1}{\sqrt{1-2\,\xi}}\,.
\eeq
Then, we can try to define a meaningful summation for the initial divergent series by transforming $\hat  g(\xi)$ back via the directional Laplace transform
\beq
{\cal L}^\theta[\hat g](z) = \int_0^{e^{i\theta}\infty}\;d\xi\;e^{-z\xi}\,\hat g(\xi)\,,
\eeq 
where the integral is taken on a half-line starting from the origin and making an angle theta with the positive real axis. Since the integrand is singular in $\xi=1/2$ the procedure has an ambiguity, in the form of a nonperturbative imaginary part emerging from the discontinuity of the directional Laplace transform as the $\theta\to 0$ limit is taken from above or from below,
\beq
\lim_{\theta\to 0^\pm} {\cal L}^\theta[\hat g](z) = \frac{e^{-z/2}}{\sqrt{\frac{\pi}{2 z}}}\left(\mathrm{Erfi}\left(\sqrt{\frac{z}{2}}\right)\pm i \right)\,
\eeq
which leads to the possible identification

\beq
\sum_{n=1}^\infty c_{2n}(0) \sigma_A^{2n}\simeq  e^{-\frac{1}{2\sigma_A^2}}\sqrt{\frac{\pi}{2\sigma_A^2}}\left(\mathrm{Erfi}\left(\sqrt{\frac{1}{2\sigma_A^2}}\right)+ i\, C \right)  -1\,,
\label{CCB}
\eeq
where the $C$ constant contains the ambiguity of the procedure. It can be checked that the PT expansion of the expression at the RHS reproduces \re{PTd} at all orders, and moreover that it is finite for any value of $\sigma_A^2$. 

\begin{figure}[t]
\centering 
\includegraphics[width=.65\textwidth,clip]{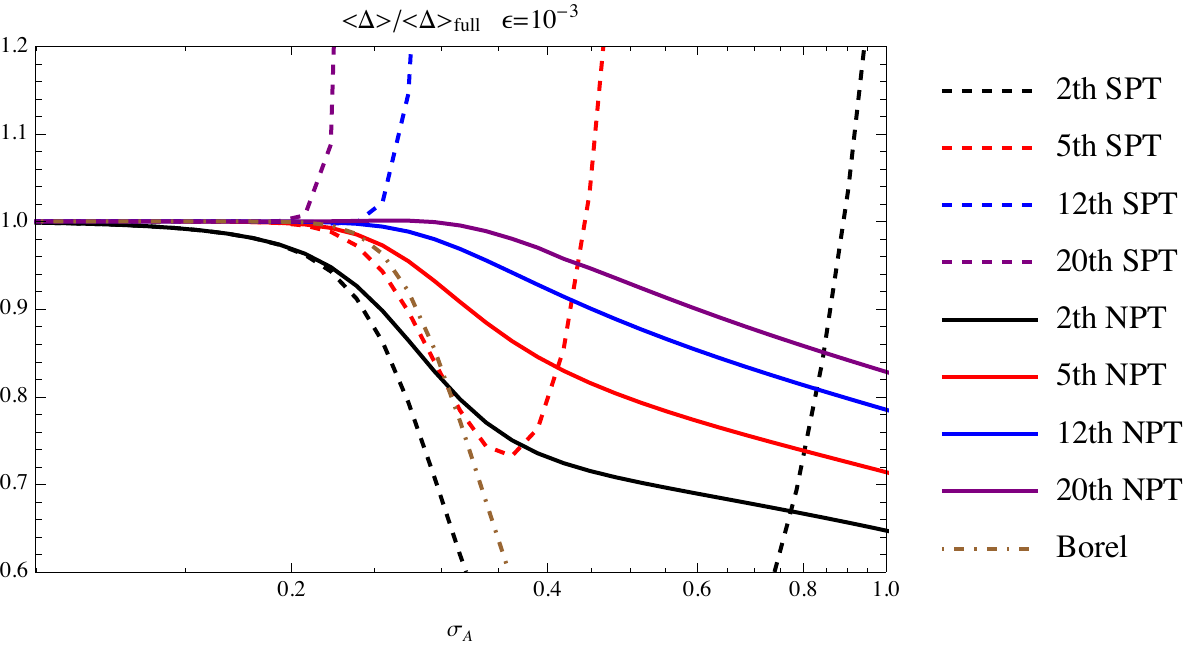}
\caption{The ratio between  $\Delta$ as obtained in different approximations and the full one, obtained by integrating eq.~\re{it} numerically. The order is referred to the value of $N_\mathrm{max}$   in the summations of eqs.~\re{PTd} and \re{tss}. For the dashed lines, the nonperturbative coefficients $d_{2n}$ have been set to zero, leaving only the SPT expansion. The dash-dotted line represents the Borel summation of eq.~\re{CCB}.}
\label{Ratioeps001}
\end{figure}

In Fig.~\ref{Ratioeps001} we show the ratios between different approximations to the full integral \re{it} and the integral itself, evaluated numerically. The non-convergence of the standard PT expansion (the first sum in \re{sumnpt}) is clear from the behavior of the dashed lines for increasing values of the truncation order, that is, on the value of $N_\mathrm{max}$ in the sums.

On the other hand, the full result \re{tss} not only shows convergence, but also convergence to the correct function. In order to obtain these lines we had to tune the parameter $C_0(0)$ in \re{tss}, which cannot be extracted from our considerations in \ref{transerie}. However, this is done once for all, as we checked that it does not depend neither on $\epsilon$ (as long as $\epsilon \ll \sigma_A$) nor on the truncation order of the full sum \re{sumnpt}.

We also show (by brown dash-dotted line) the result for the Borel summation procedure described above, in which we also take the freedom to fine tune the $C$ parameter in \re{CCB} to imaginary values. As we see, this procedure gives a non diverging result, but does not appear to converge to the true function \re{it}. It was to be expected, as the only input we gave are the perturbative coefficients in the $\epsilon \to 0$ limit, $c_{2n}(0)$, and therefore all the information on the logarithmic divergence of the integral was obliterated since the beginning. It would be interesting to see if this information can, at least in part, be recovered by giving instead the $c_{2n}(\epsilon)$'s, or at least some perturbative truncation in $\epsilon$ of these coefficients.

\subsection{Fourier space}
Before closing this section, we briefly outline how the post shell-crossing features discussed above should manifest in Fourier space.  Going back to the density contrast  of eq.~\re{Gmap} and taking the Fourier transform, we get,

\beq
\!\!\!\!\!\!\!\!\!\!\!\!\!\!\!\!\!\!\!\!\!\! \!\!\!\!\!\!\!\!\!\!\!\int dx\,e^{i k \bar x}\left(-1+ \int\frac{dq}{\sqrt{2\pi \sigma^2}} e^{-\frac{(q+\Psi(q)-\bar x)^2}{2\sigma^2}} \right)=-2\pi\delta_D(k) +e^{-\frac{k^2 \sigma^2}{2}}\int\frac{dq}{\sqrt{2\pi \sigma^2}} e^{i k\left(q+\Psi(q)\right)}\,,
\eeq
of which we can safely take the $\sigma\to 0$ limit. Indeed, we could have directly started from the unregolarized density contrast, eq.~\re{0map}, and we woud have obtained the same result. The point is that the Fourier transformation acts as a regulator, where the role of the spatial resolution $\sigma$ of \re{Gmap}  is played by $\sim 2\pi/k$. By following our case study of the cubic displacement field, \re{qpsi}, we expect that the role of $\epsilon$ is played, in Fourier space, by 
\beq
\epsilon \to \frac{2 \pi}{k\,\bar q}.
\eeq
This relation summarizes very well the nature of the divergences for $\epsilon\to 0$, that is, for 
\beq
k\,\bar q\gg 2\pi\,,
\eeq
 of both \re{ds} and \re{tss}. They manifest themselves as Eulerian UV divergences (large $k$) induced by Lagrangian IR effects (non-locality on scales $O(\bar q)$). 

Our result \re{tss} indicates that, when considering statistically averaged quantities, shell-crossing  should give rise to non-perturbative terms which are at most logarithmically divergent in $k$ for large $k$'s. 

\section{Exact dynamics in 1+1 dimensions}
\label{exdyn}
In this section we take into account the exact dynamics in one spatial dimension, and discuss the property of the equation of motion and its solutions after shell-crossing.

The equation of motion for the displacement field is
\beq
\ddot \Psi(q,\tau)+\Hc\, \dot \Psi(q,\tau) = -\partial_x\Phi(x(q,\tau),\tau)\,,
\label{eom1}
\eeq
where dots denote derivatives with respect to conformal time $\tau$  and ${\cal H}=\dot a/a$, where $a(\tau)$ is the scale factor.
The gravitational potential satisfies a Poisson equation
\beqra
\partial^2_x \Phi(x,\tau) &=& \frac{3}{2}{\cal H}^2 \delta(x,\tau)\nonumber\\
& =& \frac{3}{2}{\cal H}^2\, \int d q\left(\delta_D\left(x-q- \Psi(q,\tau)\right)-\delta_D\left(x-q\right)\right)\,,
\eeqra
where we have used the relation \re{0map} for the density contrast.

The force is therefore obtained by integration,
\beq
-\partial_x\Phi(x,\tau) = -\frac{3}{2}{\cal H}^2\, \int d q\left(\Theta\left(x-q- \Psi(q,\tau)\right)-\Theta\left(x-q\right)\right) + c(\tau)\,,
\label{forcef}
\eeq
where $\Theta(x)$ is the Heaviside theta function, and the possibly time-dependent quantity $c(\tau)$ is  zero in the ``CMB rest frame" in which the force vanishes everywhere when all the particles are in they unperturbed positions, i.e. $\Psi(q,\tau)=0$. In the following, we will will assume to be in that frame and will set $c(\tau)=0$.

Eq.~\re{forcef} has a very simple physical interpretation.  Since in 1 spatial dimension the force is independent on the distance, to compute the force on $x$ of a segment of matter of infinitesimal length  $dq$ that was initially in $q$, say, at the left (right) with respect to $x$,  it suffices to know if this segment is still at the left (right), in which case there is no change in the force, or it has moved to the right (left), in which case $x$ will feel an excess infinitesimal force towards the right (left) given by 
\beq
+ (-) \frac{3}{2}{\cal H}^2\, dq\,.
\eeq 
Therefore, counting the total amount of matter crossing x from its initial position is exactly what the $\Theta$ functions in \re{forcef} do, keeping track of the sign of the crossing.

The expression for the force can be written in a more useful form in terms of the roots $q_i(x,\tau)$ of eq.~\re{S}.
Let's start from the case in which there is a single root, $q_1(x,\tau)$, as in panel $a)$ of Fig.~\ref{configs} Then
\beqra
&& \!\!\!\!\!\!\!\!\!\!\!   \!\!\!\!\!\!\int d q\left(\Theta\left(x-q- \Psi(q,\tau)\right)-\Theta\left(x-q\right)\right) =\lim_{L\to\infty}  \int_{-L/2}^{q_1(x,\tau)}dq-\int_{-L/2}^{x}dq\nonumber\\
 && \!\!\!\!\!\!\!\!\!\!\!  \!\!\!\!\!\!= q_1(x,\tau)+\frac{L}{2}-\left(x+\frac{L}{2}\right)= -\Psi(q_1(x,\tau),\tau)\,\qquad\qquad \mathrm{(one\;stream\;in\;}x),
\eeqra
where we have introduced a finite box of length $L$, and then sent $L\to\infty$.

Then, let's consider the case in which the point $x$ is in a region which in which first shell crossing has occurred, as in panel $b)$ of Fig.~\ref{configs}, and there are three coexisting streams. In this case, eq.~\re{S} has three solutions, $q_i(x,\tau)$ ($i=1,\cdots,3$). We get, in this case,
\beqra
 &&\lim_{L\to \infty}\int_{-L/2}^{L/2} d q\left(\Theta\left(x-q- \Psi(q,\tau)\right)-\Theta\left(x-q\right)\right) \nonumber\\
&&=\lim_{L\to \infty}\left[  \left(q_1(x,\tau) +\frac{L}{2}\right) + \left(q_3(x,\tau)-q_2(x,\tau)\right) -\left(x+\frac{L}{2}\right)\right]\nonumber\\
&&\!\!\!\!\!\!\!\!\!\!\! \!\!\!\!\!\!\!\!\!\!\! \!\!\!\!\!\!\!\!\!\!\! =-\Psi(q_1(x,\tau),\tau)+\Psi(q_2(x,\tau),\tau)-\Psi(q_3(x,\tau),\tau)\,\;\;\qquad\quad \mathrm{(three\;streams\;in\;}x).
\eeqra
At this point, one realises that, considering an arbitrary (odd) number of roots of Eq.~\re{S}, $N_s(x,\tau)$,   gives
\beqra
 &&\!\!\!\!\!\!\!\!\!\!\!\!\!\!\!\!\!\!\!\!\!\!\!\! \lim_{L\to \infty}\int_{-L/2}^{L/2} d q\left(\Theta\left(x-q- \Psi(q,\tau)\right)-\Theta\left(x-q\right)\right) =-\sum_{i=1}^{N_s(x,\tau)} (-1)^{i+1} \Psi(q_i(x,\tau),\tau)\,,\nonumber\\
&&\;\;\qquad\quad\qquad\quad\qquad\quad\qquad\quad\qquad\quad\qquad( N_s(x,\tau)\;\mathrm{streams\;in\;}x).
\label{fullstream}
\eeqra
 The exact equation of motion \re{eom1} therefore can be also written as,
\beq
\ddot \Psi(q,\tau)+\Hc\, \dot \Psi(q,\tau) = \frac{3}{2}\Hc^2 \sum_{i=1}^{N_s(x(q,\tau),\tau)} (-1)^{i+1} \Psi(q_i,\tau)\,,
\label{eom2}
\eeq
where, for each $q$,  the roots $q_i$ are the solutions of 
\[
q_i+\Psi(q_i,\tau)=q+\Psi(q,\tau)=x(q,\tau)\,.
\]
One of the $q_i(x)$ roots is clearly coincides with  $q$.
Each root contributes to the RHS with a sign given by the sign of  $1+\Psi'(q_i,\tau)=x'(q,\tau)$.

In the case in which at point $x$ there is no shell-crossing, $N_s(x(q,\tau),\tau)=1$, the only root is $q$ itself, and eq.~\re{eom2} reduces to the equation of motion in the Zel'dovich approximation
\beq
\ddot \Psi_Z(q,\tau)+\Hc\, \dot \Psi_Z(q,\tau) = \frac{3}{2}{\cal H}^2 \Psi_Z(q,\tau)\,,
\label{eomZ}
\eeq
which is then exact in absence of multistreaming. When multistreaming is present, the Zel'dovich approximation is unable to reproduce the backreaction on the force term, and it departs from the exact dynamics, see next section, and in particular Figs.~\ref{force} and ~\ref{force2}.

The exact equation of motion, eq.~\re{eom2}, is manifestly not problematic at shell-crossing. At fixed $x$, shell crossing occurs when two new real roots of \re{S}, $q_j(x,\tau)$ and $q_{j+1}(x,\tau)$, appear. Since at the time of shell crossing, $\tau_x$, one has $q_j(x,\tau_x)=q_{j+1}(x,\tau_x)$, the contribution of the new couple to the RHS of \re{eom2} vanishes for $\tau \le \tau_x$ and is continuous in $\tau_x$ as $\lim_{\tau\to\tau_x^+}(q_j(x,\tau)-q_{j+1}(x,\tau))=0$.

On the other hand, it is clear from eq.~\re{eom2} that the force term is nonlocal in Lagrangian space after shell-crossing, and any attempt based on the expansion of it in terms of the Zel'dovich displacement field evaluated in $q$,  appears unjustified.

``Naive'' PT expansion schemes, both in Eulerian and in Lagrangian space, are therefore doomed to failure due to multistreaming. 
\section{Numerical solution}
\label{numsol}
Using as ``time'' variable  the logarithm of the scale factor,
\beq
\eta=\log\frac{a}{a_0}=-\log (1+z)\,,
\eeq
the equation of motion \re{eom2} can be written as the system
\beqra
 &&\partial_\eta \Psi(q,\eta) = \chi(q,\eta)\,,\nonumber\\
 &&\partial_\eta \chi(q,\eta) = -\frac{1}{2} \chi(q,\eta)+ \frac{3}{2} \sum_{i=1}^{N_s(x,\eta)} (-1)^{i+1} \Psi(q_i(x,\eta),\eta)\,.
 \label{syst}
\eeqra
The initial condition is given at an early redshift in which we assume the linear theory growing mode, namely
\beq
\Psi(q,\etain) = \chi(q,\etain) = \frac{v(q,\etain)}{\Hc(\etain)}  \,,
\eeq
where $v$ is the peculiar velocity.
The solution of the above system of equations can then be computed by a straightforward algorithm, which requires just a few lines of code.
At each time-step, for each $x$ we identify the subset of Lagrangian points $\{q_i(x,\eta)\}$, containing all the real roots of the equation $x-q-\Psi(q,\eta)=0$. Then, for each $q$, we compute the corresponding $x=q+\Psi(q,\eta)$, and then  the increment of  $\Psi(q,\eta)$, and $\chi(q,\eta)$, which involves, through the sum in \re{syst}, the previously identified subset $\{q_i(x,\eta)\}$ (which, of course, includes also $q$).

\begin{figure}[t]
\centering 
\includegraphics[height=.24\textwidth]{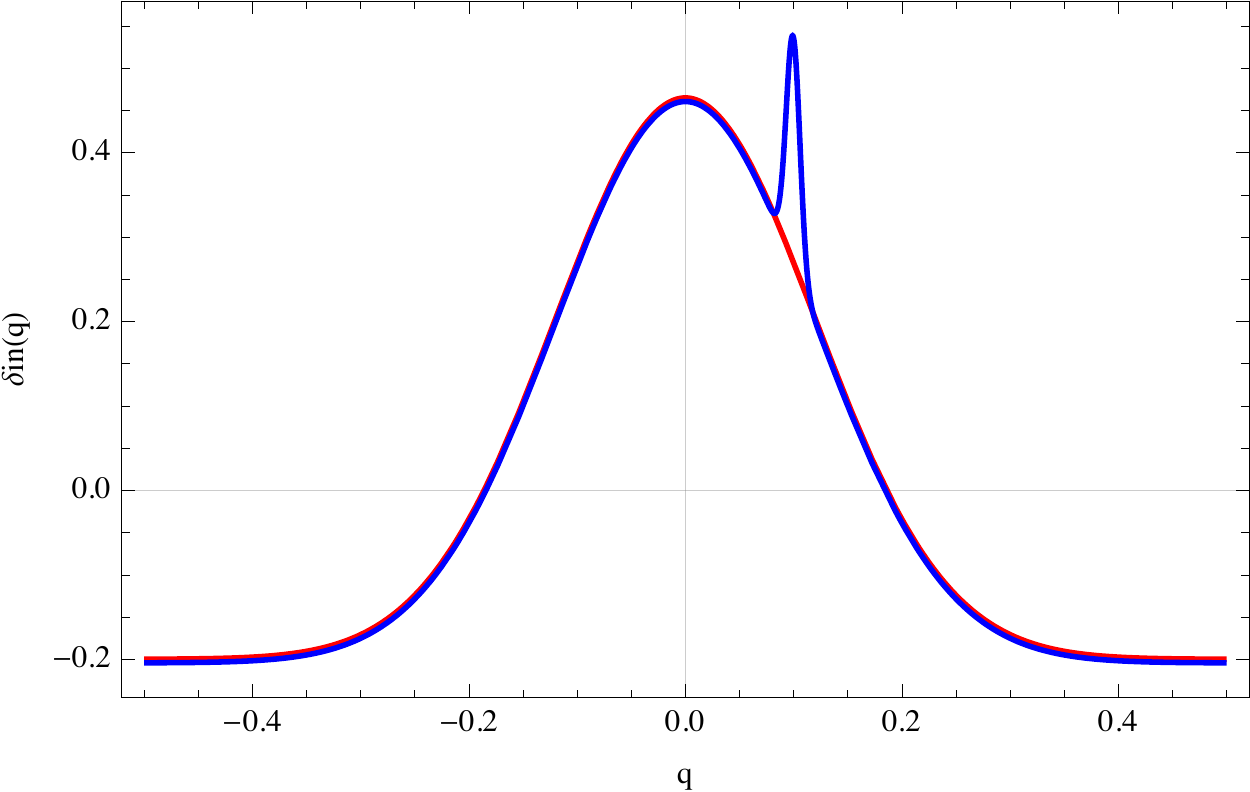}
\includegraphics[height=.24\textwidth]{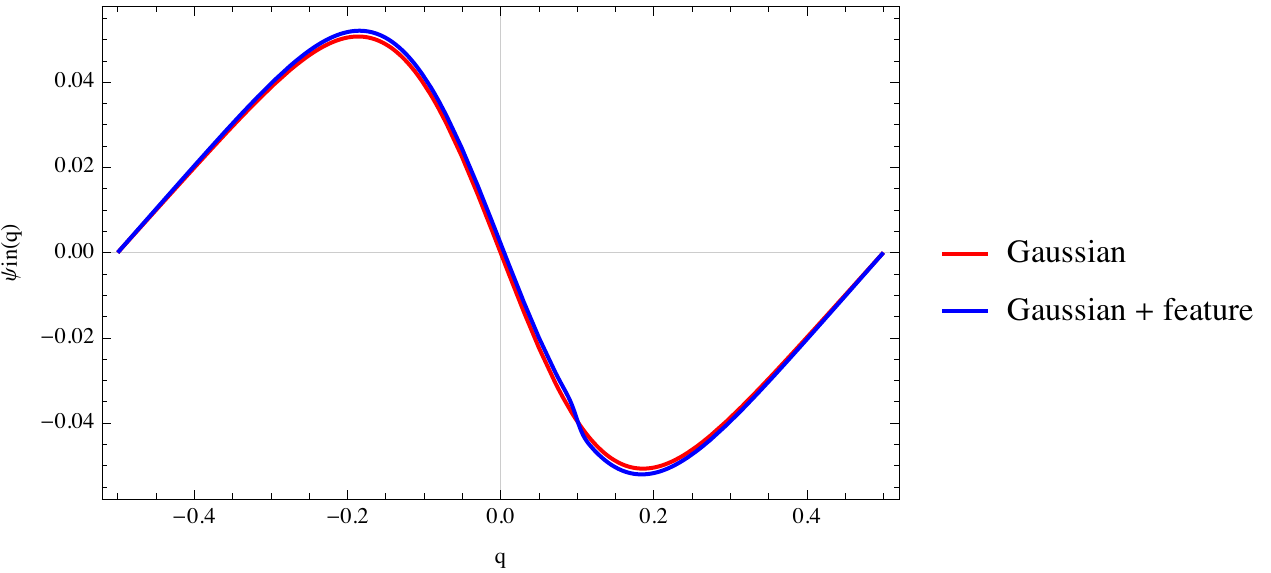}
\caption{ Gaussian initial condition for the density contrast with and without feature (left). The  corresponding initial displacement field (right).}
\label{feat}
\end{figure}

In order to familiarise with the solution, and to follow the example of  \cite{Taruya:2017ohk}, where with similar tests were performed  with  the particle-mesh code presented in that paper, one can apply the algorithm to some simple initial conditions.
The first one is a single initial gaussian overdensity, see the red curve in Fig.~\ref{feat},
\beq
\delta(q,\etain)=\frac{A}{\sqrt{2\pi \sigma^2}} e^{-\frac{q^2}{2\,\sigma^2}} + C\,,
\label{deltain}
\eeq
on a periodic segment bounded by $-\frac{L}{2}<q\le\frac{L}{2}$. The constant $C$ is tuned so that the integral of the overdensity on the full $q$ range vanishes. Later we will also consider a modified initial condition, in which we added a gaussian feature on top of \re{deltain}, by multiplying it by $1+B \exp(-(q-q_0)^2/\sigma_1^2)$, see the blue line in  Fig.~\ref{feat}.

\begin{figure}[tbp]
\centering
\includegraphics[width=.45\textwidth,clip]{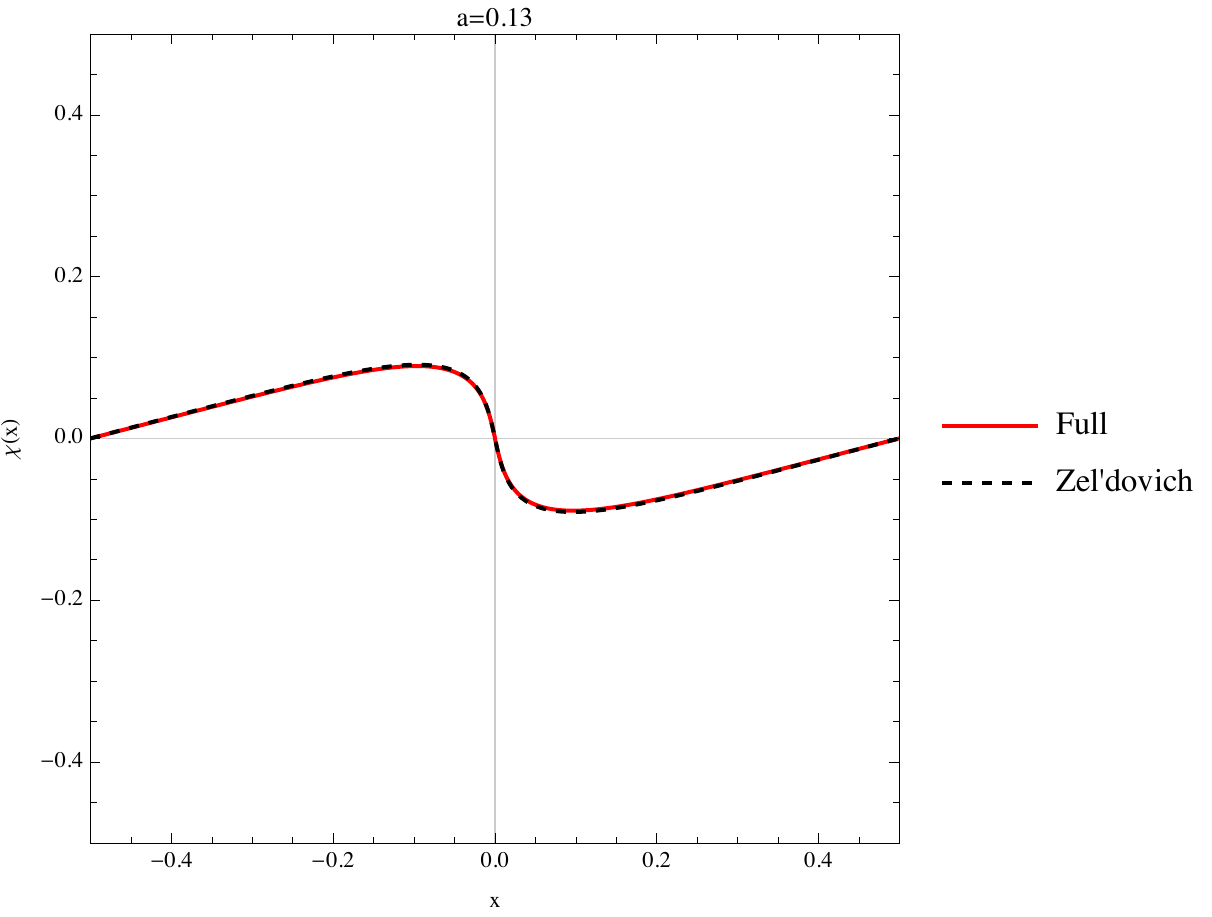}
\includegraphics[width=.45\textwidth,clip]{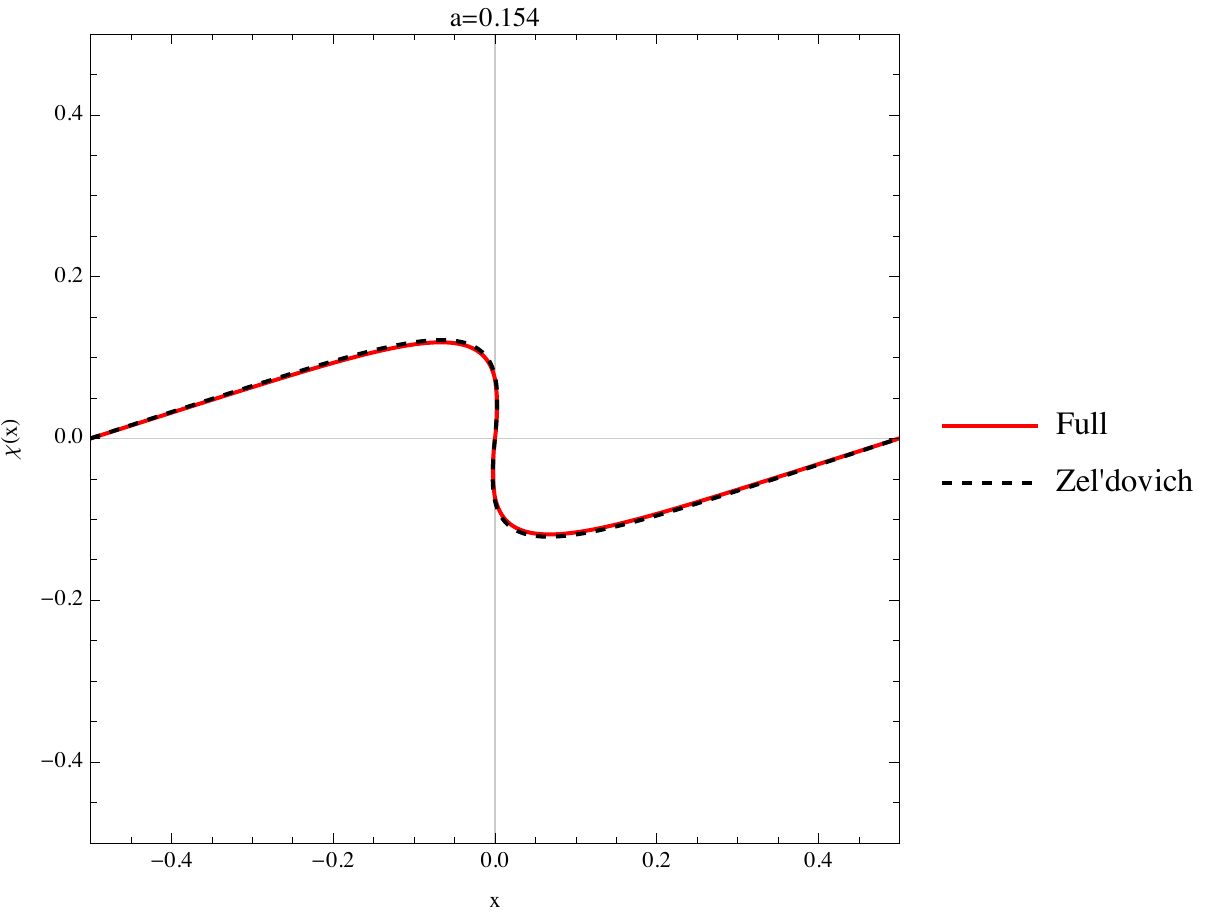}
\includegraphics[width=.45\textwidth,clip]{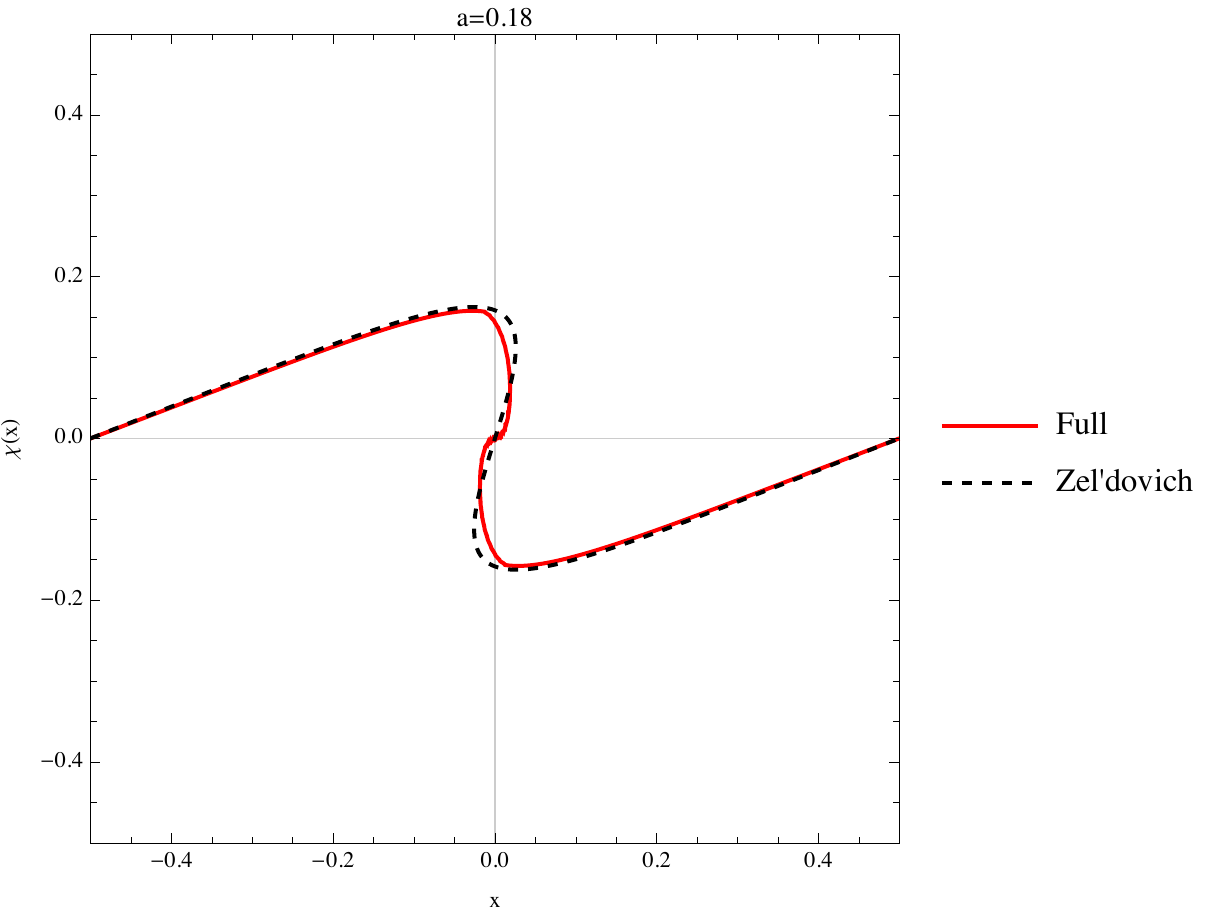}
\includegraphics[width=.45\textwidth,clip]{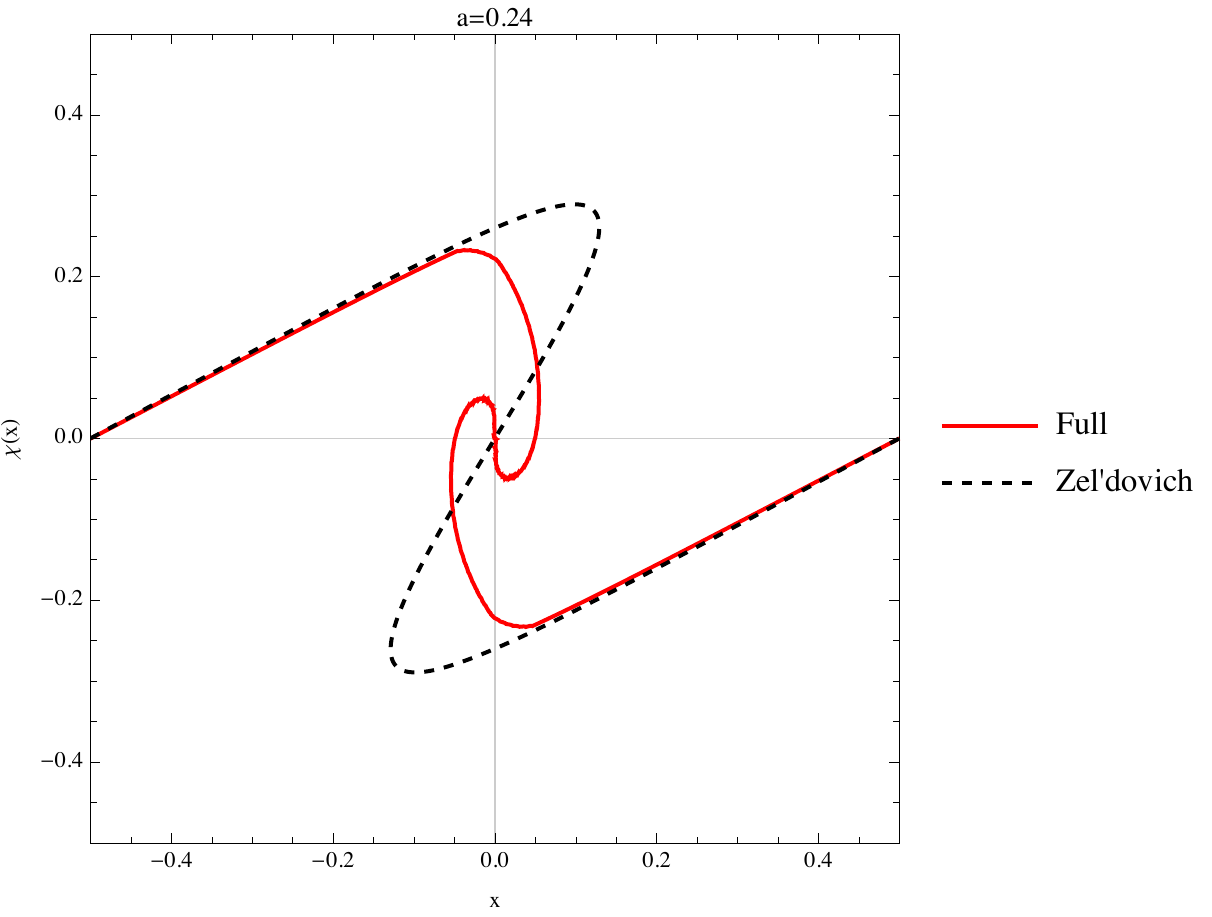}
\includegraphics[width=.45\textwidth,clip]{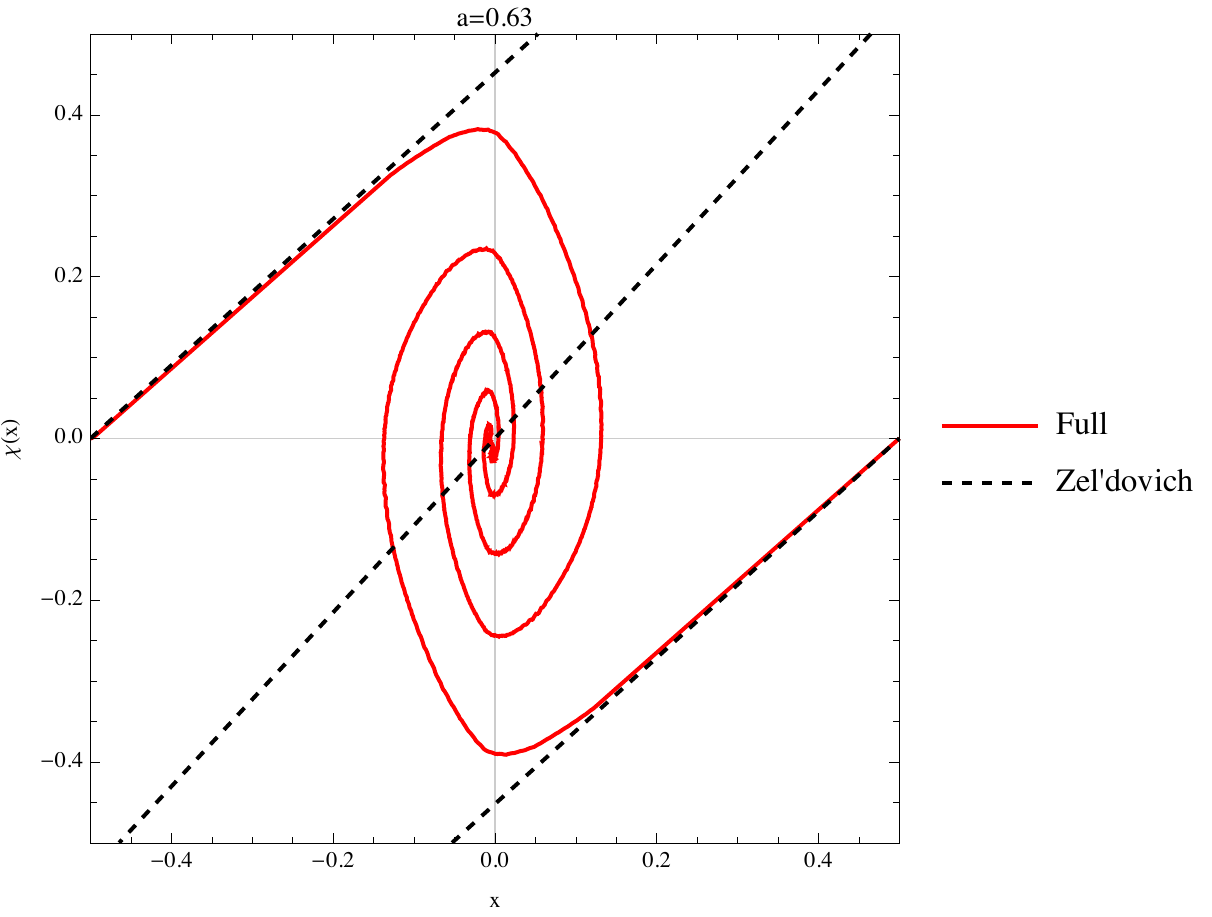}
\includegraphics[width=.45\textwidth,clip]{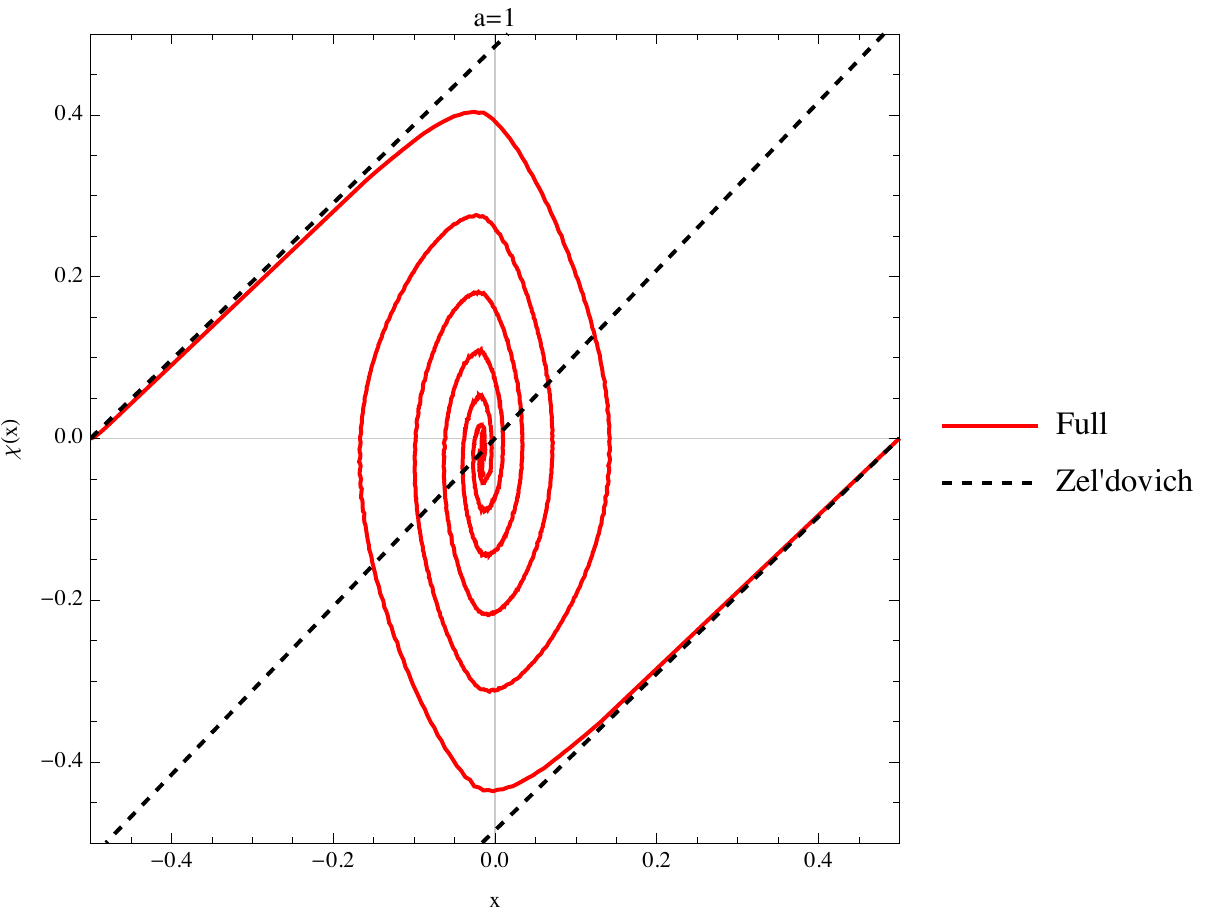}
\caption{Phase space at different epochs obtained from the  initial conditions of eq.~\re{deltain}. Continuous red lines are obtained with the exact dynamics, black dashed ones with the  Zel'dovich one.} 
\label{gaussia}
\end{figure}

\begin{figure}
\centering
\includegraphics[height=.25\textwidth]{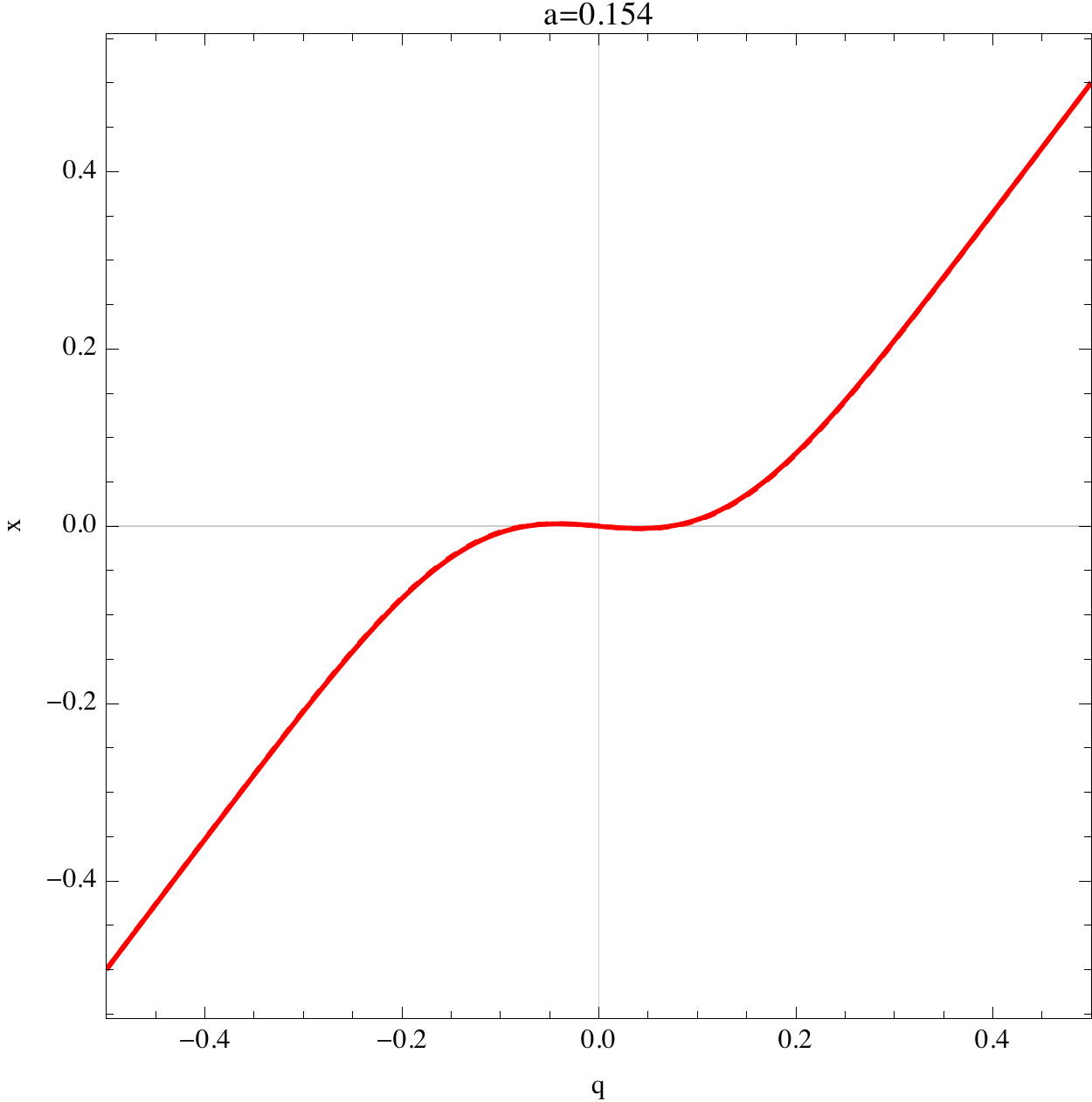}
\includegraphics[height=.25\textwidth]{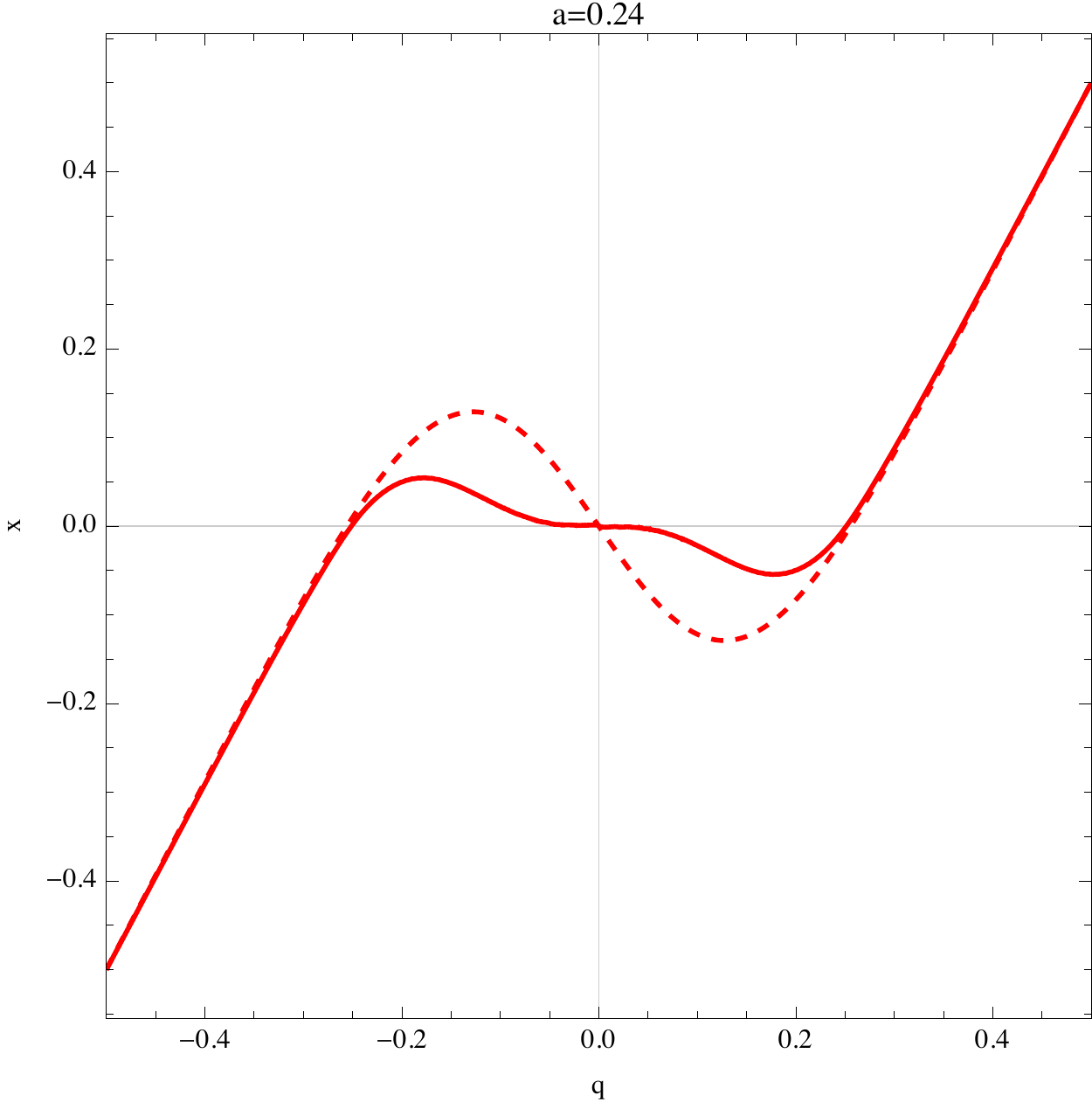}
\includegraphics[height=.25\textwidth]{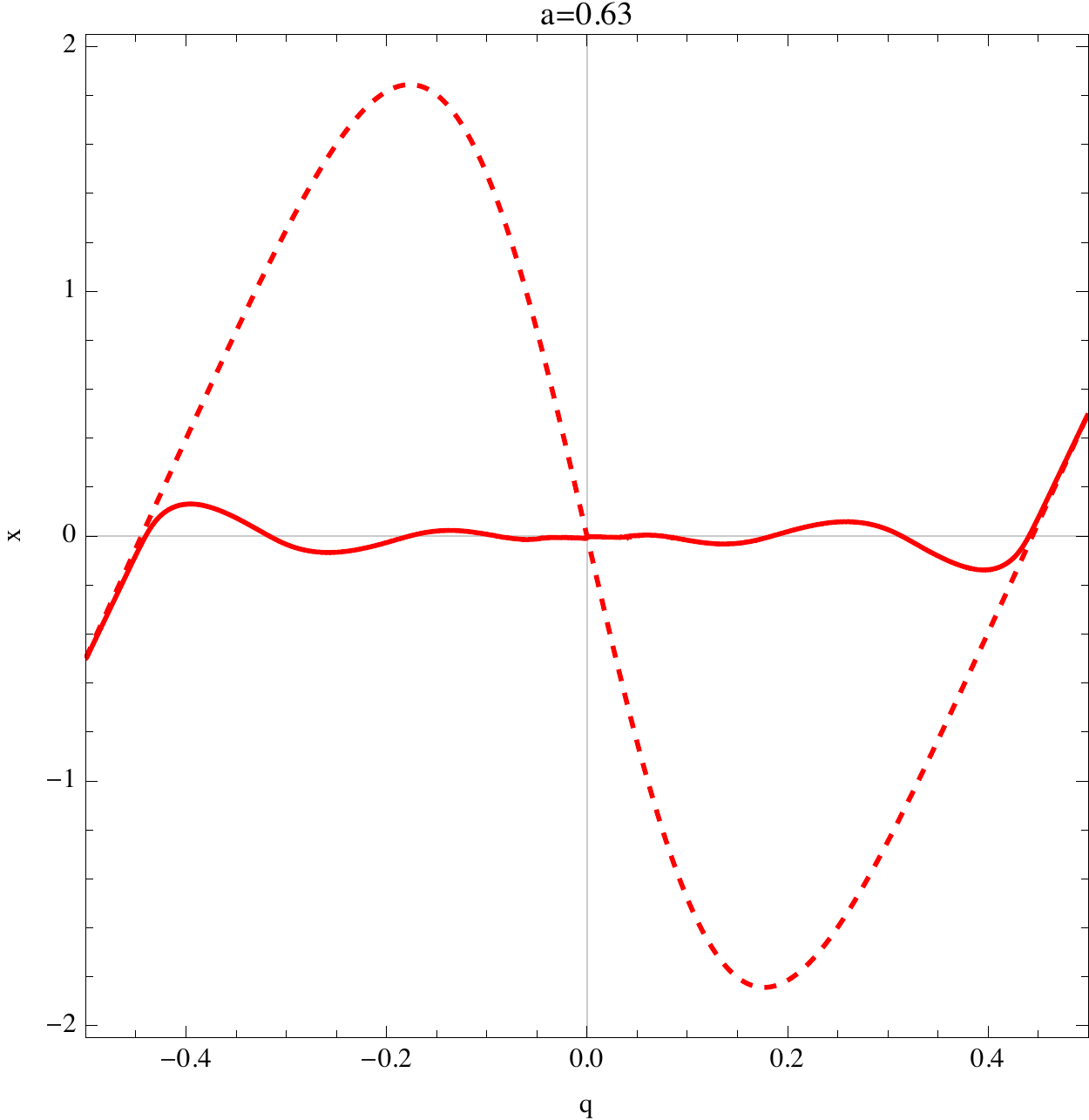}
\includegraphics[height=.32\textwidth]{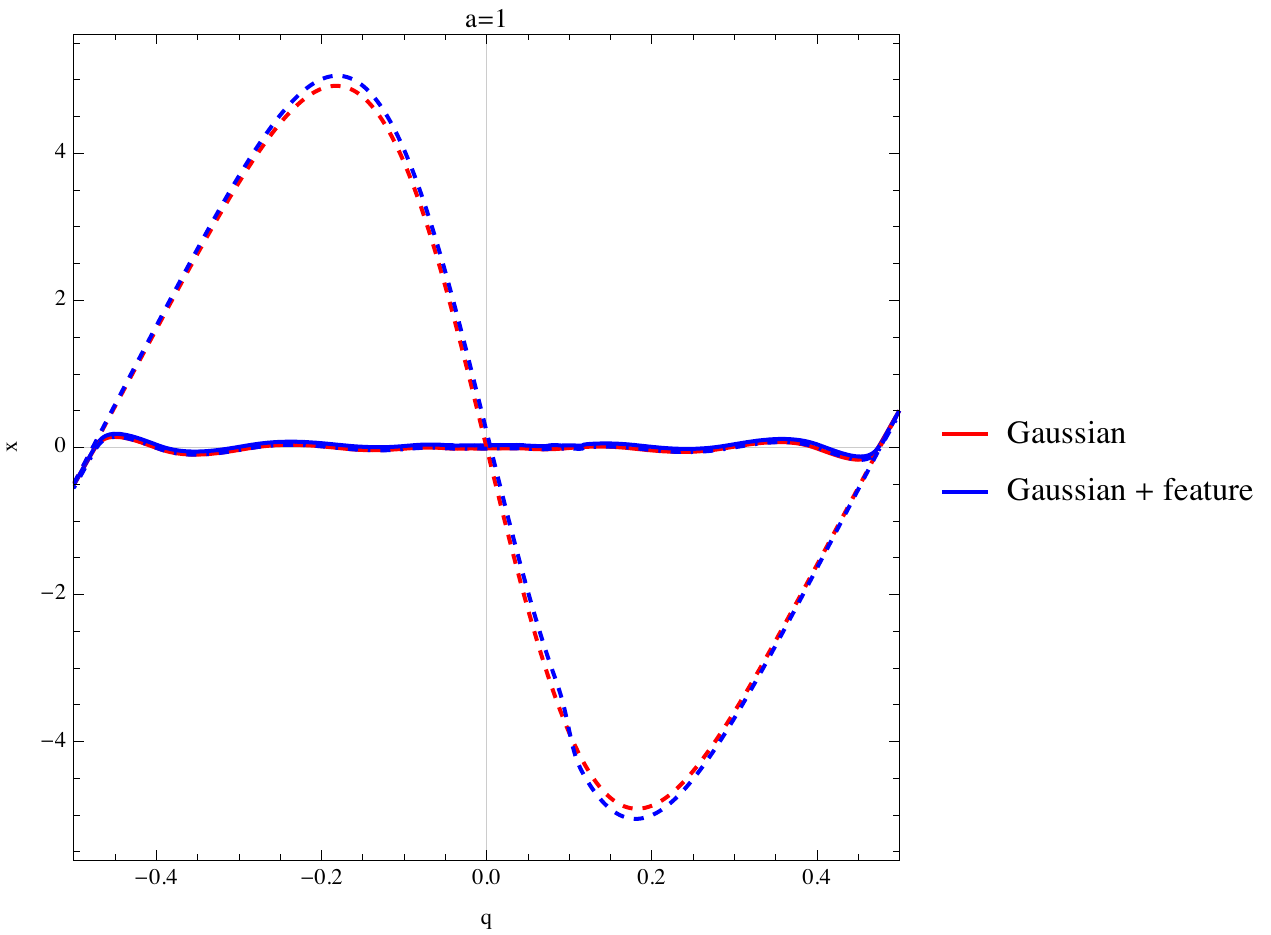}
\includegraphics[height=.32\textwidth]{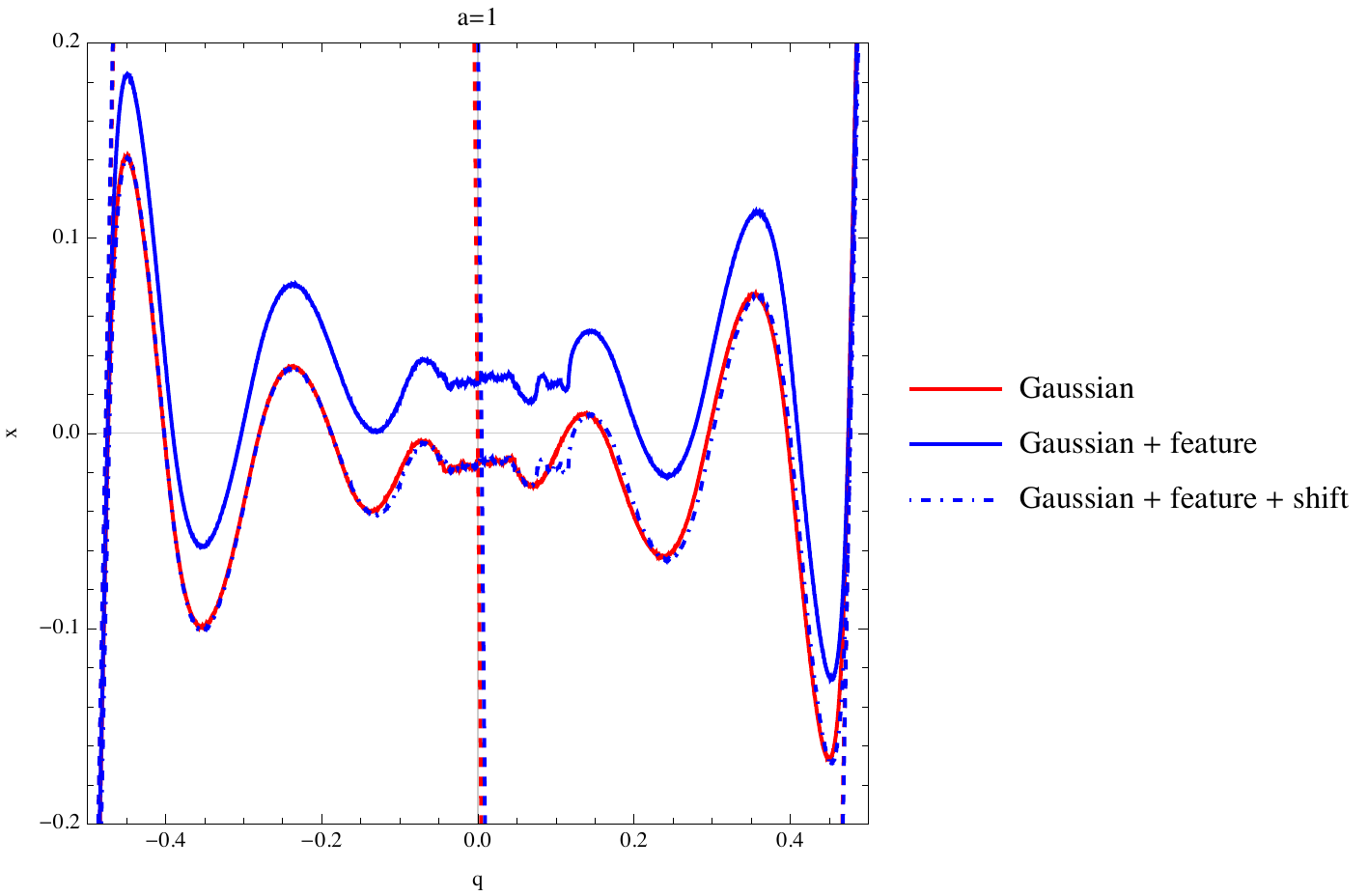}
\caption{The mapping between Lagrangian and Eulerian space  evolved from the  initial conditions of eq.~\re{deltain}  without (red) and with (blue) the feature shown in Fig.~\ref{feat}. Continuous lines are obtained with the exact dynamics, dashed ones with the  Zel'dovich one.}
\label{elshift}
\end{figure}

Using again linear theory, we find that the initial condition for $\Psi(q,\etain)$ is obtained by integrating \re{deltain},
\beq
\Psi(q,\etain)=- \int_{-L/2}^q dq'\, \delta(q',\etain) \,.
\label{psidelta}
\eeq
We set $A=0.2$, $\sigma=0.12$, $L=1$, $z_{in}=99$ and integrate the equations on a grid with $1200$ points and in $100$ logarithmic steps in time. We show in Fig.~\ref{gaussia} the results of the integration in phase space, namely, in the $(x,\chi)$ plane. The first shell-crossing occurs at about $a=0.154$ in $x=0$ and the second one at $a=0.24$. Before first shell-crossing, the full solution coincides with the Zel'dovich one everywhere, but the two rapidly diverge afterwards. Around the position $x=0$ we count $3$, $9$, and $13$ streams, respectively, in the snapshots taken at $a=0.18,\,0.63$, and $1$. Notice that, even when high order multistreaming occurs, the Zel'dovich solution is recovered very fast for $x$'s outside the multistreaming region. So, while multistreaming is very non-local in Lagrangian space, it is a local effect in Eulerian space, (see also Fig.~\ref{elshift}), where it manifests itself by the emergence of higher moments of the distribution function.  Moreover, the Zel'dovich approximation greatly overestimates the extension of the multistreaming region in Eulerian space (compare the solid and the dashed lines both in Fig.~\ref{gaussia} and \ref{elshift}) and it fails in giving  the number of streams after second shell-crossing.

\begin{figure}[tbp] 
\centering
\includegraphics[width=.45\textwidth]{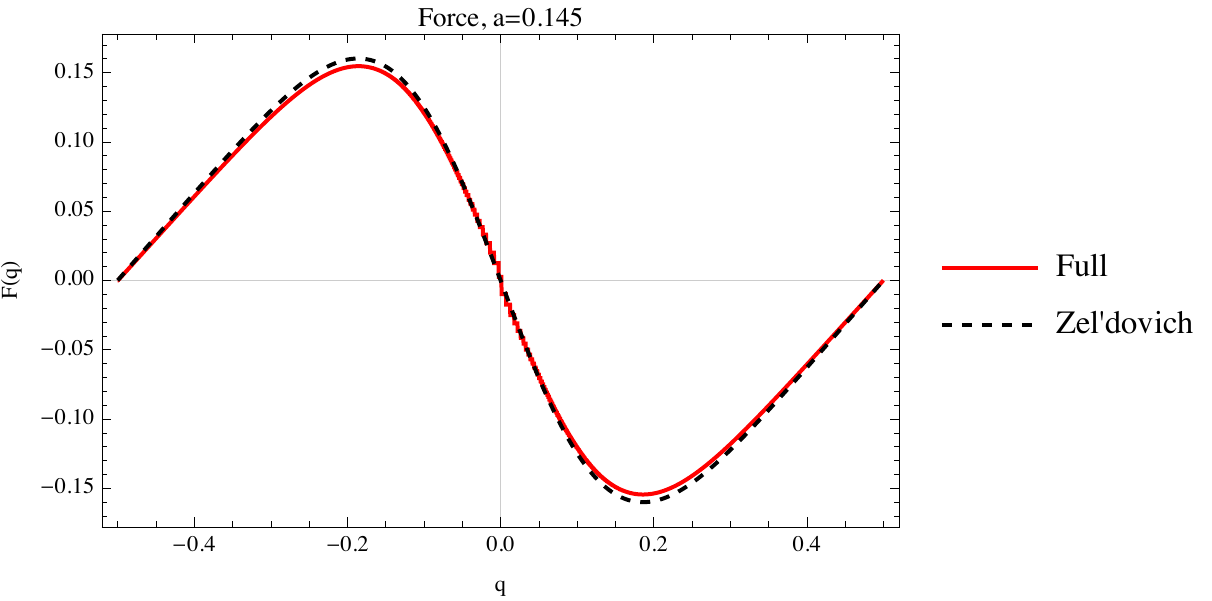}
\includegraphics[width=.45\textwidth]{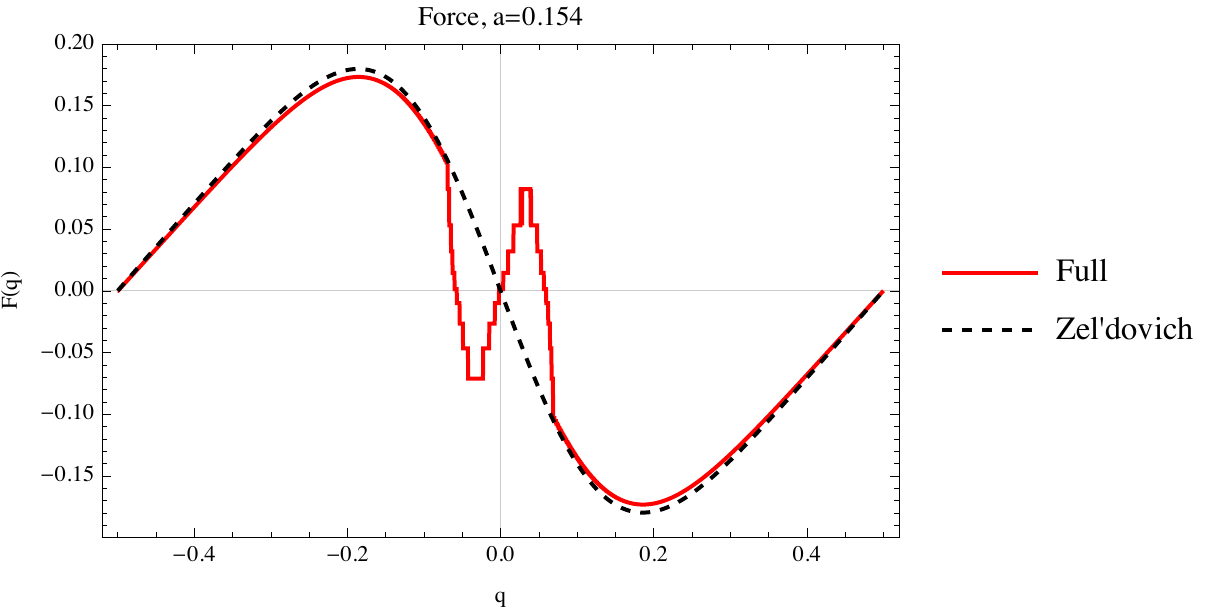}
\includegraphics[width=.45\textwidth]{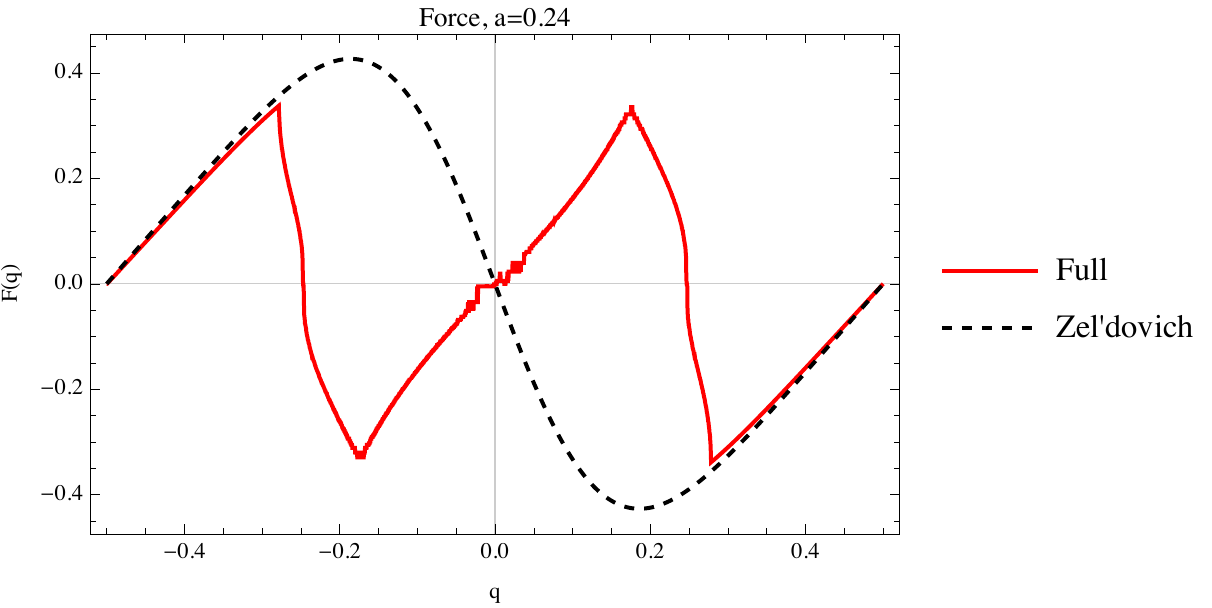}
\includegraphics[width=.45\textwidth]{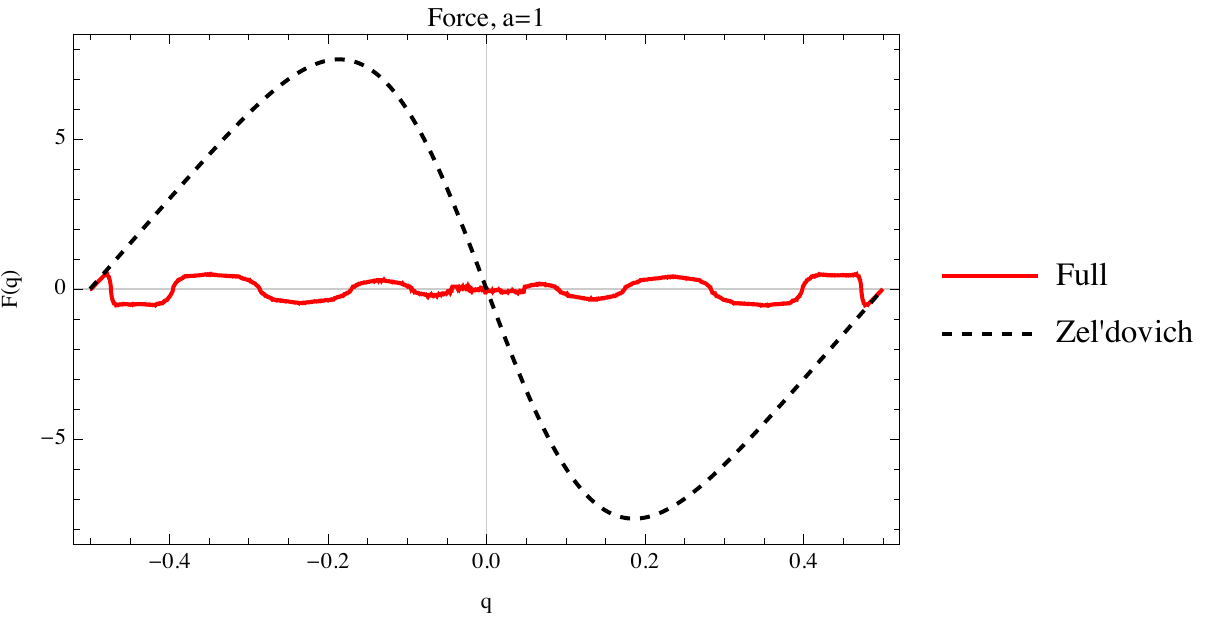}
\caption{The exact force term, RHS of eq.~\re{eom2} (continuous red lines) compared to the RHS of the Zel'dovich equations of motion, eq.~\re{eomZ} (dashed black lines).}
\label{force}
\end{figure}

To better visualize the origin of the failure of the Zel'dovich approximation, in Fig.~\ref{force} we plot the force, namely the right hand side of eq.~\re{eom2}, and compare it to the right hand side of the Zel'dovich equation of motion, eq.~\re{eomZ}. Before shell-crossing, as expected, the Zel'dovich right hand side coincides with the full force term, however as soon as shell-crossing happens, $O(1)$ deviations take place. They are to be expected: after crossing each other, the mutual attraction between two particles changes sign, whereas in the Zel'dovich approximation they proceed along their ballistic paths. While the amplitude of the Zel'dovich ``force" grows in time as $\Psi(q,\tau)$, namely, proportionally to the linear growth factor, the amplitude of the  full force stays approximately constant. 

These results clearly show that expanding around the Zel'dovich solution is not a good option to explore the post shell-crossing regime, apart for, possibly, a very short time after the first shell-crossing, along the lines explored recently in  \cite{Taruya:2017ohk, McDonald:2017ths}.

\section{Post shell-crossing attractor}
\label{pattr}
The relation between the Lagrangian and Eulerian positions  is shown in Fig.~\ref{elshift}. While before shell-crossing the Zel'dovich mapping is exact, 
soon after the first shell crossing the exact mapping deviates sensibly from the Zel'dovich one. As time passes, it flattens out over the whole shell-crossing region.

This behavior does not depend on the particular initial condition, but is a generic attractor feature of the equations of motion, as we now show. 

The equations of motion \re{syst} can be written in terms of $x(q,\eta) = q+\Psi(q,\eta)$ as
\beqra
\!\!\!\!\!\!\!\!\! \!\!\!\!\!\!\!\!\! \!\!\!\!\!\!\!\!\! \!\!\!\partial_\eta^2 x(q,\eta)+\frac{1}{2}\partial_\eta x(q,\eta)=-\frac{3}{2} \int dr\left[\Theta(x(q,\eta)-r-\Psi(r,\eta))-\Theta(x(q,\eta)-r)\right]\,,
\label{att1}
\eeqra
taking the first derivative with respect to $q$, we get the equation for
\beq
x'(q,\eta)= 1+\Psi'(q,\eta)\,,
\label{xq}
\eeq
\beqra
&&\!\!\!\!\!\!\!\!\! \!\!\!\!\!\!\!\!\! \!\!\!\!\!\!\!\!\!  \partial_\eta^2 x'(q,\eta)+\frac{1}{2}\partial_\eta  x'(q,\eta)\nonumber\\
&&\;\;\;\;\;= -\frac{3}{2} x'(q,\eta) \int dr\left[\delta(x(q,\eta)-r-\Psi(r,\eta))-\delta(x(q,\eta)-r)\right]\,,\nonumber\\
&&\;\;\;\;\;=-\frac{3}{2} x'(q,\eta)\left(\sum_{i=1}^{N_s(x(q))}\frac{1}{|x'(q_i,\eta)|}-1\right)\,,
\label{att2}
\eeqra
where the summation is made over all the roots in $x(q,\eta)$. In absence of shell-crossing, $N_s(x(q))=1$, the RHS gives
\beq
-\frac{3}{2}\left(1-x'(q,\eta)\right) = \frac{3}{2} \Psi'(q,\eta)\,,
\eeq
which, using \re{xq}, coincides with the equation of motion for the first derivative of the displacement field in the Zel'dovich approximation, see eq.~\re{eomZ}.
Now, assume that first shell-crossing takes place in $q_{sc}$ at $\eta_{sc}$, that is,  $x'(q_{sc},\eta_{sc})=0$. The RHS of eq.~\re{att2} is negative for $\eta\leq\eta_{sc}$. Soon after shell-crossing, the term inside parenthesis at the RHS of \re{att2} gives 
\beq
\left(\frac{1}{|x'(q_1,\eta_{sc}^{+})|}+\frac{1}{|x'(q_3,\eta_{sc}^{+})|}+\frac{1}{|x'(q_{sc},\eta_{sc}^{+})|}-1\right) >0\,,
\eeq
 where $q_1$ and $q_3$ are the two new roots. The term inside parentheses is positive, as 
$ |x'(q_{sc},\eta)|\ll 1$ close to shell-crossing. As a consequence, $x'(q_{sc},\eta)$, which is negative soon after shell-crossing, starts to increase, as the RHS of  eq.~\re{att2} is globally positive. As it crosses zero again, two new streams are generated, whose contribution to the parentheses at the RHS is still positive, and therefore the latter is now globally negative. As a result, $x'(q_{sc},\eta)$ oscillates around zero with decreasing amplitude, and  is driven asymptotically to zero. Notice that the value $x'(q_{sc},\eta)=0$ can be safely reached both from positive and negative time directions, there is no divergence there, therefore $x'(q_{sc},\eta)=0$ represents a fixed point of the post-shell crossing evolution.

One can now investigate the fate of the {\em second derivative} of $x(q_{sc},\eta)$, $x''(q_{sc},\eta)$,  by taking the derivative of eq.~\re{att2} with respect to $q$, finding 
\beqra
&&\!\!\!\!\!\!\!\!\! \!\!\!\!\!\!\!\!\! \!\!\!\!\!\!\!\!\!  \partial_\eta^2 x''(q_{sc},\eta)+\frac{1}{2}\partial_\eta  x''(q_{sc},\eta)\nonumber\\
&& \!\!\!\!\!\!\!\!\! \!\!\!\!\!\!\!\!\! =-\frac{3}{2} x''(q_{sc},\eta)\left(\sum_{i=1}^{N_s(x(q))}\frac{1}{|x'(q_i,\eta)|}-1\right)-\frac{3}{2}(x'(q_{sc},\eta))^2\sum_{i=1}^{N_s(x(q))}\frac{ x''(q_i,\eta) }{|x'(q_i,\eta)|^3} \,.\nonumber\\
&&
\label{att3}
\eeqra
The second term at the RHS goes to zero due to the behavior of  eq.~\re{att2} discussed above, and the first term gives the same fixed point mechanism for $ x''(q_{sc},\eta)$. The same holds for all the higher order derivatives, so we conclude that, deep into the shell-crossing regime, the Lagrangian to Eulerian mapping function $x(q)$ approaches the attractor solution
\beqra
&& \!\!\!\!\!\!\!\!\!\!\!\!\! x(q,\eta)=x_Z(q,\eta), \qquad\qquad\qquad\qquad\qquad\quad \,\; \;\mathrm{for}\;  N_s(x(q,\eta),\eta)=1\,,\nonumber\\
&& \!\!\!\!\!\!\!\!\!\!\!\!\! x'(q_{sc},\eta)=x''(q_{sc},\eta)=x'''(q_{sc},\eta)\cdots \to 0, \qquad \mathrm{at\;shell-crossing\;points}\,,
\label{fpoint}
\eeqra
where $x_Z(q,\eta)$ is the Zel'dovich solution.
We still have to determine the value  $x(q_{sc},\eta)$ at shell-crossing points. It represents the coordinate of the center of mass of the mass falling in the multi-streaming region. Therefore, it can be determined by the requirement that the force in $\bar x(q,\eta)$ due to all the mass contained in the multi-streaming region vanishes
\beqra
&&\int_{q_1(\bar x)}^{q_{N_s}(\bar x)} dr\left[\Theta(\bar x-r-\Psi(r,\eta))-\Theta(r+\Psi(r,\eta)-\bar x)\right]\nonumber\\
&&=-2\left[\sum_{i=1}^{N_s(\bar x)} (-1)^{i+1} \Psi(q_i(\bar x),\eta) -\frac{\Psi(q_1(\bar x),\eta) +\Psi(q_{N_s}(\bar x),\eta) }{2}\right]=0\,.
\label{barx}
\eeqra

In other terms, we find that the asymptotic configuration is that in which all the particles in the multi-streming regions form a thin shock at the position given by their center of mass, whose position $\bar x$ is determined by solving eq.~\re{barx}. 

This asymptotic configuration is similar to  the one predicted by the `adhesion model' for structure formation \cite{Gurbatov:1989az,Dubrulle:1994psg}. In particular, introducing the potential $\varphi(q,\eta)$, defined by
\beq
x(q,\eta)=\frac{\partial \varphi(q,\eta) }{\partial q}\,,
\eeq
we find that it becomes convex everywhere and flat in the multistreaming regions. The
 ``geometrical adhesion model" of \cite{Bernardeau:2009ab,Valageas:2010uh} gives a similar prescription, by imposing that the nonlinear potential is obtained as the convex hull of the Zel'dovich one. 
As we will discuss later, this prescription does not coincide with our attractor, as the actual position of the shock cannot be inferred correctly from the Zel'dovich solution when it predicts a number of streams larger than three. 
Moreover, the attractor  is reached only asymptotically, in the deep multi-streaming regime, and deviations from it are present at any epoch.

To visualize the attractor 
behavior we added a feature on top of the gaussian initial condition of eq.~\re{deltain}, to get the blue line in Fig.~\ref{feat}, again keeping to zero the integral of the function over the full spatial interval. The resulting evolution of the Lagrangian to Eulerian mapping is shown in Fig.~\ref{elshift} at $a=1$ (lower row). The red lines are obtained with pure gaussian overdensities, while the blue lines are obtained with the feature added on top of it (continuous lines are for the full dynamics, dashed ones for Zel'dovich).  Flattening in the multistreaming Lagrangian region is evident from the left plot and, by zooming inside (right plot), one sees that it starts from the point of first shell-crossing, at $q=0$, to propagate outwards, as higher and higher derivatives approach the fixed points \re{fpoint}. 

The memory of the feature at $q=0.1$ is barely noticeable from the comparison between the shapes of the blue and red solid curves. A stronger effect is a  shift in the position $\bar x$ of the plateau in the two cases. This shift is entirely explained by the shift in the position of the center of mass due to the feature, and can be reproduced by computing $\bar x$ from eq.~\re{barx} {\em using the Zel'dovich solutions} for the two different initial conditions. Indeed, shifting the solution of the exact equations evolved from initial condition with feature by the $\delta \bar x = \bar x(\mathrm{no\; feature})- \bar x(\mathrm{feature})$ computed in the Zel-dovich approximation we get the dash-dotted line. The fact that it overlaps nearly perfectly with the solution for the featureless case shows that one can use the Zel'dovich approximation to identify the center of mass of the particles fallen in the multi-streaming region. On the other hand,  the same effect can be obtained by any other initial condition which perturbs the gaussian one by shifting the center of mass by the same amount, and therefore it carries no information on the detail of the feature that we have imposed.

\begin{figure}[t] 
\centering
\includegraphics[width=.35\textwidth]{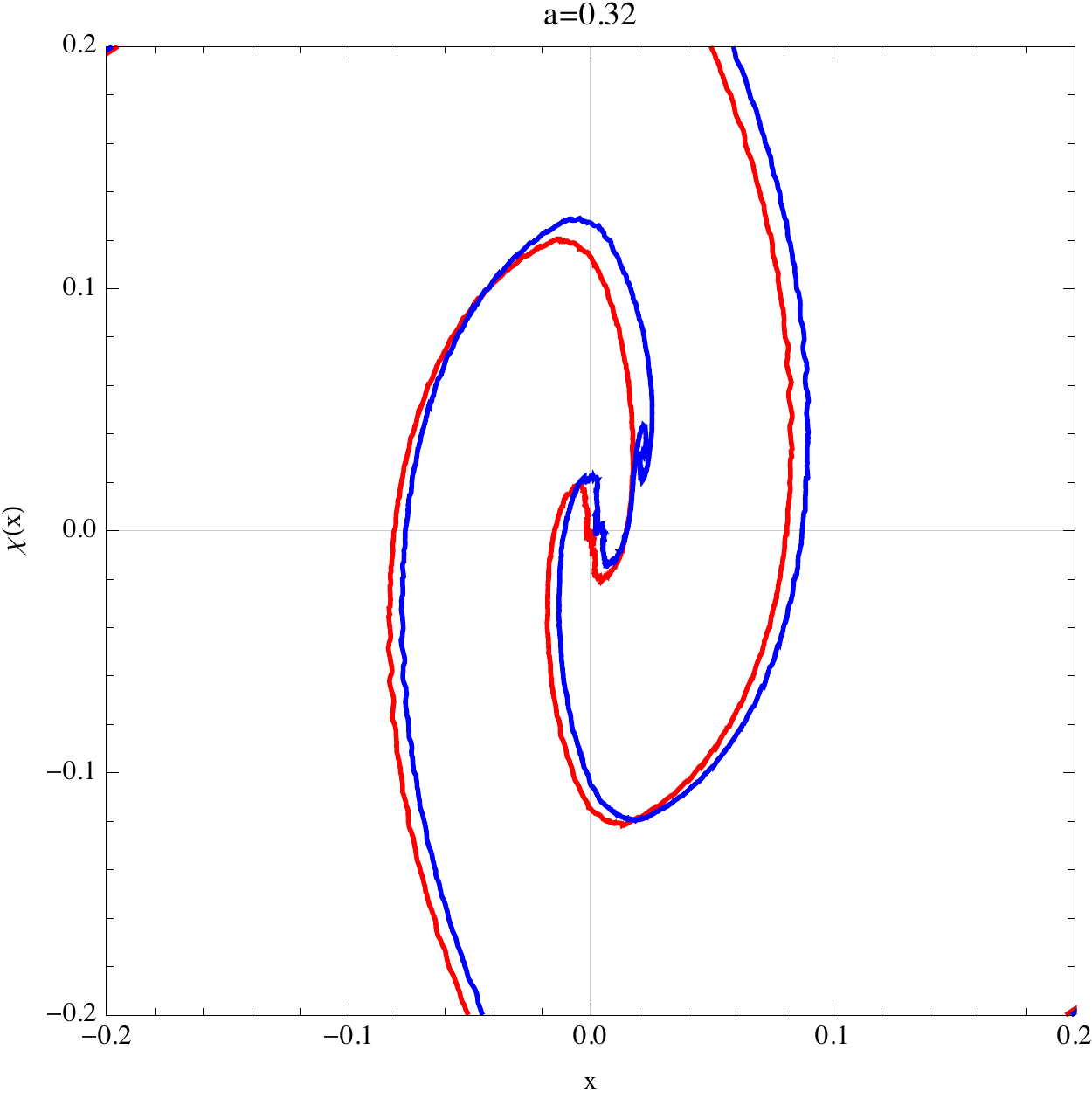}
\includegraphics[width=.35\textwidth]{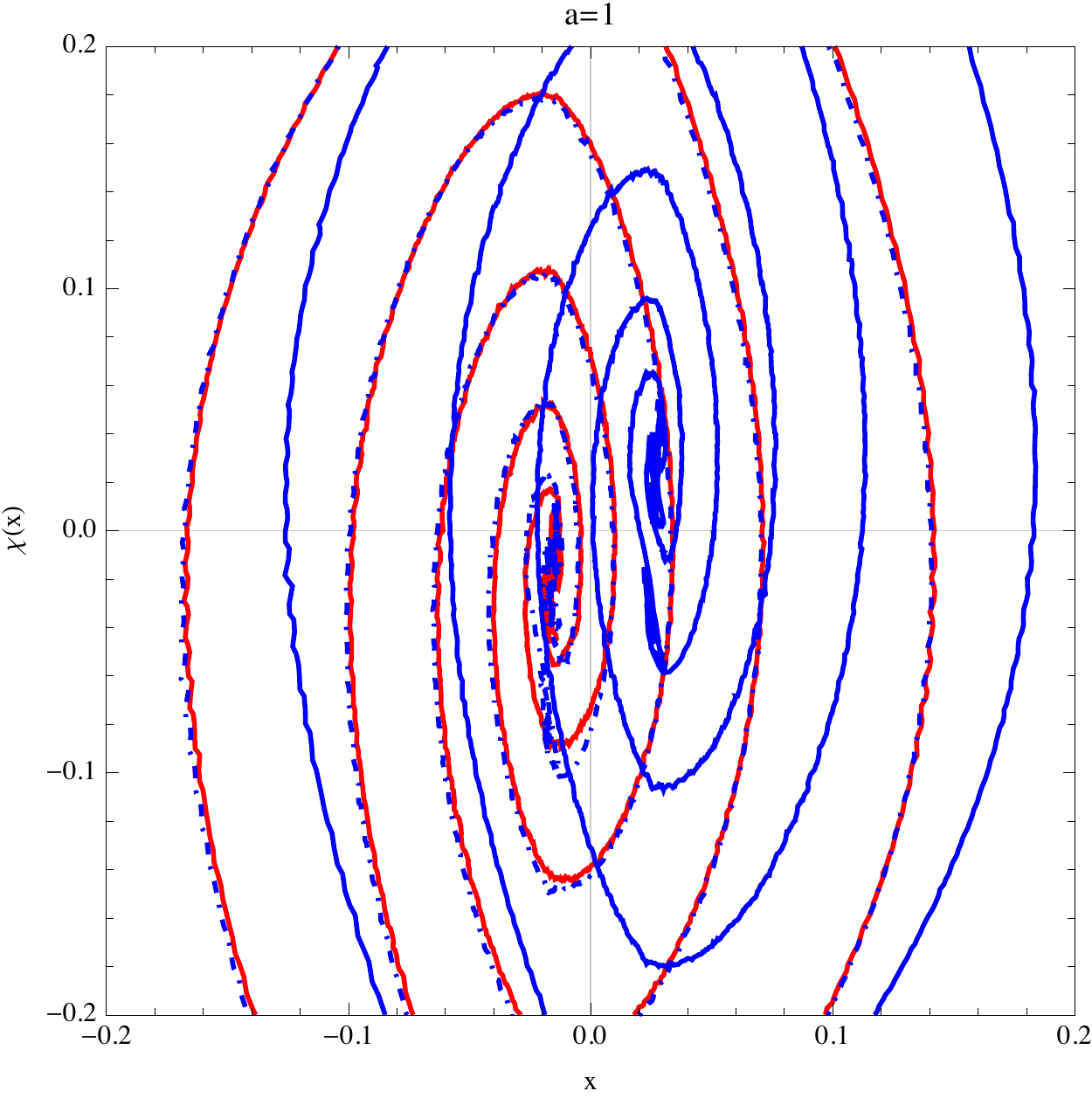}
\caption{Phase space solution for the initial condition of eq.~\re{deltain} without (red) and with (blue) the added feature. At $a=1$ we also show the effect of shifting the blue curve by the deviation of the center of mass position induced by the feature, obtained according to Eq.~\re{barx}  from the Zel'dovich solution (dash-dotted blue).
}
\label{pswf}
\end{figure}

We repeat the same operation in phase space, see Fig.~\ref{pswf}. Shifting both $x$ and $\chi$ by the same amount as we do in Fig.~\ref{elshift}, we get the blue dash-dotted line, which has a nearly perfect overlap with the featureless red curve. Notice the remnant of the initial feature, in the form of the little spiral in the blue line at $x\sim 0.02$, $\chi\sim 0.03$ at $a=0.32$ and at $x\sim 0.02$, $\chi\sim -0.04$ at $a=1$: the information is still there, but it is diving deeper and deeper in the multistreaming region and becoming more and more difficult to recover.

\begin{figure}%[tbp] 
\centering
\includegraphics[height=.25\textwidth]{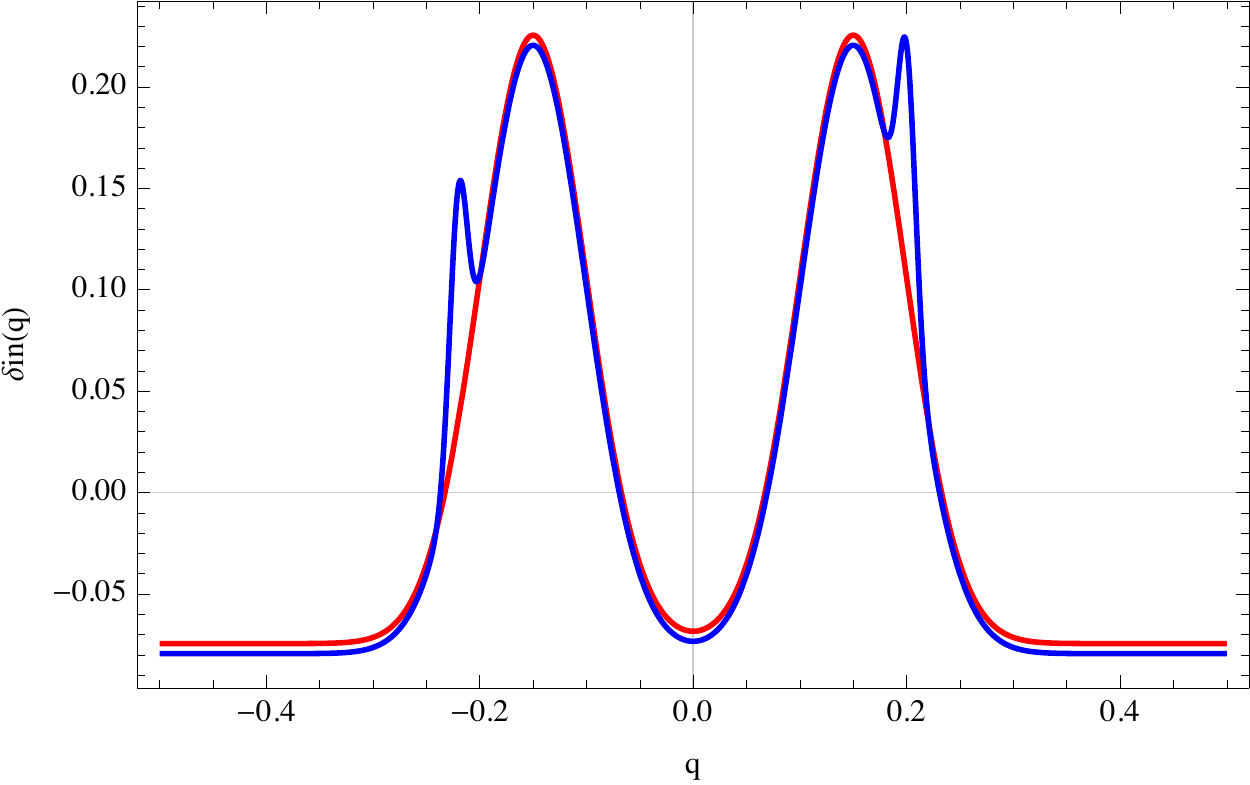}
\includegraphics[height=.25\textwidth]{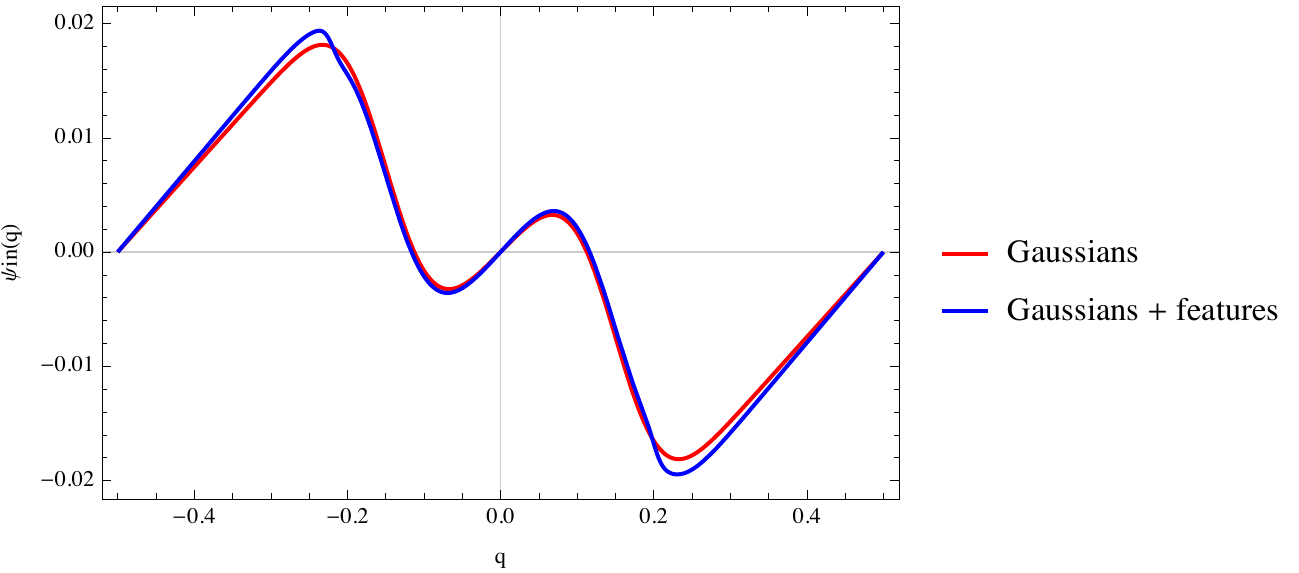}
\caption{Initial condition with two gaussians without (red) and with (blue) features.}
\label{feature2}
\end{figure}

As a second test we consider initial conditions given by a pair of gaussian overdensities, as the red lines in Fig.~\ref{feature2}, on top of which we added gaussian features asymmetrically (blue lines).

\begin{figure}%[tbp]
\centering
\includegraphics[width=.45\textwidth,clip]{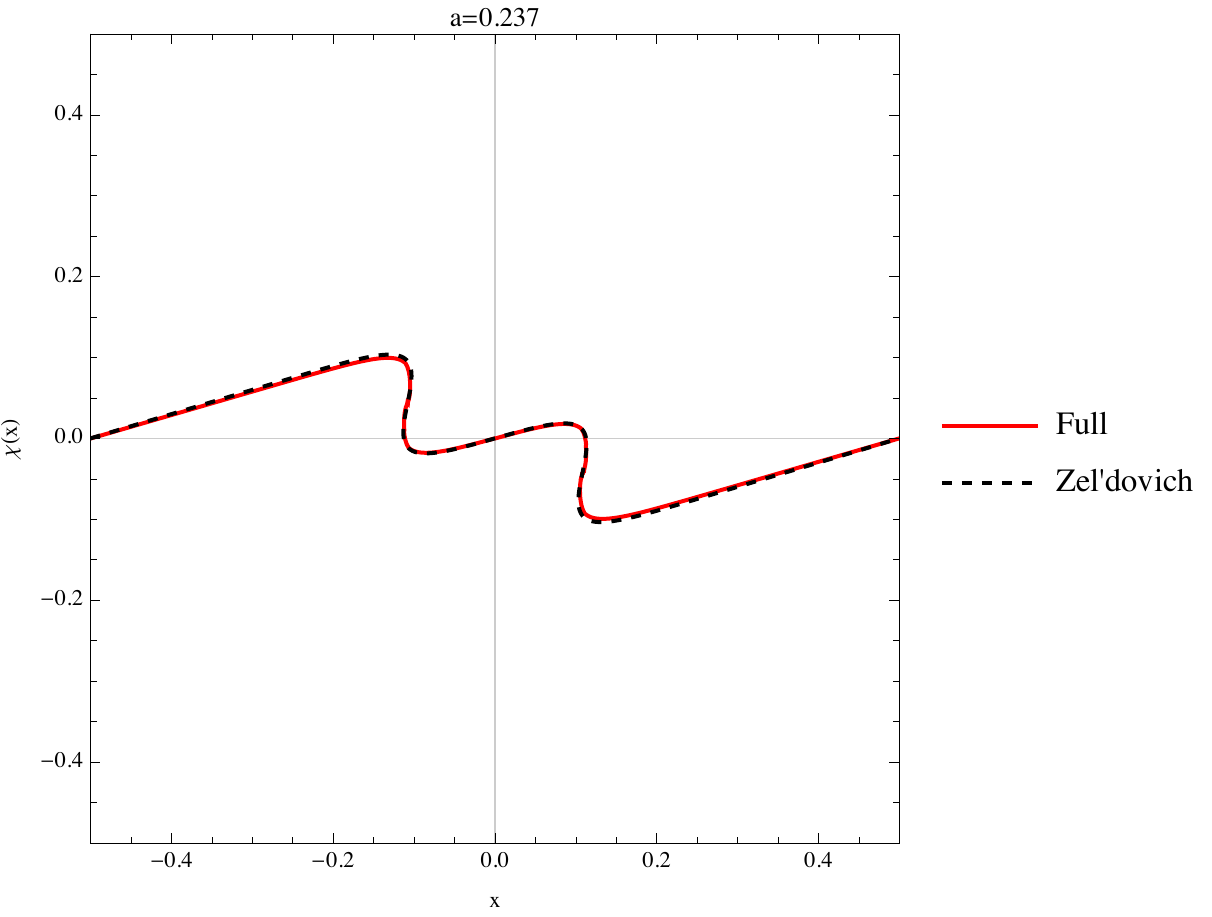}
\includegraphics[width=.45\textwidth,clip]{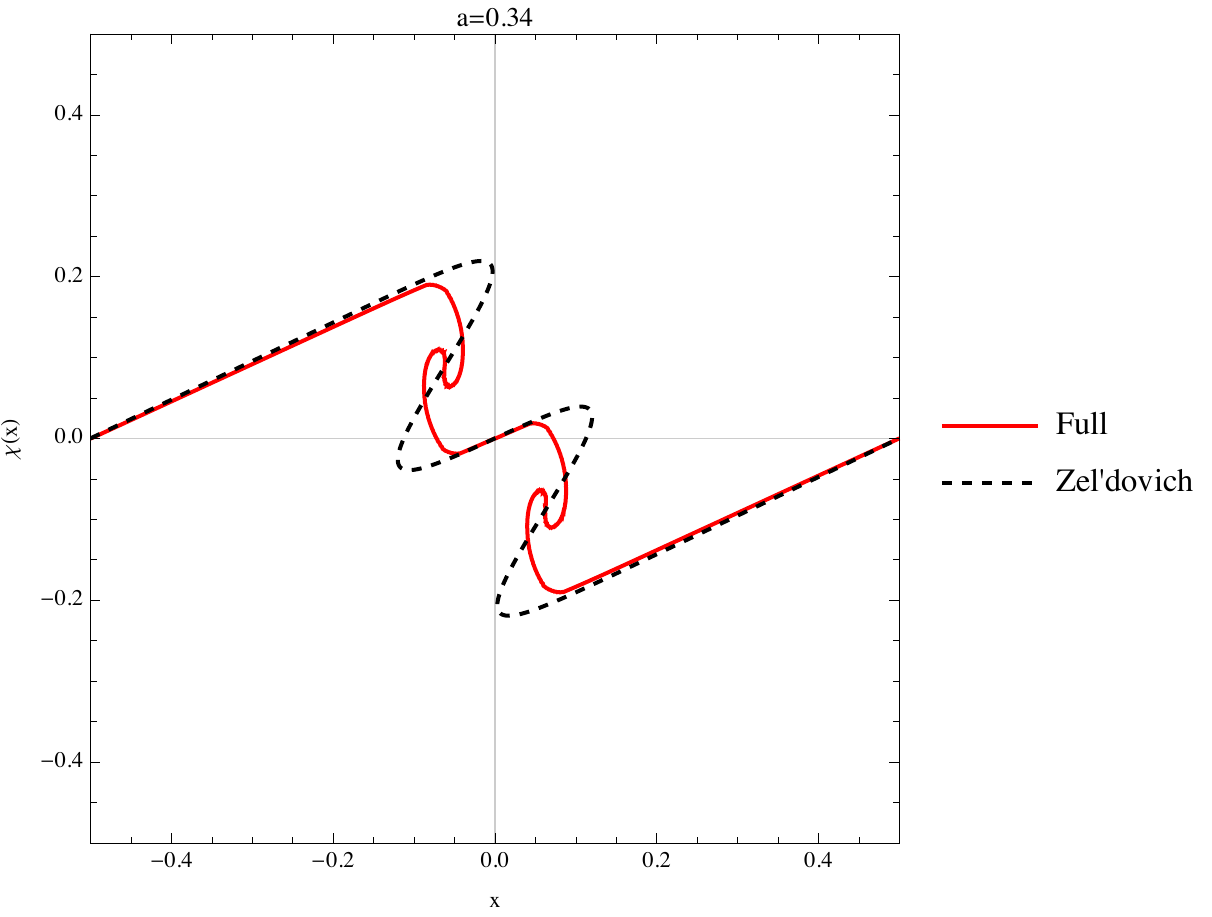}
\includegraphics[width=.45\textwidth,clip]{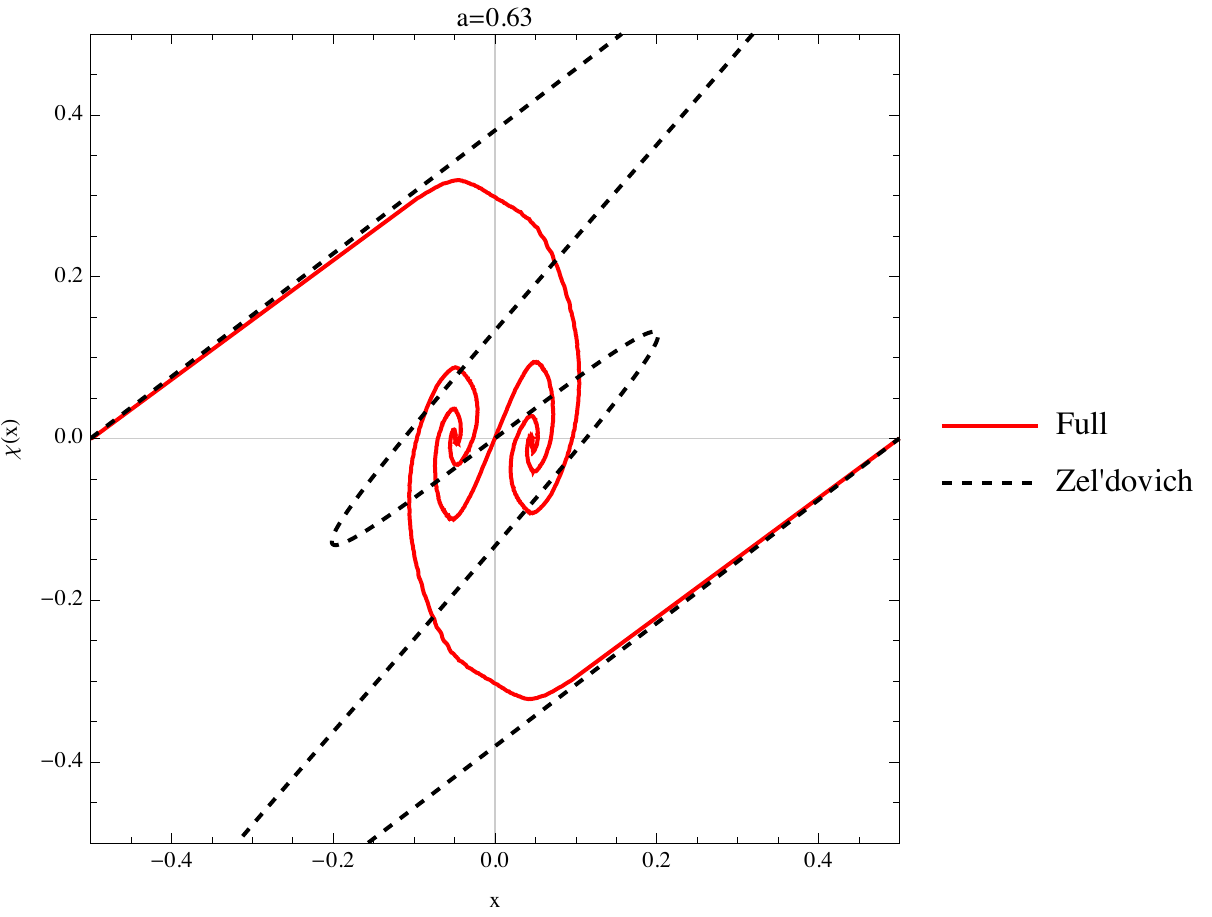}
\includegraphics[width=.45\textwidth,clip]{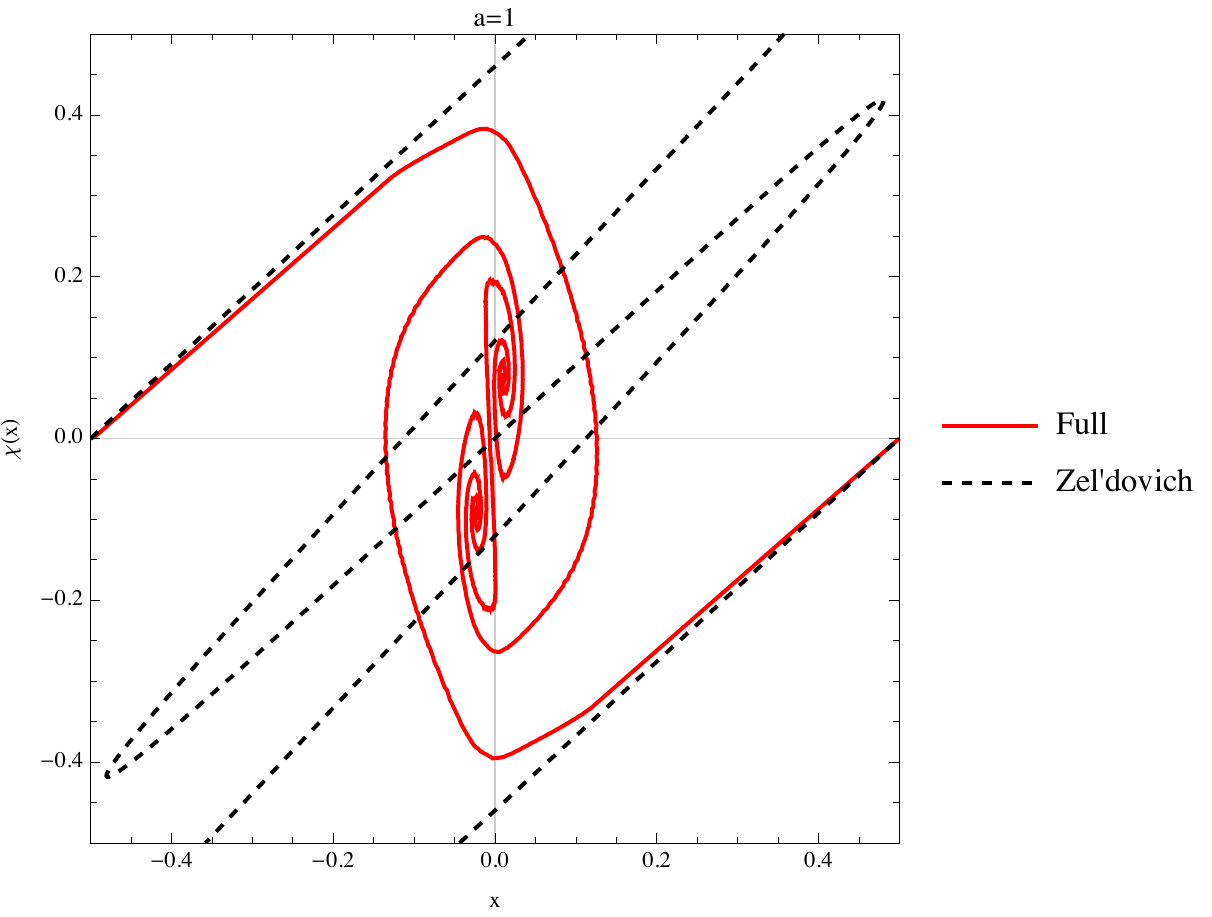}
\caption{Phase-space snapshots of the evolution of the initial conditions of Fig.~\ref{feature2} without features, in the exact dynamics (continuous red) and in Zel'dovich approximation (dashed black).  }
\label{gauss}
\end{figure}

\begin{figure}%[tbp]
\centering
\includegraphics[width=.48\textwidth,clip]{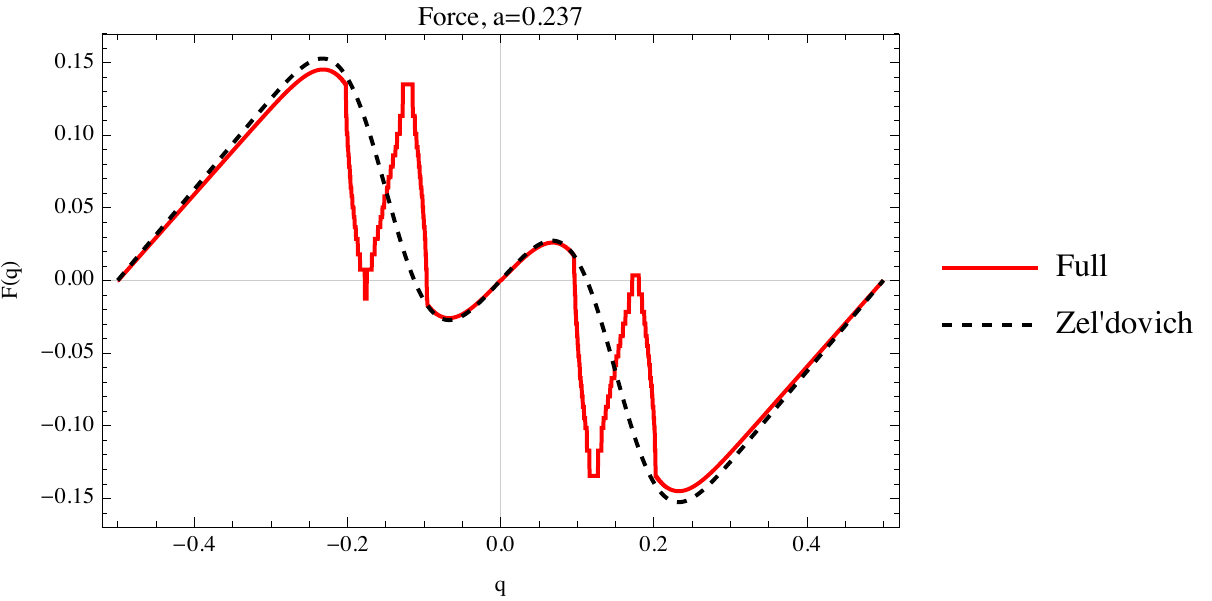}
\includegraphics[width=.48\textwidth,clip]{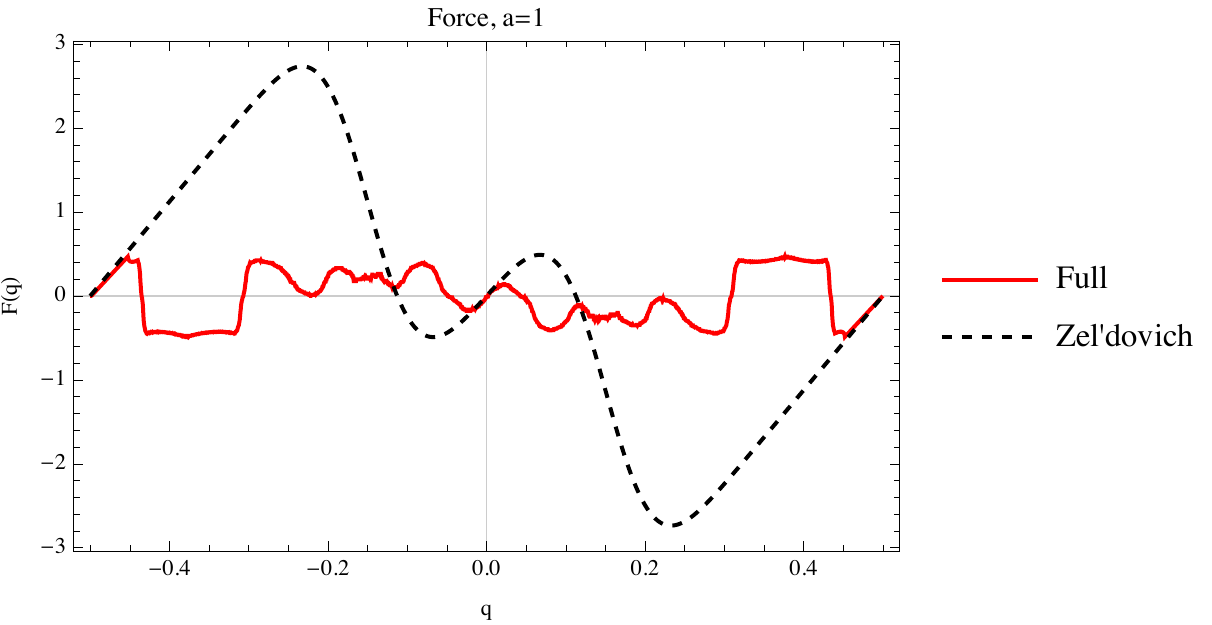}
\caption{The evolution of the exact force (continuous red) and the Zel'dovich one (dashed black)  for the initial conditions Fig.~\ref{feature2} without features. }
\label{force2}
\end{figure}

The evolution in phase space and the evolution of the force in the featureless case is given in Figs.~\ref{gauss} and \ref{force2}, respectively. 

\begin{figure}%[tbp] 
\centering\includegraphics[height=.29\textwidth]{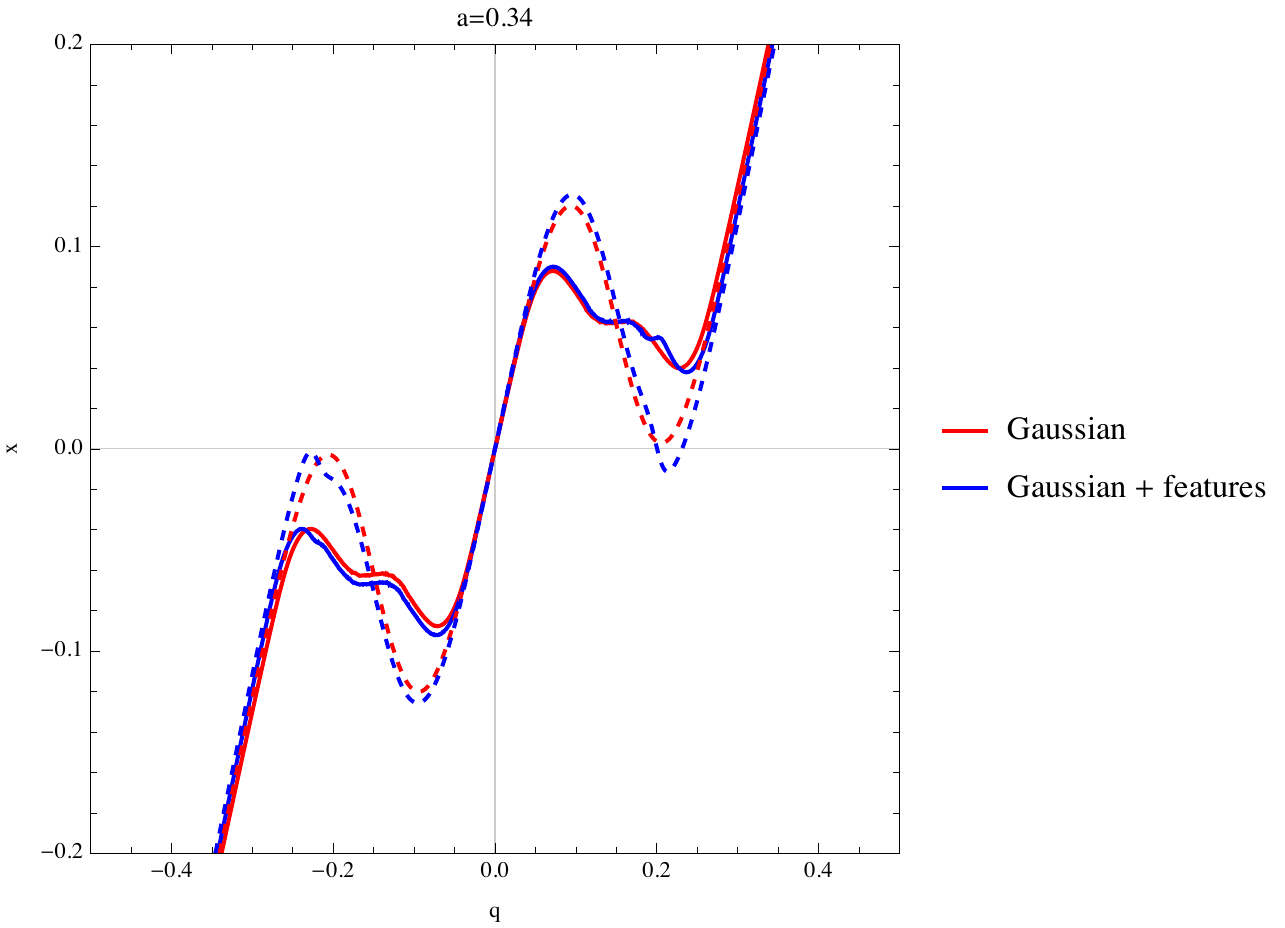}
\includegraphics[height=.29\textwidth]{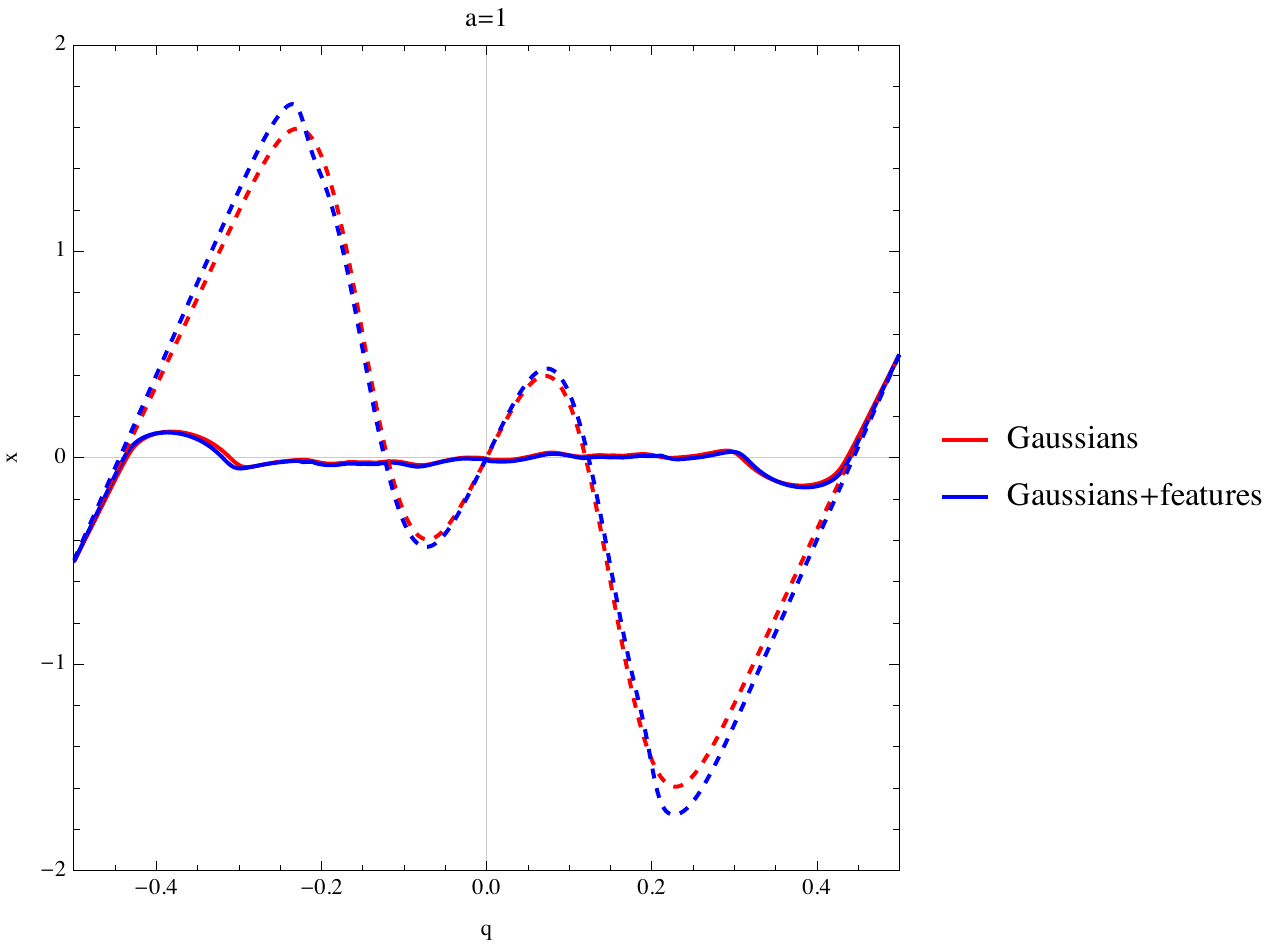}
\includegraphics[height=.29\textwidth]{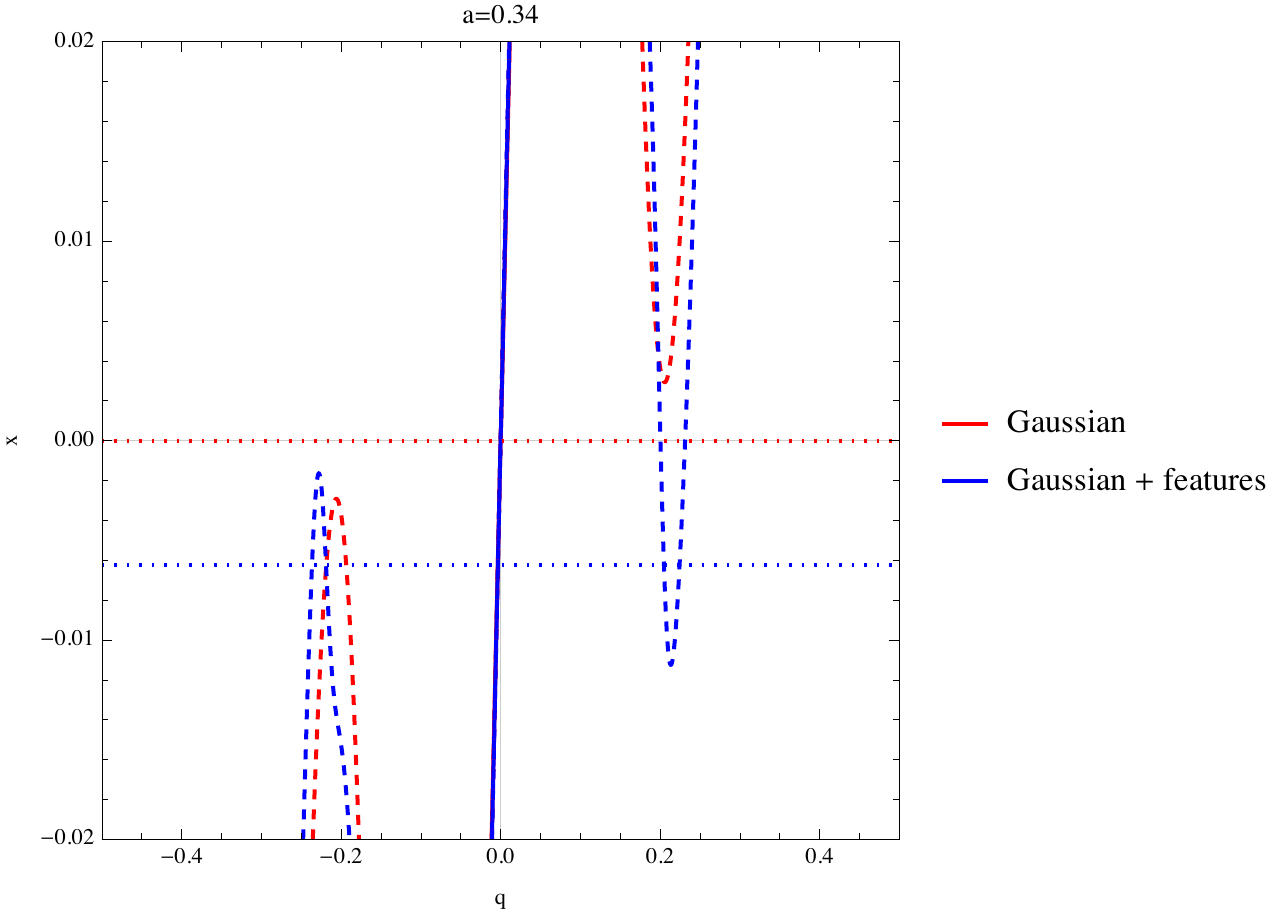}
\includegraphics[height=.29\textwidth]{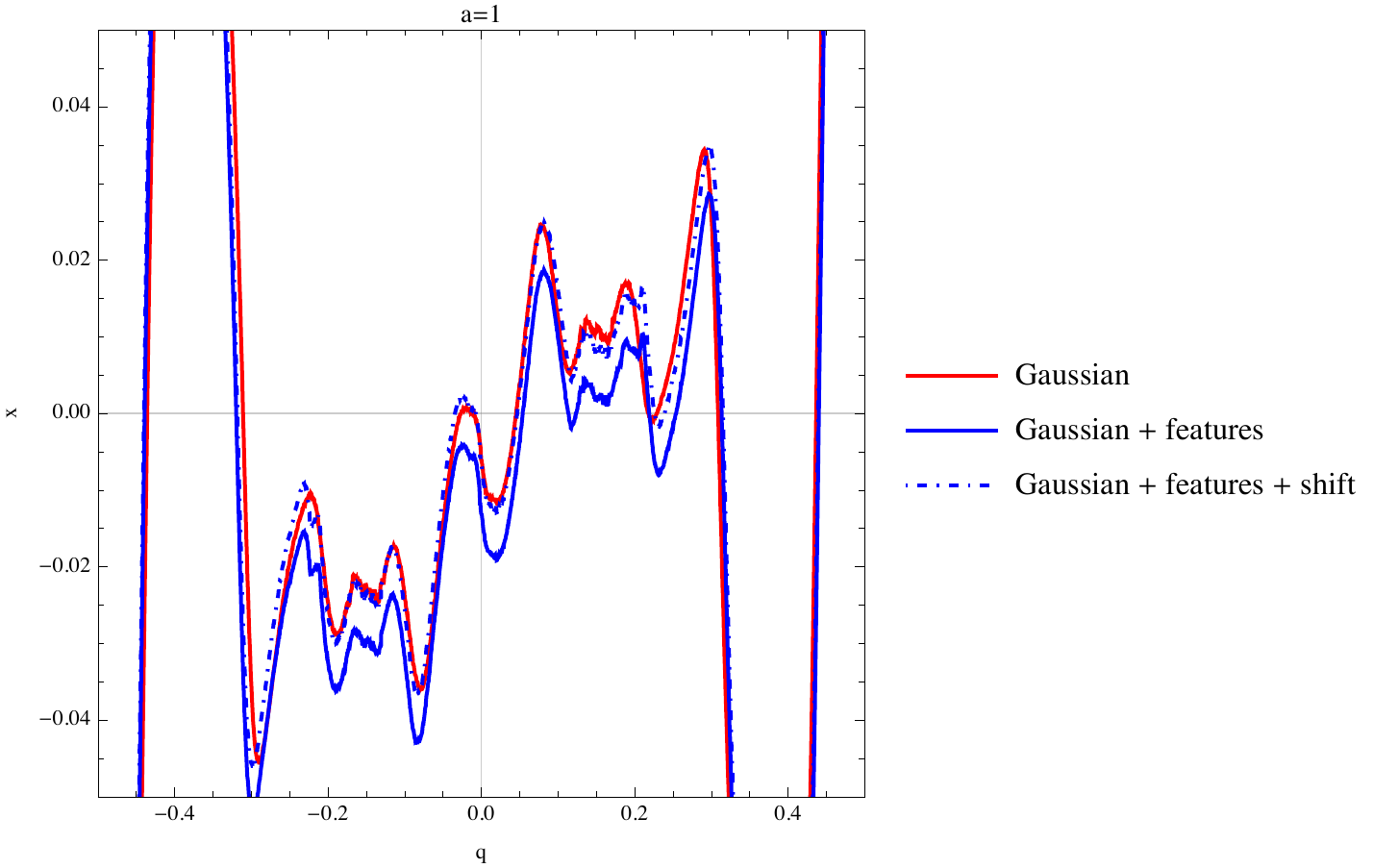}
\caption{Lagrangian to Eulerian space mapping for the initial conditions of Fig.~\ref{feature2} without (red) and with features (blue). Solid lines are for the exact dynamics and dashed ones for Zel'dovich. In the lower left panel we also show the effect of shifting the blue line by the center of mass deviation induced by the features, evaluated in the Zel'dovich solution before second shell-crossing. }
\label{eulag2}
\end{figure}

The flattening inside the shell-crossing region is evident also in this case, see Fig.~\ref{eulag2}, and the main difference induced by the features is, again, a shift in the position of the plateau. In order to compute this shift, we apply eq.~\re{barx} to the Zel'dovich solution at the time at which the featureless initial conditions evolve to shell-crossing at $q=0$, namely at $a=0.34$, and compare it with the Zel'dovich solution with features at the same redshift. In this way, the information on the shift of the center of mass induced by the features is preserved, and the latter is given by the difference between the horizontal dotted lines in the lower-left plot in Fig.~\ref{eulag2}. Had we computed the shift from the Zel'dovich solution at $z=0$ we would have got it wrong, as the multiple streams predicted by this approximation are completely irrealistic and cannot account for the true position of the center of mass of the shock. 

\begin{figure}[t] 
\centering
\includegraphics[height=.32\textwidth]{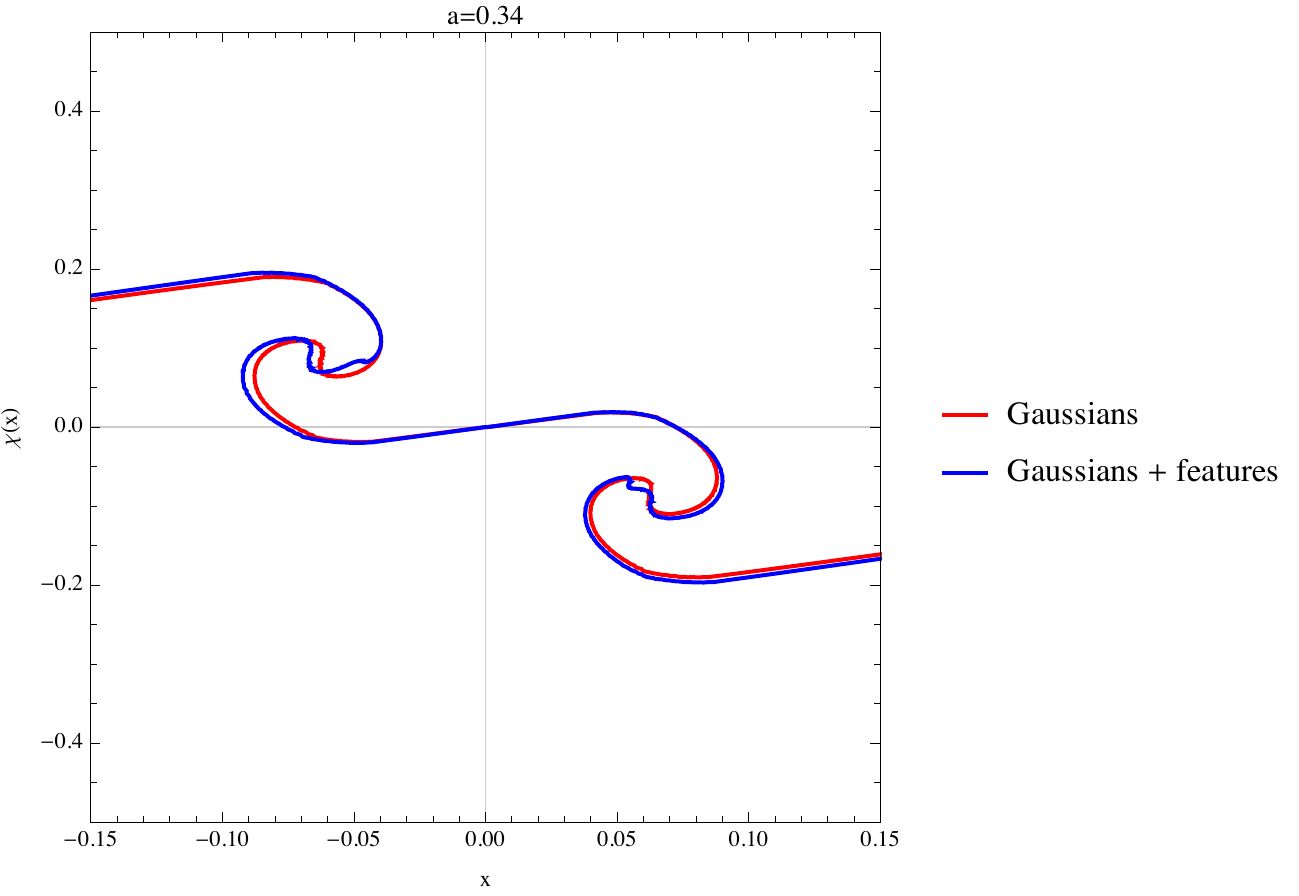}
\includegraphics[height=.32\textwidth]{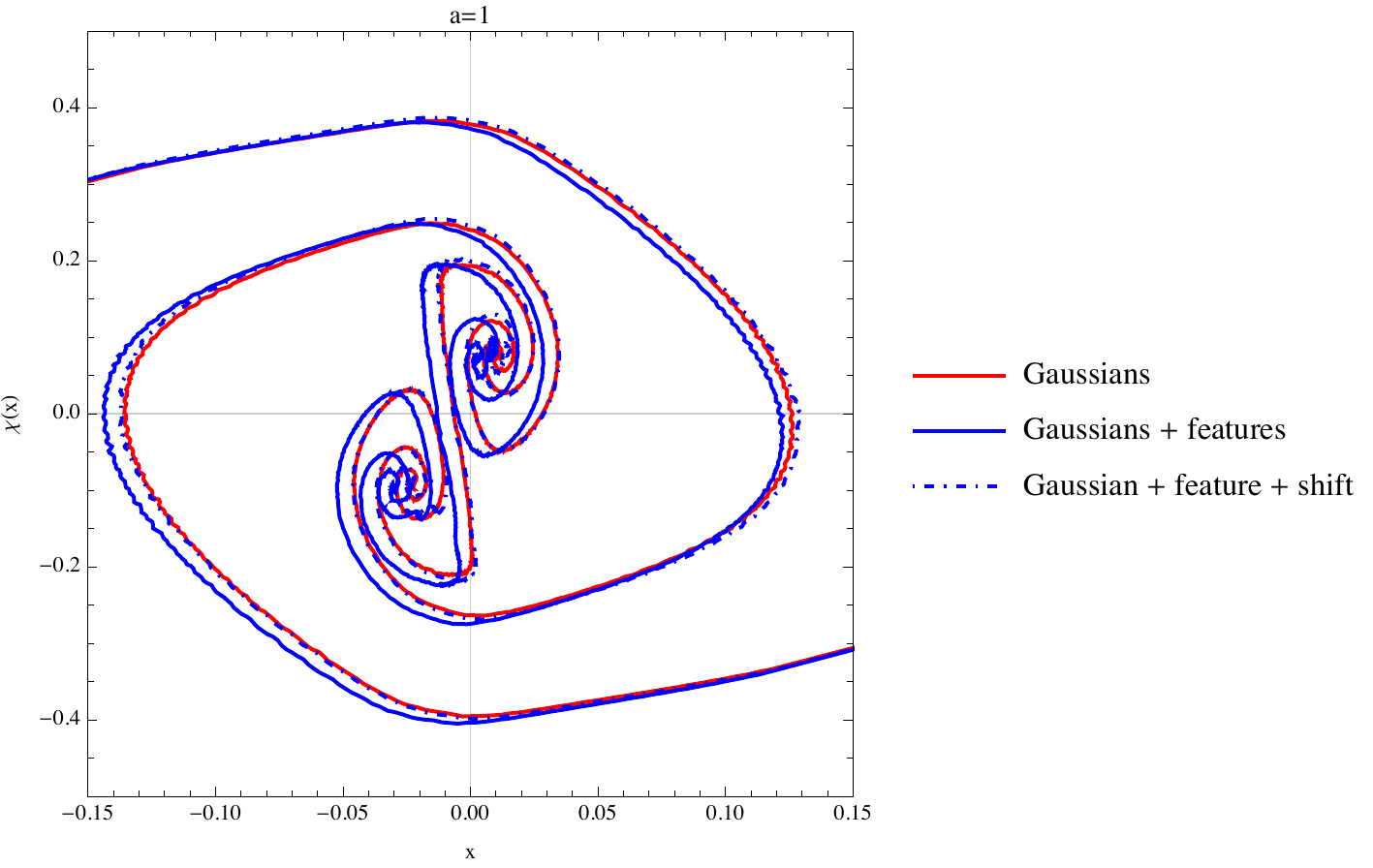}
\caption{Phase space solution for the initial condition of Fig.~\ref{feature2}  without (red) and with (blue) the added feature. At $a=1$ we also show the effect of shifting the blue curve by the deviation of the center of mass position induced by the feature, obtained according to Eq.~\re{barx}  from the Zel'dovich solution before second shell-crossing (dash-dotted blue).}
\label{pswf2}
\end{figure}

On the right we show that this effect accounts for the  shift between the exact solutions at $a=1$. The same is shown in phase space in Fig.~\ref{pswf2}, where, again, we see that the direct deformation induced by the features on the phase-space curves is getting swallowed by the multistreaming spirals. 

\section{Attractors in a cosmological setting}
\label{cosmosym}
Finally, we describe what would be a cosmological simulation if the world were 1+1 dimensional. We will consider a power spectrum $P_{1D}(k)$ related to the $3D$ one by
\beq
P_{1D}(k)=\frac{k^2}{2\pi}P_{3D}(k)\,,
\eeq
where the linear 3D PS has been obtained by the Class Boltzmann code \cite{Blas:2011rf}. The relation above results in the same variance in the density per interval in $k$ as in 3D CDM, as well as the same linear-order parallel RMS displacement \cite{McQuinn:2015tva}. We choose $n_s=0.966$, $\Omega_b h^2=0.02269$, $\Omega_m h^2=0.134$, $h=0.703$, and scalar amplitude $A_s=2.42\cdot10^{-9}$.  The linear PS is scaled back at  $z=99$ using the growth factor computed from the above cosmological parameters,  and then used to give the gaussian initial conditions in Fourier space
\beq
\Psi(q,\etain)=\chi(q,\etain)=\frac{1}{L}\sum_{m=1}^{N_p} c_m \cos\left(q \,p_m+\phi_m\right)\,,
\eeq
with $p_m=2\pi\, m  L^{-1}$.
The $c_m$'s are taken from a Rayleigh distribution with
\beq
\sigma_m=\sqrt{ \frac{L P_{1D}(p_m)}{2 p_m^2} }\,,
\eeq
where the $1/p_m^2$ factor comes from the relation \re{psidelta}, which in Fourier space reads,
\beq
\tilde\Psi(p_m,\etain) = i\frac{\tilde\delta(p_m,\etain)}{p_m}\,.
\eeq
The phases $\phi_m$ are extracted randomly from the $[0,2\pi)$ interval.

We perform a run on a one-dimensional line of $L=1\,\mathrm{Gpc\,h^{-1}}$, sampled in 4000 points. 
We run from $a_{in}=1/100$ to $a=1$ in 100 logarithmic steps. The evolution forward in time assumes an Einstein de Sitter cosmology.

\begin{figure}[t]
\centering
\includegraphics[width=.48\textwidth,clip]{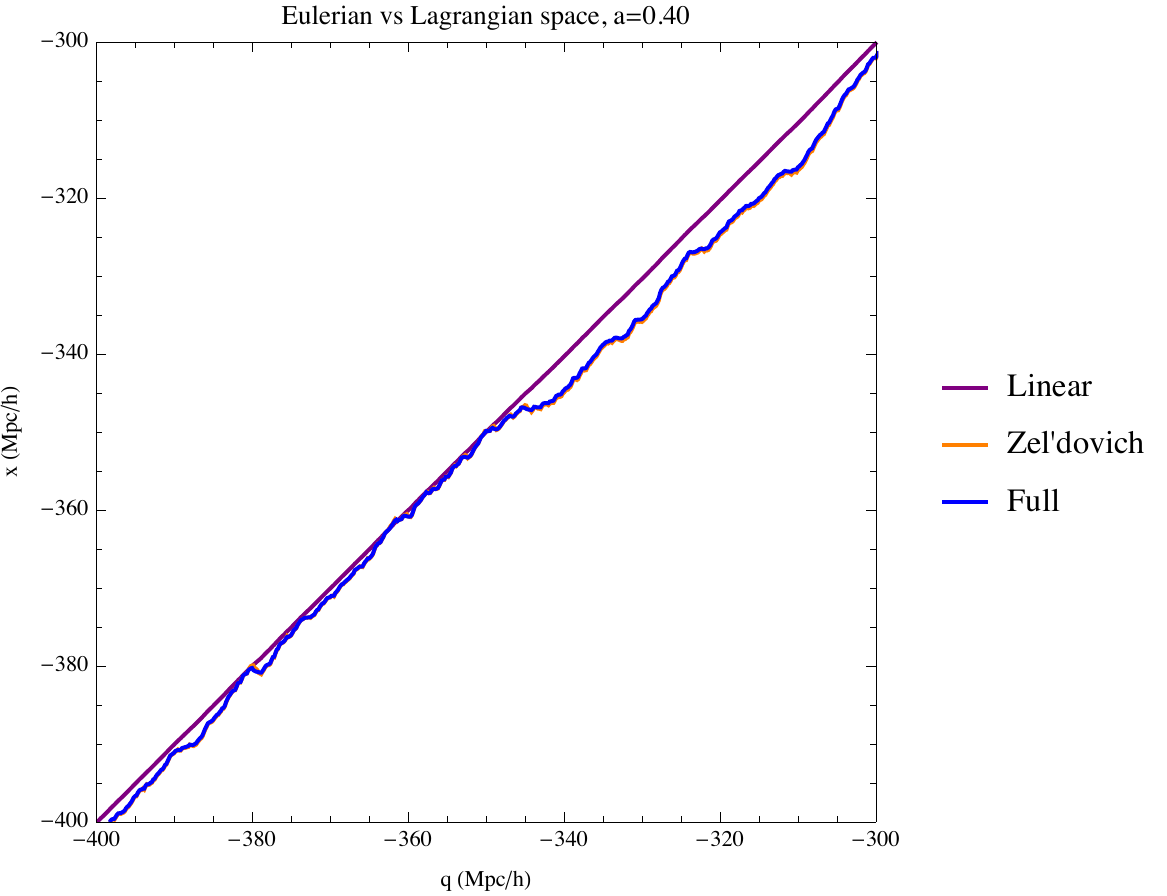}
\includegraphics[width=.48\textwidth,clip]{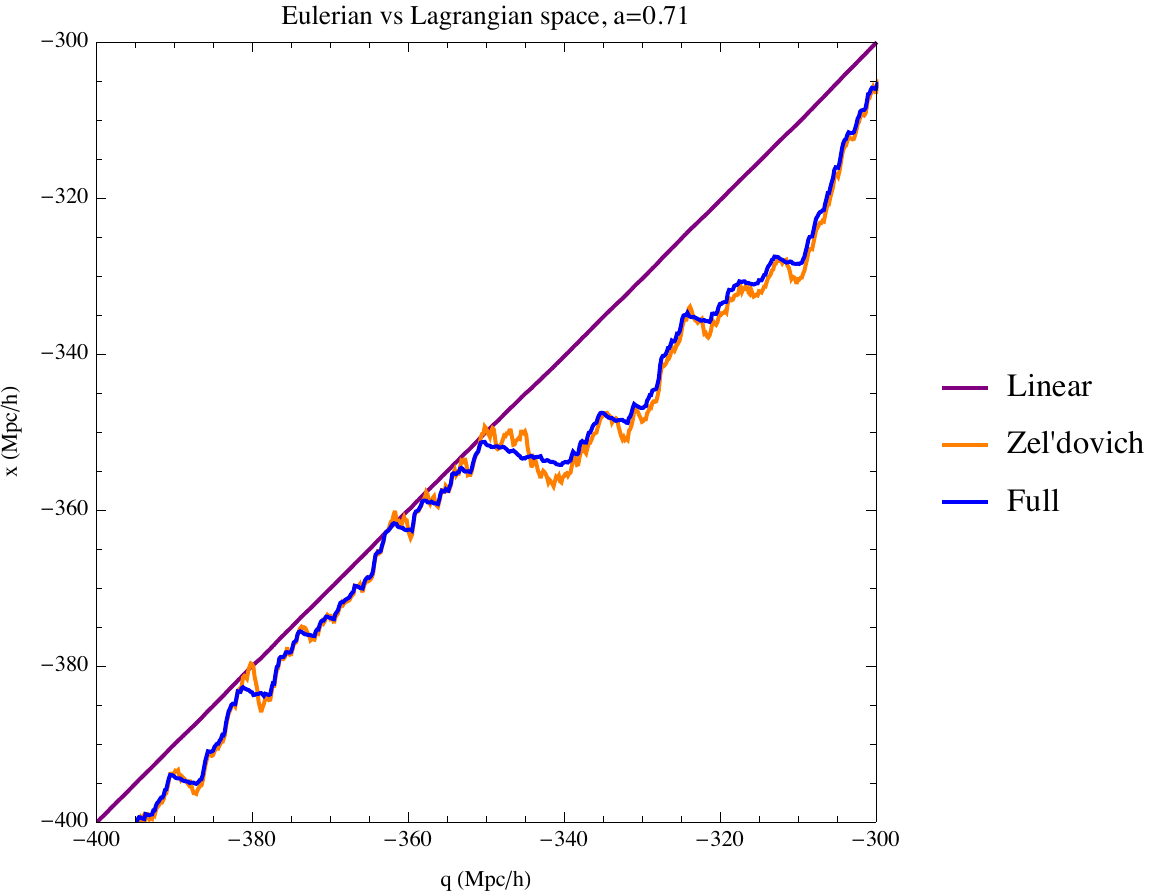}
\includegraphics[width=.48\textwidth,clip]{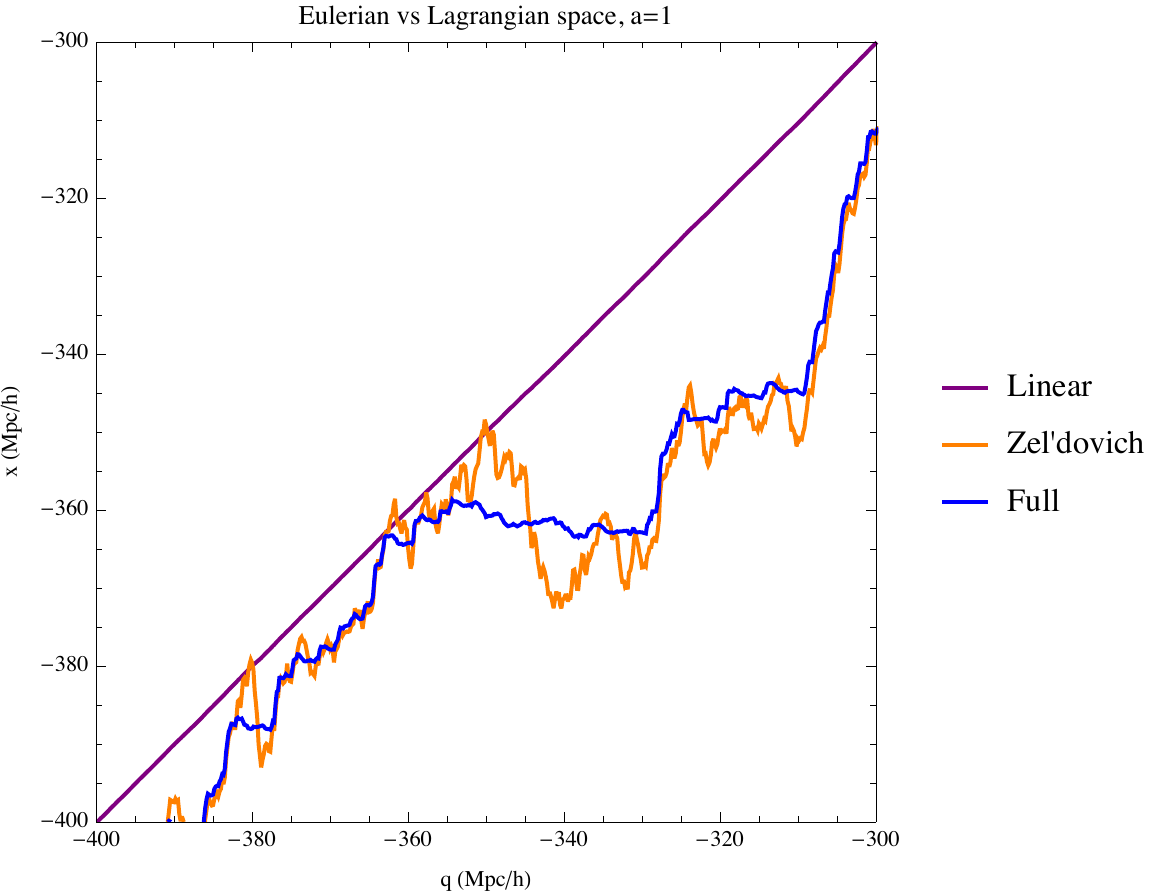}
\caption{Lagrangian to Eulerian mapping for a portion of our cosmological simulation. The purple line is the initial condition, the orange one the Zel'dovich solution, and the blue line is obtained from the exact dynamics.  }
\label{cosmosim}
\end{figure}

\begin{figure}%[tbp] 
\centering
\includegraphics[height=.30\textwidth]{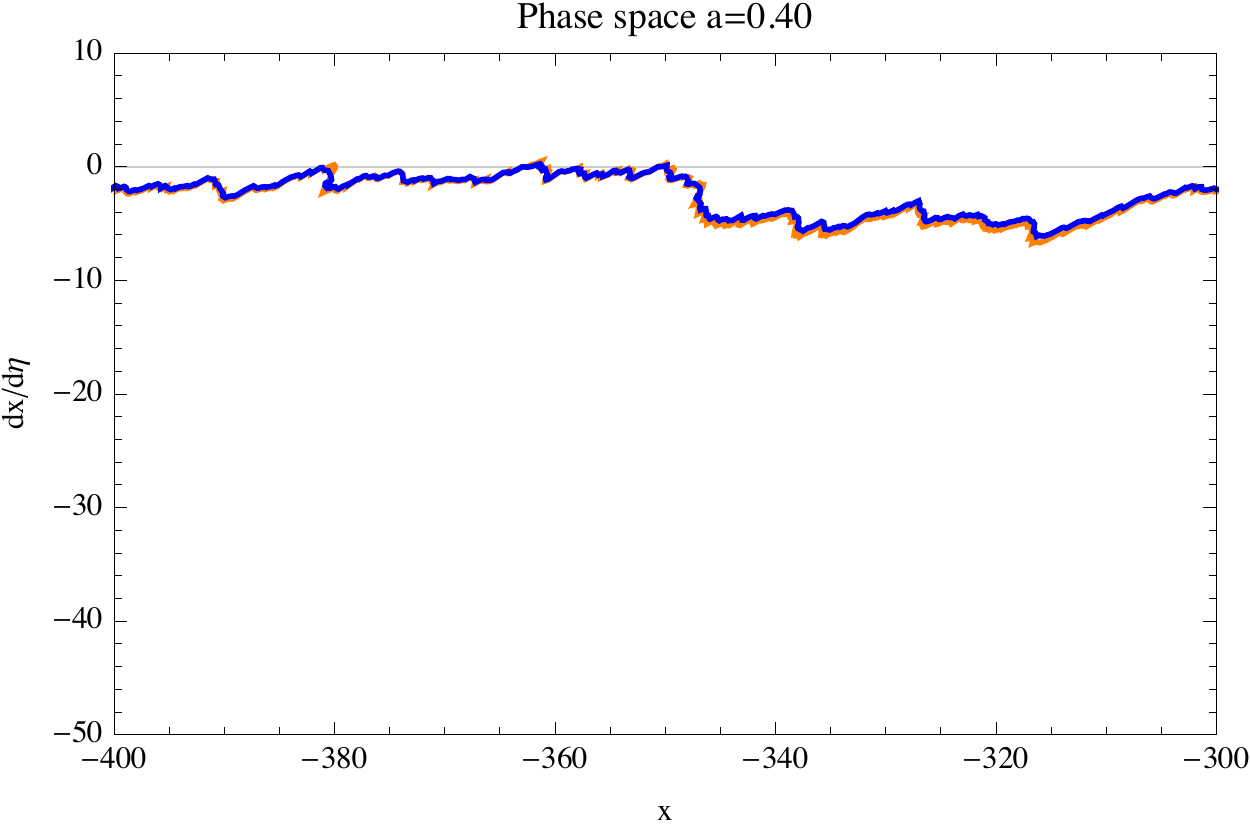}
\includegraphics[height=.30\textwidth]{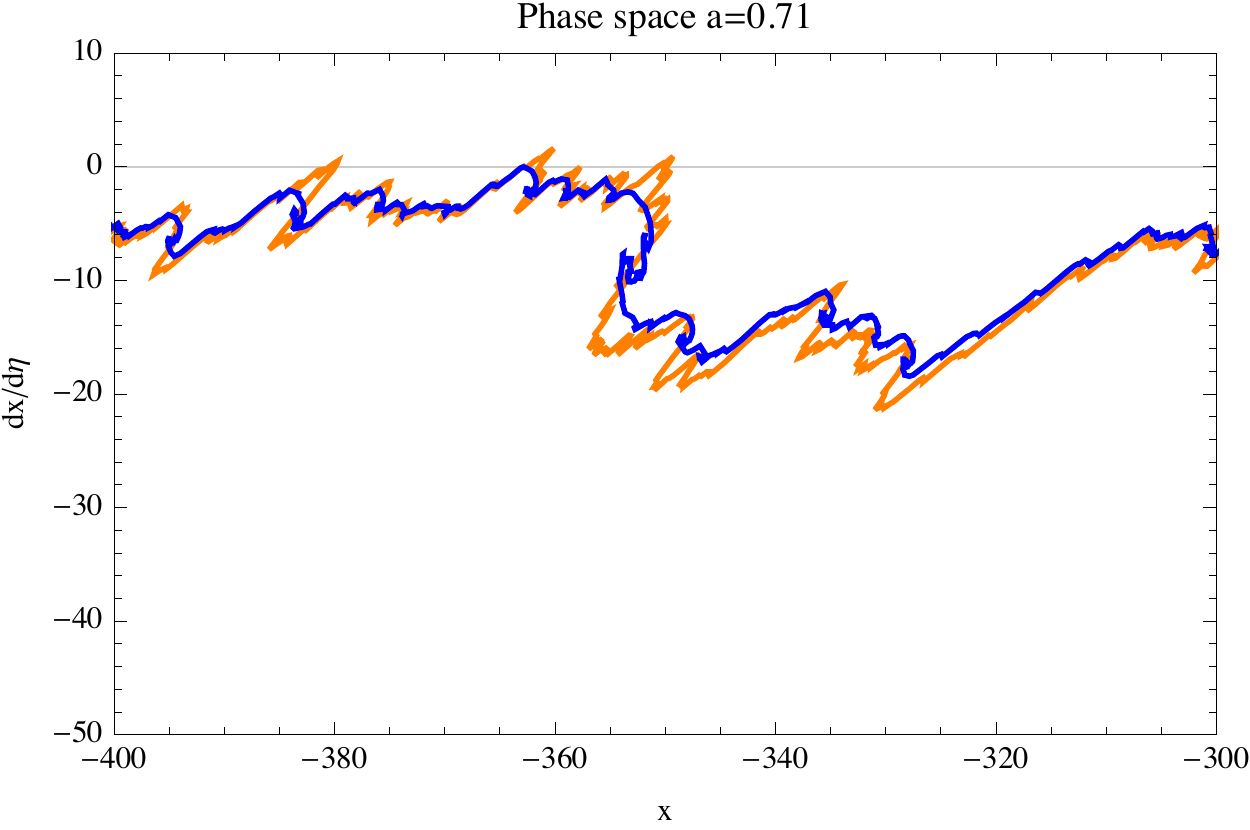}
\includegraphics[height=.30\textwidth]{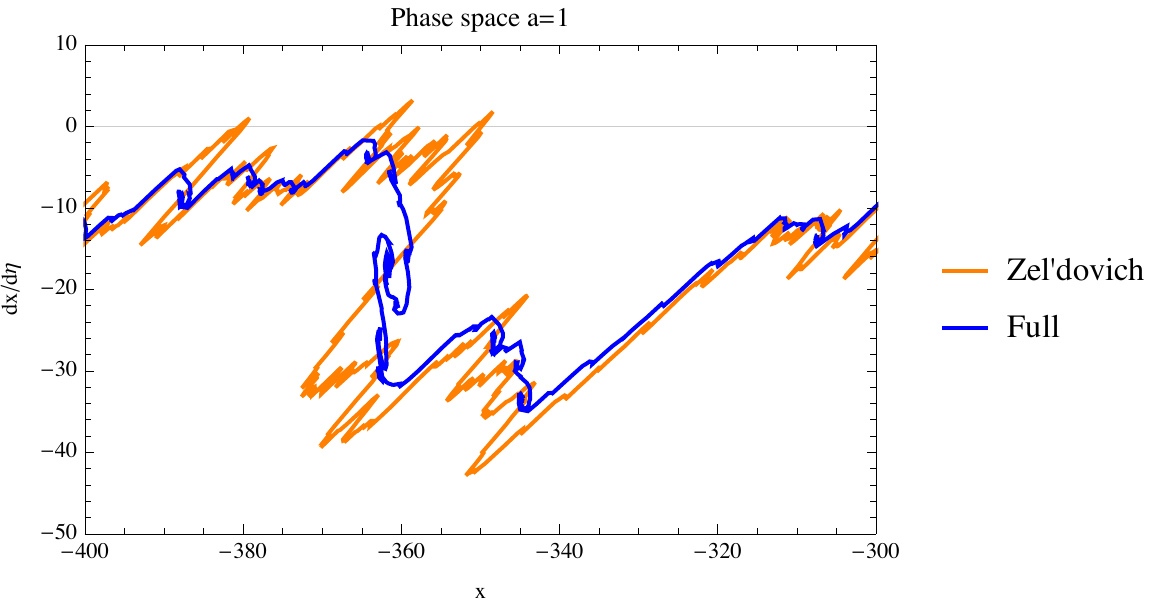}

\caption{Phase space for a portion of our cosmological simulation. The  the orange line is the Zel'dovich solution, and the blue line is obtained from the exact dynamics.   }
\label{cosmosimphase}
\end{figure}

In Fig.~\ref{cosmosim} we show our results on the Lagrangian to Eulerian mapping zooming on  a portion of space of $100\,\mathrm{Mpc \;h^{-1}}$. We confirm the general trend observed for simple initial conditions in the previous sections: the Zel'dovich approximation is able to roughly identify the regions undergoing multistreaming in Lagrangian space, but greatly overestimates their extension in Eulerian space. The failure of the Zel'dovich approximation after shell-crossing is even more evident in phase space, see Fig.~\ref{cosmosimphase}. 

\begin{figure}%[tbp] 
\centering
\includegraphics[width=.62\textwidth]{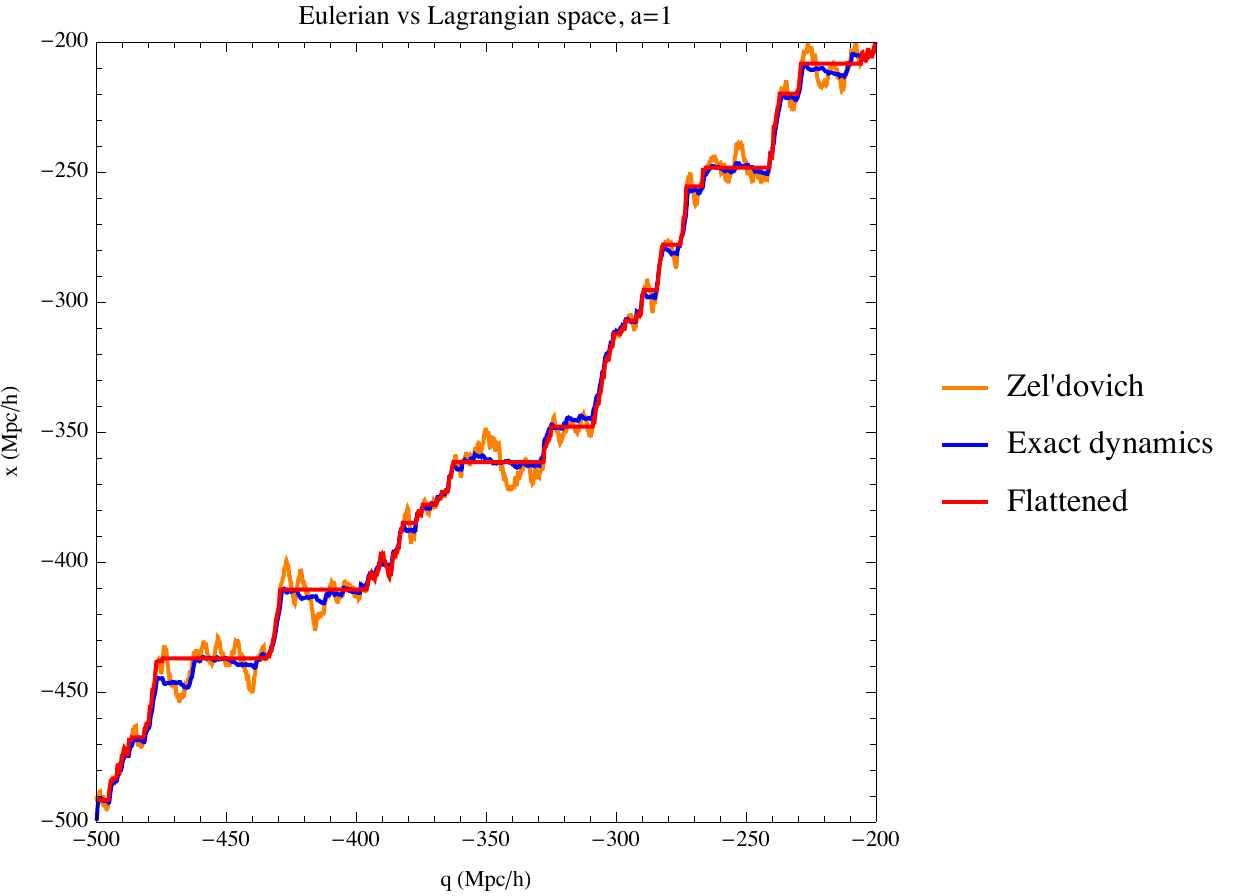}
\caption{Again, Lagrangian to Eulerian mapping for a portion of our cosmological simulation. We also show the result of the ``flattening" algorithm discussed in the text (red line), constracted from the Zel'dovich approximation at $z=0$. }
\label{eulag}
\end{figure}

Following the discussion of the previous section, we now investigate to what extent the attractor behavior can describe the displacement field at late times.  First, we implement a simple algorithm to ``flatten out'' the Lagrangian to Eulerian mapping inside multistreaming regions, in order to reproduce the asymptotic state. To do so, we consider the Zel'dovich approximation at $a=1$ and find the points $\bar x$'s such that eq.~\re{barx} is fullfilled. This procedure has a number of shortcomings, which should be addressed in order to improve it. First of all, as we discussed in the two-gaussians case, in order to properly evaluate $\bar x$,  the Zel'dovich approximation should be used in each region only between the first and the second shell-crossing, as, after that, it becomes unreliable in estimating the extension of the Lagrangian region falling in a given Eulerian multistreaming region. Therefore, we should implement the plateau-finding algorithm \re{barx} in time, extracting the position of each ``halo''  from the Zel'dovich approximation after the first shell crossing epoch for that particular halo, and not at the same epoch ($a=1$ in our case) for all halos. Second, and connected to the previous point, we have to artificially cut off the maximum possible length of the multistreaming intervals, otherwise our operation on the final Zel'dovich solution would end up in a large number of supermassive ``halos".
Finally, the attractor is exact only asymptotically, therefore the flattening is never completely accomplished at finite times.
Nevertheless our flattened displacement field (red lines in Fig.~\ref{eulag})  provides a reasonable localization of real ``halos", failing mainly for haloes corresponding to a large number of streams in the Zel'dovich approximation.

\begin{figure}%[tbp] 
\centering
\includegraphics[width=.45\textwidth]{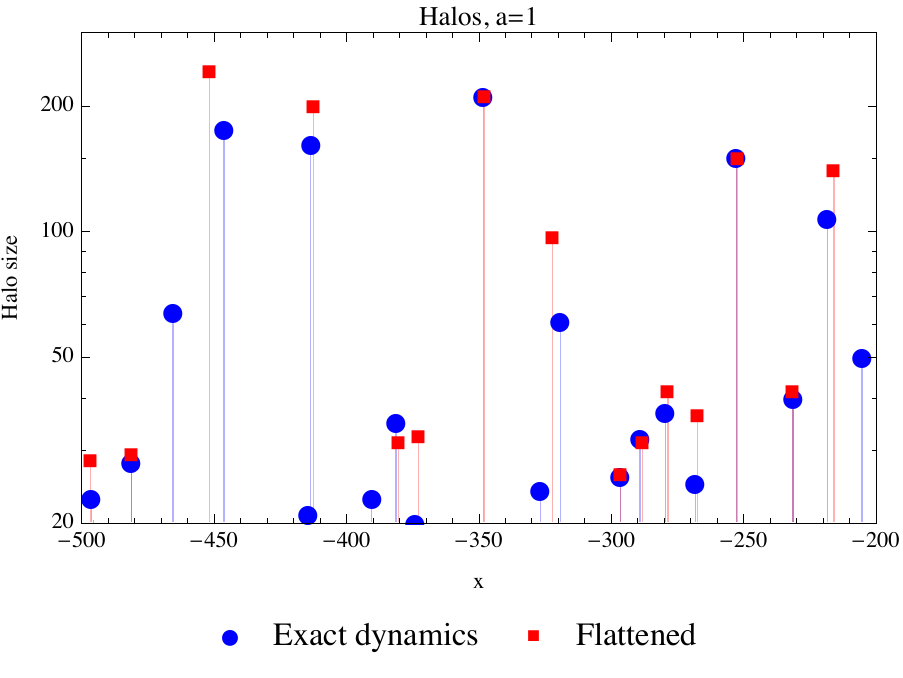}\\
\includegraphics[width=.45\textwidth]{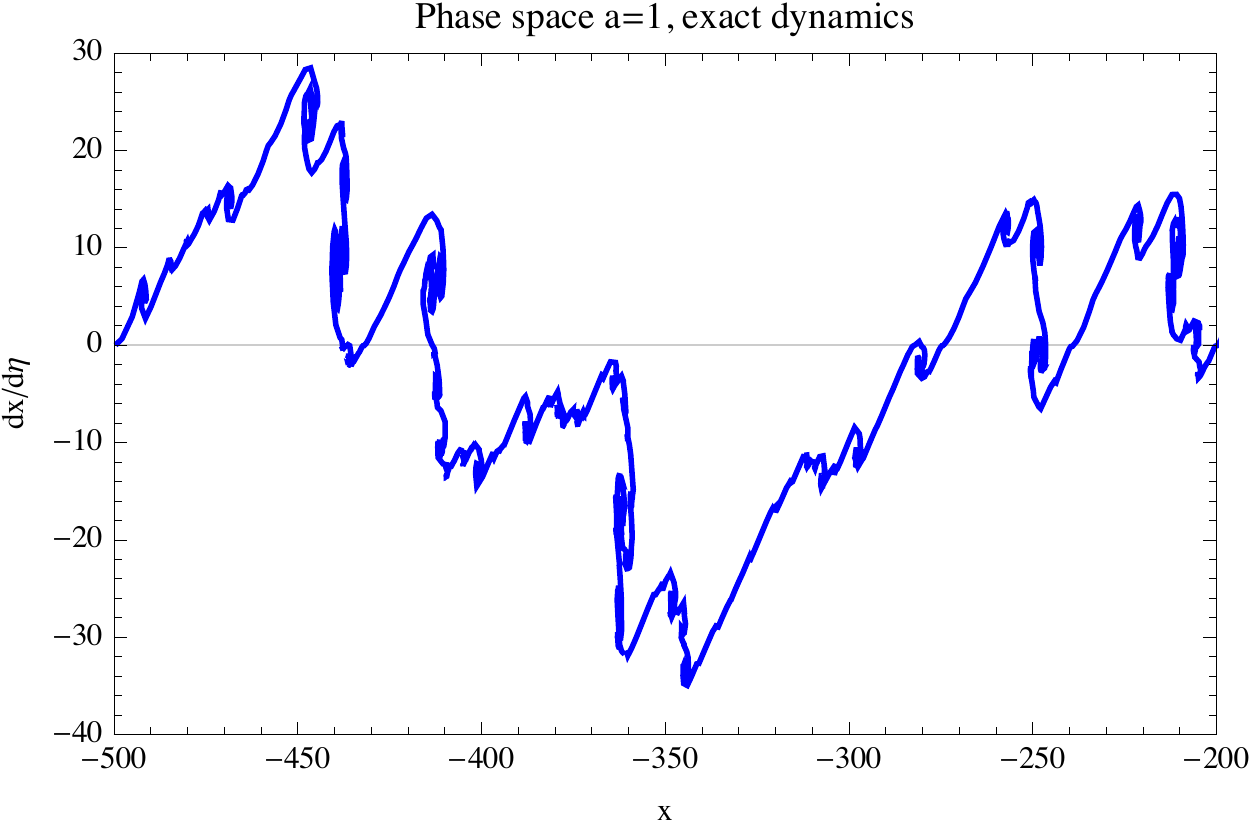}
\caption{(Upper panel) Location and sizes of the ``halos'' identified from the cosmological simulation with the exact dynamics (blue circles) and from the flattened Zel'dovich approximation discussed in the text (red squared). (Lower panel) Phase space in the same portion of Eulerian space, obtained from the exact dynamics. }
\label{halos}
\end{figure}

A complementary view of this result is given in Fig.~\ref{halos}, where the squares and circles represent the positions and sizes of halos obtained by a friends of friend algorithm for the real dynamics and the flattened result, respectively. The flattening prescription is able to reproduce the location and sizes of the halos of the real simulation. It typically overestimates the size of the halos, mainly due to the exact flat limit. This problem can be mitigated by introducing some form of smoothing on the halo profiles of the flattened Zel'dovich solution.

\section{Outlook}
\label{out}
What is the link between the first and the second part of this paper? In section \ref{npterms} we showed, using a toy model (Zel'dovich dynamics in 1+1 dimensions) that the origin of the failure of the PT expansion, namely, shell-crossing, manifests itself as a divergent contribution, which acts as a ``bridge'' between the perturbative and the nonperturbative sectors. In concrete, eq.~\re{sumnpt} shows that one can obtain the terms of the PT expansion, at any order, by taking the $\sigma_A\to \infty$ limit of the corresponding term in the nonperturbative  expansion. Of course, for practical reasons, one would prefer to be in  the inverse situation (get the nonperturbative expansion from the perturbative one), nevertheless, the connection exhibited by this simple computation is remarkable. 

In the second part of this work we considered the real dynamics and showed that, inside multistreaming regions, the mapping between Lagrangian and Eulerian spaces becomes flatter and flatter, that is, more and more singular. Therefore, it is not unreasonable that also the statistical averages of the exact density field can be represented by transseries, and that  eq.~\re{tss} would then be just one of the simplest examples of a large class. In this class of expansions, the presence of singularities forces a correlation between PT and nonperturbative terms. Exploring these connections and considering nonperturbative expansions can, in principle, shed light on how to extend the range of validity of analytic methods beyond shell-crossing.

Another pressing question is, of course, the extension of these 1+1 results to the 3+1 world. We believe that some of our results are generic, even if the possibility to express them in manageable analytical terms in 3+1 dimensions is still to be proved. A generic result is that post shell-crossing effects cannot be expressed as power laws in the PT expansion parameter which is, ultimately, proportional to the linear growth factor.  Therefore, one cannot expect power law dependence on the linear growth  factor of fully nonperturbative quantities, such as the UV sources of coarse grained PT \cite{Pietroni:2011iz,Manzotti:2014loa}, or the counterterms of the effective field theory of the large scale structure \cite{Carrasco:2012cv}. Indeed, as shown in \cite{Manzotti:2014loa}  the time dependence of these source terms appears to be much steeper than predicted in PT and cannot be expressed as a power law. It would be of great interest to see if it can be expressed by  nonperturbative terms such as $\sim \exp(-\alpha/D^2)$ which become sizable only at low redshifts.

The other generic feature is, probably, the existence of attractors in the multistreaming regime. A complete characterization of these should be obtained in phase space, also for the 1+1 case. In this connection, our examples of modifications of the initial conditions adding some ``features'' shows that, es expected, most of the information gets practically lost as the attractor is approached. The language of the renormalization group could be fruitfully employed to tell apart  the `relevant' content of the initial conditions (in our example, the position of the center of mass before second shell crossing) from the `irrelevant' one (as the detailed shape of the added features). This type of analysis could be of practical use to extend the reconstruction procedures presently used to recover linear information from the nonlinear configurations.

\section*{Acknowledgments}
It is a pleasure to thank D. Comelli, L. Griguolo, S. Matarrese and C. Pavlidis for many useful discussions and comments.
The author acknowledges support from the European Union Horizon 2020 research and innovation programme under the Marie Sklodowska-Curie grant agreements Invisible- sPlus RISE No. 690575, Elusives ITN No. 674896 and Invisibles ITN No. 289442.
\appendix
\section{Computation of eq.~\re{tss}}
\label{transerie}

Our purpose is to express the integral defined in \re{it}
\beq
I[\sigma_A,\epsilon]\equiv  \int\frac{dy}{\sqrt{2 \pi}} \frac{e^{-\frac{y^2(1+\epsilon^2y^2)^2}{2\left(1+y^2 \sigma_A^2\right)}}}{\sqrt{1+y^2 \sigma_A^2}}\,,
\label{it2}
\eeq
as a transseries containing, besides powers of the expansion parameter $\sigma_A$, also nonperturbative terms of the type $e^{-1/\sigma_A^2}$ and its powers, and, as we will see, logarithmic terms. We are interested in the $\epsilon \ll \sigma_A$ regime, in practice, we will take the $\epsilon\to 0$ limit whenever possible. On the other hand, we will not make any assumption about the size of $\sigma_A$, in particular, we will not assume it to be less than unity.

We can split the integration interval in three regions, $0<|y|<\alpha/\epsilon$, $\alpha/\epsilon<|y|<\beta/\epsilon $ and $|y|>\beta/\epsilon$, with $\alpha\alt 1$ and $\beta\agt 1$, so that we can approximate
\beq
I[\sigma_A,\epsilon]\simeq I_1[\sigma_A,\epsilon]+I_2[\sigma_A,\epsilon]+I_3[\sigma_A,\epsilon]\,,
\label{Isplit}
\eeq
with
\beqra
&&
I_1[\sigma_A,\epsilon] = \int_{|y|<\alpha/\epsilon} \frac{dy}{\sqrt{2 \pi}} \;\frac{e^{-\frac{y^2}{2\left[1+y^2 \sigma_A^2\right]}}}{\sqrt{1+y^2 \sigma_A^2}}\,,\nonumber\\
&& I_2[\sigma_A,\epsilon] = \int_{\alpha/\epsilon<|y|<\beta/\epsilon} \frac{dy}{\sqrt{2 \pi}} \;\frac{e^{-\frac{ (1+\epsilon^2y^2)^2}{2 \sigma_A^2}}}{y\,\sigma_A}\,,\nonumber\\
&& I_3[\sigma_A,\epsilon] = \int_{|y|>\beta/\epsilon} \frac{dy}{\sqrt{2 \pi}} \;\frac{e^{-\frac{ \epsilon^4y^4}{2 \sigma_A^2}}}{y\,\sigma_A}\,.
\eeqra
We perform the following change of variable in the integral $I_1$,
\beq
y=\frac{z}{\sqrt{1-z^2\sigma_A^2}}\,,
\eeq
to get
\beq
I_1[\sigma_A,\epsilon] =2\int_0^{\frac{1}{\sigma_A}\left(1-\frac{1}{2}\frac{\epsilon^2}{\alpha^2\sigma_A^2}\right)}\;\frac{dz}{\sqrt{2 \pi}} \,\frac{e^{-\frac{z^2}{2}}}{1-z^2 \sigma_A^2}\,,
\label{A5}
\eeq
where in the upper extreme of integration we have expanded in $\epsilon$. Notice that the integral is logarithmically divergent in the $\epsilon\to 0$ limit, as can be seen by integrating by parts,
\beqra
&&I_1[\sigma_A,\epsilon] =\int_0^{\frac{1}{\sigma_A}\left(1-\frac{1}{2}\frac{\epsilon^2}{\alpha^2\sigma_A^2}\right)}\;\frac{dz}{\sqrt{2 \pi}} \,e^{-\frac{z^2}{2}}\left(\frac{1}{1-z\sigma_A}+\frac{1}{1+z\sigma_A}\right),\nonumber\\
&&= \frac{e^{-\frac{1}{2\sigma_A^2}}}{\sqrt{2\pi \sigma_A^2}}\left(\log\left(\frac{\sigma_A^2}{\epsilon^2}\right)+\log(2\alpha^2) \right) + I_1^{\mathrm{finite}}[\sigma_A,\epsilon] \,,
\label{logdiv}
\eeqra
where the $\log(2\alpha^2)$ term depends on the details of the splitting of the integral and of the initial regularization of the mapping in eq.~\re{Gmap} (that is, it would be different for a top-hat instead of gaussian regularization, for instance), and $ I_1^{\mathrm{finite}}[\sigma_A,\epsilon] $ is finite in the $\epsilon\to 0$ limit.

In order to compute the finite terms one can go back to \re{A5} and expand the integrand in powers of $\sigma_A^2$. Notice that the expansion is guaranteed to converge in the whole integration domain as long as $\epsilon>0$. Truncating the summation at some $N_\mathrm{max}$, performing the integral and then letting $\epsilon\to 0$, we get
\beq
I_1^{\mathrm{finite}}[\sigma_A,\epsilon\to0] = \sum_{n=0}^{N_\mathrm{max}} \frac{(2\sigma_A^2)^n}{\sqrt{\pi}} \left[\Gamma\left(n+\frac{1}{2}\right)-\Gamma\left(n+\frac{1}{2},\frac{1}{2\sigma_A^2}\right)\right]\,,
\eeq
where $\Gamma(n,x)=\int_x^\infty dt \;t^{n-1}e^{-t}$ is the incomplete gamma function, and $\Gamma(n+1/2)=\Gamma(n+1/2,0)=\sqrt{\pi}(2n-1)!!/2^n$ for $n$ integer.

The contributions from the first Gamma function reproduce the standard perturbative expansion in $\sigma_A^2$, and 
can be obtained also by expanding the integrand with respect to $\sigma_A^2$ in \re{it2} and then integrating in $y$. The contributions from the second Gamma function carry a common nonperturbative factor, $e^{-\frac{1}{2 \sigma_A^2}}$, as and give, via \re{largecoeff}, the full nonperturbative sector of the transseries \re{tss}, besides the logarithmically divergent  and the indeterminate ones of eq.~\re{logdiv}.

We now examine the remaining  terms in  \re{Isplit}.  The second term can be integrated by parts, to give
\beqra
&&I_2[\sigma_A,\epsilon] =\frac{1}{\sqrt{2\pi \sigma_A^2}}\left(e^{-\frac{(1+\beta^2)^2}{2\sigma_A^2}}\log\frac{\beta^2}{\epsilon^2}-e^{-\frac{(1+\alpha^2)^2}{2\sigma_A^2}}\log\frac{\alpha^2}{\epsilon^2} \right)\,,
\eeqra
plus contributions vanishing in the $\epsilon \to 0$ limit. Both contributions vanish faster than $\sim e^{-\frac{1}{2\sigma_a^2}}/\sigma_A$ and are therefore subdominant nonperturbative contributions with respect to those in \re{logdiv}.

The third term finally gives,
\beq
I_3[\sigma_A,\epsilon] = -\frac{1}{4 \sqrt{2 \pi \sigma_A^2}} \mathrm{Ei}\left(-\frac{\beta^4}{2\sigma_A^2}\right) \sim e^{-\frac{\beta^4}{2 \sigma_A^2}} \frac{\sigma_A}{2\sqrt{2\pi} \beta^4}\left(1+ O(\sigma_A^2 \beta^{-4})\right)\,,
\eeq
where $ \mathrm{Ei}(x)=-\int_{-x}^\infty \,dt e^{-t}/t$, and at the rightmost hand side we have expanded for $\beta^2/\sigma_A\gg1$, to show that also this contribution is subleading in the small $\sigma_A$ limit. 

In summary, the integral \re{it} can be expressed as the sum of perturbative and non-perturbative terms in \re{tss}, given by the $I_1[\sigma_A,\epsilon]$ contribution, plus subdominant nonperturbative terms. All the coefficients of the expansion for the perturbative and the dominant nonperturbative terms are fixed, besides that for the  $\sim e^{-\frac{1}{2\sigma_a^2}}/\sigma_A$ term, which depends on the details of the integral and is therefore to be considered as a ($\epsilon$-independent) parameter to fit. 

\section*{References}
\bibliographystyle{JHEP}
\bibliography{/Users/massimo/Bibliografia/mybib.bib}

\providecommand{\href}[2]{#2}\begingroup\raggedright\begin{thebibliography}{10}

\bibitem{PT}
F.~Bernardeau, S.~Colombi, E.~Gaztanaga and R.~Scoccimarro, {\it {Large-scale
  structure of the universe and cosmological perturbation theory}},  {\em Phys.
  Rept.} {\bf 367} (2002) 1--248
  [\href{http://arXiv.org/abs/astro-ph/0112551}{{\tt astro-ph/0112551}}].
%%CITATION = ASTRO-PH/0112551;%%

\bibitem{Bernardeau:2013oda}
F.~Bernardeau, {\it {The evolution of the large-scale structure of the
  universe: beyond the linear regime}},  in {\em {Proceedings, 100th Les
  Houches Summer School: Post-Planck Cosmology: Les Houches, France, July 8 -
  August 2, 2013}}, pp.~17--79, 2015.
\newblock \href{http://arXiv.org/abs/1311.2724}{{\tt 1311.2724}}.
%%CITATION = ARXIV:1311.2724;%%

\bibitem{Pietroni:2011iz}
M.~Pietroni, G.~Mangano, N.~Saviano and M.~Viel, {\it {Coarse-Grained
  Cosmological Perturbation Theory}},  {\em JCAP} {\bf 1201} (2012) 019
  [\href{http://arXiv.org/abs/1108.5203}{{\tt 1108.5203}}].
%%CITATION = ARXIV:1108.5203;%%

\bibitem{Manzotti:2014loa}
A.~Manzotti, M.~Peloso, M.~Pietroni, M.~Viel and F.~Villaescusa-Navarro, {\it
  {A coarse grained perturbation theory for the Large Scale Structure, with
  cosmology and time independence in the UV}},  {\em JCAP} {\bf 1409} (2014),
  no.~09 047 [\href{http://arXiv.org/abs/1407.1342}{{\tt 1407.1342}}].
%%CITATION = ARXIV:1407.1342;%%

\bibitem{Peloso:2016qdr}
M.~Peloso and M.~Pietroni, {\it {Galilean invariant resummation schemes of
  cosmological perturbations}},  {\em JCAP} {\bf 1701} (2017), no.~01 056
  [\href{http://arXiv.org/abs/1609.06624}{{\tt 1609.06624}}].
%%CITATION = ARXIV:1609.06624;%%

\bibitem{Noda:2017tfh}
E.~Noda, M.~Peloso and M.~Pietroni, {\it {A Robust BAO Extractor}},
  \href{http://arXiv.org/abs/1705.01475}{{\tt 1705.01475}}.
%%CITATION = ARXIV:1705.01475;%%

\bibitem{Nishimichi:2017gdq}
T.~Nishimichi, E.~Noda, M.~Peloso and M.~Pietroni, {\it {BAO Extractor: bias
  and redshift space effects}},  {\em JCAP} {\bf 1801} (2018), no.~01 035
  [\href{http://arXiv.org/abs/1708.00375}{{\tt 1708.00375}}].
%%CITATION = ARXIV:1708.00375;%%

\bibitem{Pueblas:2008uv}
S.~Pueblas and R.~Scoccimarro, {\it {Generation of Vorticity and Velocity
  Dispersion by Orbit Crossing}},  {\em Phys.Rev.} {\bf D80} (2009) 043504
  [\href{http://arXiv.org/abs/0809.4606}{{\tt 0809.4606}}].

\bibitem{Carrasco:2012cv}
J.~J.~M. Carrasco, M.~P. Hertzberg and L.~Senatore, {\it {The Effective Field
  Theory of Cosmological Large Scale Structures}},  {\em JHEP} {\bf 1209}
  (2012) 082 [\href{http://arXiv.org/abs/1206.2926}{{\tt 1206.2926}}].
%%CITATION = ARXIV:1206.2926;%%

\bibitem{Valageas:2010rx}
P.~Valageas, {\it {Impact of shell crossing and scope of perturbative
  approaches in real and redshift space}},  {\em Astron. Astrophys.} {\bf 526}
  (2011) A67 [\href{http://arXiv.org/abs/1009.0106}{{\tt 1009.0106}}].
%%CITATION = 1009.0106;%%

\bibitem{McQuinn:2015tva}
M.~McQuinn and M.~White, {\it {Cosmological perturbation theory in 1+1
  dimensions}},  {\em JCAP} {\bf 1601} (2016), no.~01 043
  [\href{http://arXiv.org/abs/1502.07389}{{\tt 1502.07389}}].
%%CITATION = ARXIV:1502.07389;%%

\bibitem{Taruya:2017ohk}
A.~Taruya and S.~Colombi, {\it {Post-collapse perturbation theory in 1D
  cosmology -- beyond shell-crossing}},  {\em Mon. Not. Roy. Astron. Soc.} {\bf
  470} (2017), no.~4 4858--4884 [\href{http://arXiv.org/abs/1701.09088}{{\tt
  1701.09088}}].
%%CITATION = ARXIV:1701.09088;%%

\bibitem{McDonald:2017ths}
P.~McDonald and Z.~Vlah, {\it {Large-scale structure perturbation theory
  without losing stream crossing}},  {\em Phys. Rev.} {\bf D97} (2018), no.~2
  023508 [\href{http://arXiv.org/abs/1709.02834}{{\tt 1709.02834}}].
%%CITATION = ARXIV:1709.02834;%%

\bibitem{Pajer:2017ulp}
E.~Pajer and D.~van~der Woude, {\it {Divergence of Perturbation Theory in Large
  Scale Structures}},  \href{http://arXiv.org/abs/1710.01736}{{\tt
  1710.01736}}.
%%CITATION = ARXIV:1710.01736;%%

\bibitem{Rampf:2017jan}
C.~Rampf and U.~Frisch, {\it {Shell-crossing in quasi-one-dimensional flow}},
  {\em Mon. Not. Roy. Astron. Soc.} {\bf 471} (2017), no.~1 671--679
  [\href{http://arXiv.org/abs/1705.08456}{{\tt 1705.08456}}].
%%CITATION = ARXIV:1705.08456;%%

\bibitem{Gurbatov:1989az}
S.~N. Gurbatov, A.~I. Saichev and S.~F. Shandarin, {\it {The large-scale
  structure of the universe in the frame of the model equation of non-linear
  diffusion}},  {\em Mon. Not. Roy. Astron. Soc.} {\bf 236} (1989) 385--402.
%%CITATION = MNRAA,236,385;%%

\bibitem{Dubrulle:1994psg}
B.~Dubrulle, U.~Frisch, A.~Noullez and M.~Vergassola, {\it {The Mass Function
  In The Adhesion Model}},  in {\em {Clusters of Galaxies}}, (Gif-sur-Yvette),
  pp.~259--264, Edition Frontieres, Edition Frontieres, 1994.
%%CITATION = INSPIRE-1648333;%%

\bibitem{Bernardeau:2009ab}
F.~Bernardeau and P.~Valageas, {\it {Merging and fragmentation in the Burgers
  dynamics}},  {\em Phys. Rev.} {\bf E82} (2010) 016311
  [\href{http://arXiv.org/abs/0912.3603}{{\tt 0912.3603}}].
%%CITATION = ARXIV:0912.3603;%%

\bibitem{Valageas:2010uh}
P.~Valageas and F.~Bernardeau, {\it {Density fields and halo mass functions in
  the Geometrical Adhesion toy Model}},  {\em Phys. Rev.} {\bf D83} (2011)
  043508 [\href{http://arXiv.org/abs/1009.1974}{{\tt 1009.1974}}].
%%CITATION = ARXIV:1009.1974;%%

\bibitem{Aniceto:2018bis}
I.~Aniceto, G.~Basar and R.~Schiappa, {\it {A Primer on Resurgent Transseries
  and Their Asymptotics}},  \href{http://arXiv.org/abs/1802.10441}{{\tt
  1802.10441}}.
%%CITATION = ARXIV:1802.10441;%%

\bibitem{Blas:2011rf}
D.~Blas, J.~Lesgourgues and T.~Tram, {\it {The Cosmic Linear Anisotropy Solving
  System (CLASS) II: Approximation schemes}},  {\em JCAP} {\bf 1107} (2011) 034
  [\href{http://arXiv.org/abs/1104.2933}{{\tt 1104.2933}}].
%%CITATION = ARXIV:1104.2933;%%

\end{thebibliography}\endgroup
\end{document}